\documentclass[fleqn,10pt]{wlscirep}
\usepackage[utf8]{inputenc}
\usepackage{amsmath}
\usepackage[T1]{fontenc}
\usepackage{subcaption}
\usepackage{cite}

\title{Wideband High Gain Metasurface-Based 4T4R MIMO Antenna with Highly Isolated Ports for Sub-6 GHz 5G Applications}

\author[1*]{Mahdi Salehi}
\author[1]{Homayoon Oraizi}
\affil[1]{Iran University of Science and Technology, School of Electrical Engineering, Tehran, Iran}

\affil[*]{slehimahd@gmail.com}

\begin{abstract}
This study presents the design of four $178\times178$ $(mm)^{2}$ wideband, high gain, highly efficient metasurface-based 4T4R MIMO (Multiple-Input Multiple-Output) antennas with highly isolated ports, covering the middle and a portion of the upper bands of the sub 6 GHz 5G frequency spectrum for 5G-based systems, such as IoT (Internet of Things) applications, vehicular communications (e.g., rooftop antennas of cars or trains), smart industries (e.g., farms and factories). The radiating elements of these antennas use the aperture-coupled feeding technique with a dumbbell-shaped slot, a truncated square patch with two U-shaped slots, and a metasurface layer. The proposed MIMO structures place four identical radiating elements like a $2\times2$ matrix with $90^\circ$ successive rotations to produce orthogonal electromagnetic waves, improving the isolation between ports. Six-millimeter spaces are added between these elements, and two vertical and horizontal strip slots are carved on the ground as the decoupling structure to decrease the mutual coupling. Simulation results show that Antenna\_{1}, Antenna\_{2}, and Antenna\_{3} achieve gain values of 6.2 to 9.4 dBi, 8.2 to 11.6 dBi, 6.2 to 9.5 dBi, below -35, -25, and -33 isolation and almost 10 dB diversity gain from 2.8 to 4.7 GHz, 2.8 to 4.5 GHz, and 2.7 to 4.9 GHz, respectively. As a prototype, Antenna\_{4} is manufactured, and measurements are performed. It achieves 6.28 to 10.45 dBi gain values, below -23 dB isolation, and 0.001 envelope correlation coefficient over 2.7 to 4.3 GHz. The results confirm that the proposed MIMO antennas are compatible with the 5G essential requisites.
\end{abstract}

\begin{document}

\flushbottom
\maketitle
%
%
\thispagestyle{empty}


\section*{Introduction}

The need for reliable and secure high-speed data transfer (1 to 20 Gbps) with low latency (about one millisecond), low power, high capacity, high mobility, and data handovers without a break up are satisfied by applying the Fifth Generation Mobile Network (5G)\cite{ref1,ref2,ref3,ref4,ref5,ref6,ref7,ref8}. 5G can provide high-quality service and support almost one million devices per square kilometer, which causes significant improvements in IoT applications \cite{ref1,ref2,ref5}. These distinctions have caused the progressive usages of the 5G technology in smart homes, farms,cities, and factories, IoT and artificial intelligence (AI) systems, video streaming, video conferencing, monitoring systems, telepresence systems, remote healthcare networks, and autonomous vehicles.
The operational bandwidth of the 5G technology covers 600 MHz to 6 GHz and 24 to 28 GHz, which are higher and wider than 4G (600 MHz to 2.6 GHz), allowing to experience higher data rates \cite{ref1}. However, the 5G range is short and cannot infiltrate obstacles (e.g., walls, glasses, and trees), as well as 4G. Therefore, the 5G needs many 5G towers with short distances between to maintain its reliability. Many countries (e.g., China, Saudi Arabia, United Emirates, Japan, South Korea, European Unions, Canada, USA, Australia, Brazil, South Africa, Nigeria, United Kingdom, India, and Malaysia) offer their 5G services in the 3 to 4 GHz frequency band, which is the most crowded and favorite band among the possible frequency bands for 5G technology. However, the 24-28 GHz frequency band can provide higher data transfer rates due to its significantly larger bandwidth, although it is shorter and has significant penetration problems, which indeed necessitates the use of antennas with beam-scanning capabilities to recoup for the distance loss and guarantee the reliability of this band. It is worth mentioning that an antenna cannot increase the penetration rate, as it mainely depends on the wavelength, but can reduce the loss of the achievable penetration rate.
\par
High data rates, broad bandwidth, increased channel capacity, high data throughput, and less multi-path fading effects are fundamental requisites of the 5G technology, which cannot be met by the Single Input Single Output (SISO), and Single Input Multiple Output (SIMO) systems (e.g., patch array antenna) \cite{ref9,ref10,ref11,ref12,ref13,ref14,ref15,ref16,ref17}. Multiple Inputs Multiple Outputs (MIMO) technology is applied to solve these issues by providing multiple paths for sending and receiving data without increasing the input power \cite{ref18}, which allows the 5G technology to maintain its reliability and coverage area. These merits have made MIMO antennas indispensable to 5G applications like IoT and vehicular communications, as moving vehicles need multiple antenna elements with low latency and increased channel capacity to guarantee prominent connectivity \cite{ref19}. A MIMO antenna compatible with the 5G technology should achieve high gain, wide operational bandwidth, stable far-field radiation, high isolation, low ECC, and increased diversity gain. Isolation level, ECC, and DG determine how the radiating elements in a MIMO structure work independently. If the coupling among the radiating elements is high, the channel capacity and data rates depreciate significantly.

\par Until now, many studies have been done on MIMO antennas to overcome the above challenges.
In Refs.\citen{ref20,ref21,ref22,ref23}, four ultra-wideband (UWB) $2\times2$  MIMO antennas are proposed operating from 3.6 to 10 GHz, 2.7 to 12 GHz, 3.1 to 10.6 GHz, and 3.5 to 11 GHz, respectively. Antennas presented in Refs.\citen{ref20,ref21} spurn 5.3 to 6.4 GHz and 5.1 to 5.9 GHz by employing a mushroom shape Electromagnetic Band Gap (EBG) structure and a band-stop microstrip filter to the ground, respectively, to avoid interference with the Wireless Local Area Network (WLAN) applications. These studies achieve 1 to 7 dBi, 2.5 to 6 dBi, -3 to 4 dBi, and 3.5 to 5.7 dBi utmost gain values, almost 10 dB diversity gain, below -15, -17, -20, and -20 dB isolation, and less than 0.0015, 0.2, and 0.3 ECC values. In addition, Refs.\citen{ref22,ref23} attain 96\% and 70 to 90\% radiation efficiency. It is worth mentioning that Ref.\citen{ref22} employs decoupling structures for the ground and the radiating elements planes to achieve high isolation. In contrast, Ref.\citen{ref23} does not use any decoupling structures and only employs four octagonal patches, which are rotated consecutively by $90^\circ$ regarding each other to produce orthogonal polarization waves. Two UWB $1\times1$ MIMO antennas operating from 3 to 11.5 GHz are presented in Refs.\citen{ref24,ref25}. Ref.\citen{ref24} rejects the 3.3 to 3.9 GHz frequency band allotted to the Worldwide Interoperability for Microwave Access (WiMAX) by adding two folded stubs to the radiating elements, which are placed with a $180^\circ$ rotation angle. On the contrary, Ref.\citen{ref25} does not utilize any decoupling structure and attains high isolation by employing the asymmetric coplanar strip (ACS) feeding technique and adding two half-cut elliptical radiators. Refs.\citen{ref24,ref25} achieve below -18 and -15 dB isolation, almost 10 dB diversity gain, and ECC values below 0.0003 and 0.01 over their -10 dB impedance bandwidth, respectively. Besides, Ref.\citen{ref24} attains -3 to 4 dBi highest gain values over its operating frequency. Refs.\citen{ref26,ref27} put forward two narrow bandwidth MIMO antennas working from 1.66 to 2.17 GHz and 2.5 GHz, respectively. Ref.\citen{ref26} uses the defected ground plane technique and places its four elements with consecutive $180^\circ$ rotation shifts to achieve highly independent performance without using any decoupling structure. However, Ref.\citen{ref27} uses an EBG decoupling structure between the two elements in the opposite direction to attain very high isolation at 2.5 GHz. These studies achieve above 2.5 dBi peak gain values, 10 and 8.12 dB diversity gain, below 0.23 and 0.03 ECC values in their operational bandwidth. Moreover, Ref.\citen{ref26} achieves above 96\% radiation efficiency. Ref.\citen{ref28} gives a novel 5G MIMO antenna. The MIMO antenna innovatively uses only one radiating patch antenna excited by four ports to generate four isolated waves operating from 3.3 to 4.5 GHz with isolations better than -15 dB and an ECC value less than 0.3. In addition, the antenna achieves 6.1 to 7.5 dBi peak gain values and above 80\% radiation efficiency. For medical and WLAN applications, two $2\times2$ MIMO antennas using P-shaped monopole radiating elements operating at 2.4 GHz are proposed in Refs.\citen{ref29,ref30}. These studies use the defected ground technique and place the their elements at $0^\circ$, $90^\circ$, $180^\circ$, and $270^\circ$ rotation angles regarding each other to achieve -25 and -58.87 dB isolation, 2.4 and 2.84 dBi maximum gain values, 0.03 and 0.0054 ECC, and almost 10 dB diversity gain, respectively. Innovatively, the use of metal rims of a mobile phone as the radiating elements of an 8-port MIMO antenna is implemented in Ref.\citen{ref31}. The antenna operates from 3.4 to 3.7 GHz. It employs four dual-fed radiating elements, each consisting of a planer invented-F antenna (PIFA) and a loop antenna, which generate two nearly orthogonal radiation patterns, ensuring high isolation between ports. The antenna achieves less than -15 dB isolation, 0.3 ECC, and 50 to 68\% radiation efficiency.
 \par Ref.\citen{ref32} presents a triple band $2\times2$  MIMO antenna operating from 2.4 to 2.52 GHz, 3.66 to 4 GHz, and 4.62 to 5.54 GHz. The antenna attains very low mutual coupling by applying the defected ground technique and employing the coplanar feeding technique to feed four ring-shaped elements placed like a cross. Ref.\citen{ref33} uses two incomplete circular patch joints with two L-shaped stubs to achieve -10 dB impedance bandwidth covering 2.34 to 2.71 GHz and 3.72 to 5.1 GHz. It defects the ground and engraves a slot to achieve high isolation. Refs.\citen{ref32,ref33} attain almost 1 and 3.8 dBi highest gain values, below -30 and -18 dB isolation, nearly 10 dB diversity gain, less than 0.001 and 0.005 ECC, and 85 and 67\% radiation efficiency over their operational bandwidth, respectively. In Ref.\citen{ref34}, a metasurface-based $2\times2$ MIMO antenna working from 3.27 to 3.82 GHz is developed. The antenna employs a metasurface structure to enhance the performance of the radiating elements regarding gain, impedance bandwidth, and radiation efficiency. In addition, it uses a decoupling system consisting of slots, strips, and shorting vias to achieve less than -32 dB isolation, above 9.98 dB diversity gain, and less than 0.001 ECC. The antenna achieves almost 8.7 dBi peak gain and 92 to 96\% radiation efficiency. In Ref.\citen{ref35}, two $1\times2$ transparent MIMO antennas are designed. The first antenna employs two circular patches, placed side by side with separated ground planes, and achieves almost 1.83 dBi peak gain at its operational bandwidth from 4.65 to 4.97 GHz. The second antenna uses a common ground for the two circular patches, which are placed at $180^\circ$ rotation concerning each other and achieves nearly 1.65 dBi peak gain value from 4.67 to 4.94 GHz. Both antennas experience less than -15 dB isolation, 0.02 ECC, and almost 9.8 dB diversity gain. A $1\times2$ metasurface-based MIMO antenna, given in Ref.\citen{ref36}, achieves 3 to 4.1 dBi peak gain values from 3.7 to 4.3 GHz. In Ref.\citen{ref37}, a $1\times2$ flower-shaped MIMO antenna is proposed to operate from 3.296 to 5.962 GHz. The antenna uses two separated ground planes and a decoupling stub to achieve less than -50 dB isolation, 0.05 ECC, and above 9.8 dB diversity gain. In addition, the antenna achieves -1 to 6.22 dBi peak gain values and 42 to 85\% radiation efficiency in the operating frequency band.

Refs.\citen{ref38,ref39,ref40,ref41,ref42} propose decoupling structures to reduce the mutual coupling among radiating elements in SAR (Synthetic Aperture Radar) and MIMO systems. They confirm the capability of the proposed isolators by designing MIMO antennas, which embed the decoupling structures. Ref.\citen{ref38} uses a cross-shaped microstrip line with periodic circular slots as a metamaterial PBG (Photonic Band-gap) to boost the isolation between the radiating elements. Ref.\citen{ref39} applies a cross-shaped metasurface with meander-shaped slots to reduce the mutual coupling between antenna elements in a MIMO structure. A cross-shaped microstrip line with double outward E-shaped slots on each of its arms is introduced in Ref.\citen{ref40} as a metamaterial electromagnetic band-gap to lessen the mutual coupling among adjacent elements in a MIMO system. In Ref.\citen{ref41}, a fractal decoupler, comprised of four connected Y-shaped slots separated by two upturned T-shaped slots, is employed, which can reduce the mutual coupling between radiating elements of a MIMO system without having adverse effects on the bandwidth and gain. Ref.\citen{ref42}, carves two outward E-shaped slots in a rectangular patch with an open-ended $\frac{\lambda}{4}$ stub to maximize the isolation between the radiating elements of MIMO systems. Ref.\citen{ref43} puts forward a four-port MIMO antenna that achieves lower than -20 dB isolation and 0.017 ECC, above 68\% radiation efficiency, and 3 to 4 dBi peak gain values from 3.3 to 5 GHz and 8.9 to 9.2 GHz. The antenna applies the successive rotation technique and puts 5 mm spaces between its radiating elements to obtain high isolation between ports. Moreover, it uses a defective ground, which is connected with an I-shaped strip, to increase the bandwidth. A four-port MIMO antenna that radiates LP (Linearly Polarized) waves from 4.98 to 5.9 GHz and CP (Circularly Polarized) waves from 2.38 to 2.62 GHz compatible with 5G, WLAN, and Wi-Fi applications is proposed in Ref.\citen{ref44}. The antenna achieves below -20 dB isolation and 0.04 ECC, from 4 to 4.7 dBi gain, and nearly 85\% radiation efficiency. The antenna arranges its radiating elements in the mirror image mode with respect to each other, employs two Interlaced Lozenge structures (ILS), and carves a hexagon-shaped slot in the middle of the substrate to minimize the isolation between ports while using a connected ground. Another quad-port MIMO antenna is proposed in Ref.\citen{ref45}. The antenna puts 8 mm spaces between the radiating elements, which are arranged in the mirror image mode, and employs a modified plus-shaped structure as a band-stop filter to minimize the isolation. The antenna achieves below -18 dB isolation and 0.01 ECC, 70 to 78\% radiation efficiency, 2 to 3.5 dBi peak gain values, and 3-dB axial ratio bandwidth from 7.9 to 9.59 GHz.

Until now, most of the MIMO studies have proposed antennas for cell phones, WLAN, Wi-Fi, and medical applications, and rarely have worked on vehicular communications, such as rooftop antennas of cars and trains. As high-speed moving objects entail multiple send and receive paths with high data rate, high channel capacity, high signal quality, low multi-pass fading effects, and low loss in the achievable penetration range, designing MIMO antennas suitable with vehicular communications is challenging. The antenna must offer very high gain, very low ECC, very high isolation, very low CCL (Channel Capacity Loss), and high diversity gain over a wideband -10 dB impedance bandwidths to support the vehicular communications requirements. This study fulfills the present research gap by proposing MIMO structures suitable for 5G Vehicular communications, IoT applications, and smart industries (e.g., smart factories and farms).

This study presents a comprehensive design and simulation of four $178\times178$ ${mm}^2$ high gain, highly efficient metasurface-based 4T4R MIMO antennas with highly isolated ports working from 2.8 to 4.7 GHz (Antenna\_{1}) , 2.8 to 4.5 GHz (Antenna\_{2}), 2.8 to 4.8 (Antenna\_{3}), and 2.8 to 4.3 GHz (Antenna\_{4}), encompassing the middle and a segment of upper bands of the sub 6 GHz 5G spectrum where many technologies provide their 5G services. The proposed radiating element uses the aperture-coupled feeding technique with a dumbbell-shaped slot on the ground plane, a truncated rectangular patch with two U-shaped slots, and a metasurface layer. The proposed MIMO structure puts four identical radiating elements like a $2\times2$ matrix and rotates them at $0^\circ$, $90^\circ$, $180^\circ$, and $270^\circ$ angles to produce orthogonal electromagnetic waves to reduce the mutual coupling between the radiating elements. In addition, it puts 6 mm spaces between radiating elements and carves two strip slots on the ground as the decoupling structure to minimize the mutual coupling. Antenna\_{1} is made of two RO4003C dielectric layers with 1.5 mm height and an unconnected ground plane. The CST and HFSS simulation results show that Antenna\_{1} achieves 6.2 to 9.41 dBi and 6.55 to 9.7 dBi peak gain from 2.8 to 4.5 GHz, respectively. In addition, it obtains less than -35 dB isolation, almost 10 dB diversity gain, below and below 0.0002 ECC, and MEG values (Mean Effective Gain) between -8.5 and -6 dB over its -10 dB impedance bandwidth. However, this antenna suffers from undesired back lobes, which are killed by putting a $178\times178$ ${mm}^2$ reflector plane at 20 mm beneath Antenna\_{1}, creating Antenna\_{2}. Although the reflector plane increases the profile of the antenna in the simulation step, there is no need for it in reality, as most vehicles have metal bodies that work as reflectors for mounted MIMO antennas. Antenna\_{2}, like Antenna\_{1}, uses an unconnected ground plane. According to the CST and HFSS simulation results, it achieves 8.23 to 11.6 dBi and 9.58 to 11.7 dBi utmost gain values from 2.8 to 4.5 GHz, respectively. Besides, it attains below -25 dB isolation, 0.001 ECC, nearly 10 dB diversity gain, and MEG between -10 and -6 over its -10 dB impedance bandwidth. Because some applications need a MIMO antenna with a connected ground plane, Antenna\_{3} is designed, connecting the ground of Antenna\_{1} with three small rectangular metal connectors. Based on the CST simulation results, Antenna\_{3} achieves 6.2 to 9.5 dBi peak gain values. Furthermore, it obtains less than -33 dB isolation, 0.00016 ECC, almost 10 dB diversity gain, and MEG between -8 to -6 dB over its -10 dB bandwidth. As a prototype, Antenna\_{4} is manufactured and tested. It comprises two RO4350B dielectric layers with 0.508 mm height and achieves  6.28 to 10.45 dBi gain values at $(\theta=0, \phi=0)$ from 3 to 4 GHz, according to the measurement results. Moreover, it obtains below -23 dB isolation, 0.001 ECC, virtually 10 dB diversity gain, and MEG values between -9 to -6 from 2.8 to 4.3 GHz. These achievements underly reliable wireless communication with high data rate, low latency, increased channel capacity, high signal quality, low power, high data throughput, and low loss in the achievable penetration rate, which are the 5G essential requisites.

This research paper is classified in this way: Section 2 explains the design and presents the simulation results of a radiating element using HFSS and the time domain solver of CST. It applies the aperture-coupled feeding technique with a dumbbell-shaped slot engraved on the ground plane, a ring-shaped metasurface layer, and a truncated square patch with two U-shaped slots. Dimensions of the proposed radiating element are optimized to obtain the high gain and wideband performance using the CST optimizer tool. In addition, a circuit model to study the performance of the radiating element using the AWR software is provided. Section 3 creates a $2\times2$ matrix of identical radiating elements with successive $90^\circ$ rotations to develop a MIMO structure with reduced mutual coupling among the radiating elements. It also embeds two vertical and horizontal strip slots on the ground and 6 mm gaps between the radiating elements to maximize the isolation. Three MIMO antennas are presented in this section, dubbed Antenna\_1, Antenna\_{2}, and Antenna\_{3}, and their performances are elaborated. In the first step, Antenna\_{1} is designed, and its performance is analyzed. Because this antenna experiences undesired back lobes, a reflector plane is put at 20 mm beneath Antenna\_{1} to suppress the back lobes, creating Antenna\_{2} in the second step. Finally, as some applications entail a MIMO antenna with a connected ground plane, Antenna\_{3} connects the ground plane of  Antenna\_{1} by three small rectangular metal connectors. As a validation of the principle, a prototype made of two $178\times178$ ${mm}^2$ RO4350B dielectric layers with an air gap of 13 mm is manufactured and measured in section 4. The scattering parameters of Antenna\_{4} are measured from 2.5 to 4.5 GHz; its radiation patterns are measured at 3, 3.5, 3.8, and 4 GHz, and its gain values at ($\theta=0,\phi=0$) are measured from 3 to 4 GHz. The measurement results are in good agreement with the CST simulation results. This section also provides a thorough analysis of the measurement and simulation results and compares the achievements of the fabricated antenna with other studies, reflecting that the proposed MIMO antenna is among the best choices for 5G vehicular communications, smart industries, and IoT applications. Finally, section 5 offers the conclusion of this study.

\section*{Radiating Element Design}

\begin{figure}[!t]
\centering
	\begin{subfigure}[b]{0.4\textwidth}
		\includegraphics[width = \textwidth]{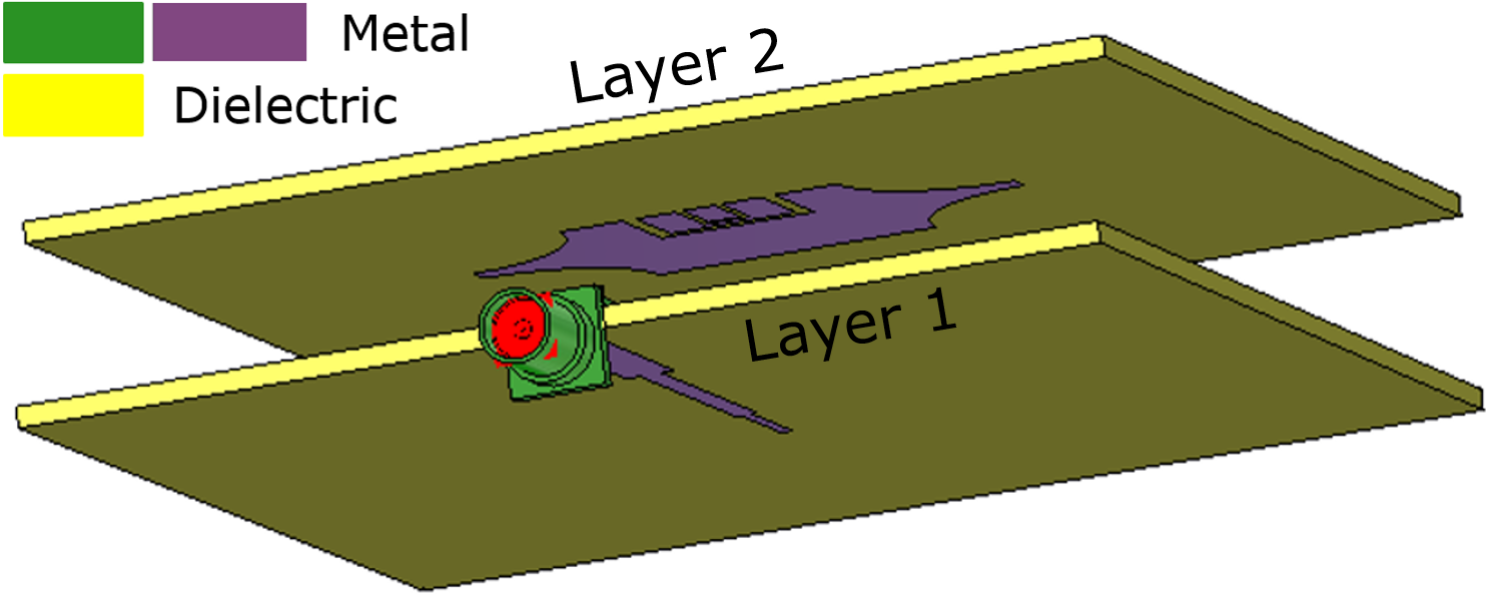}
		\caption{}
		\label{fig:fig1(a)}
	\end{subfigure}
	\begin{subfigure}[b]{0.4\textwidth}
		\includegraphics[width = \textwidth]{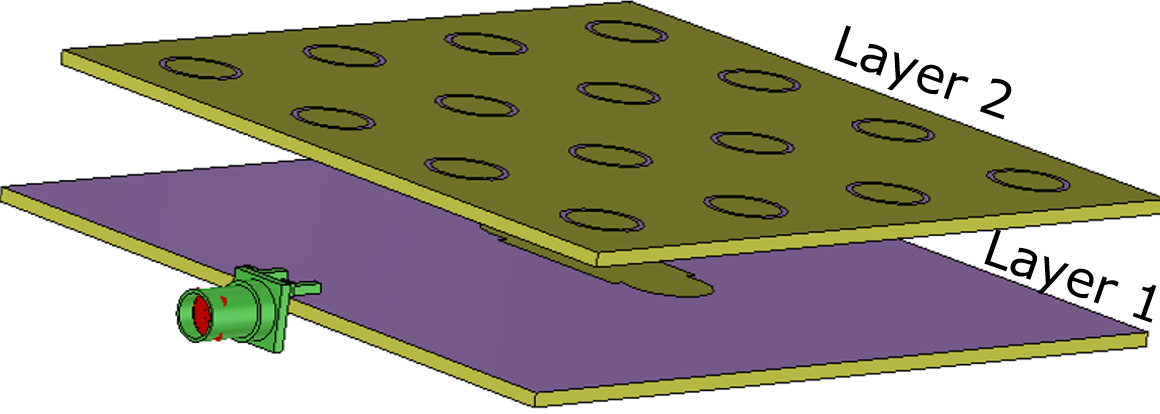}
		\caption{}
		\label{fig:fig1(b)}
	\end{subfigure}
	\begin{subfigure}[b]{0.5\textwidth}
		\includegraphics[width = \textwidth]{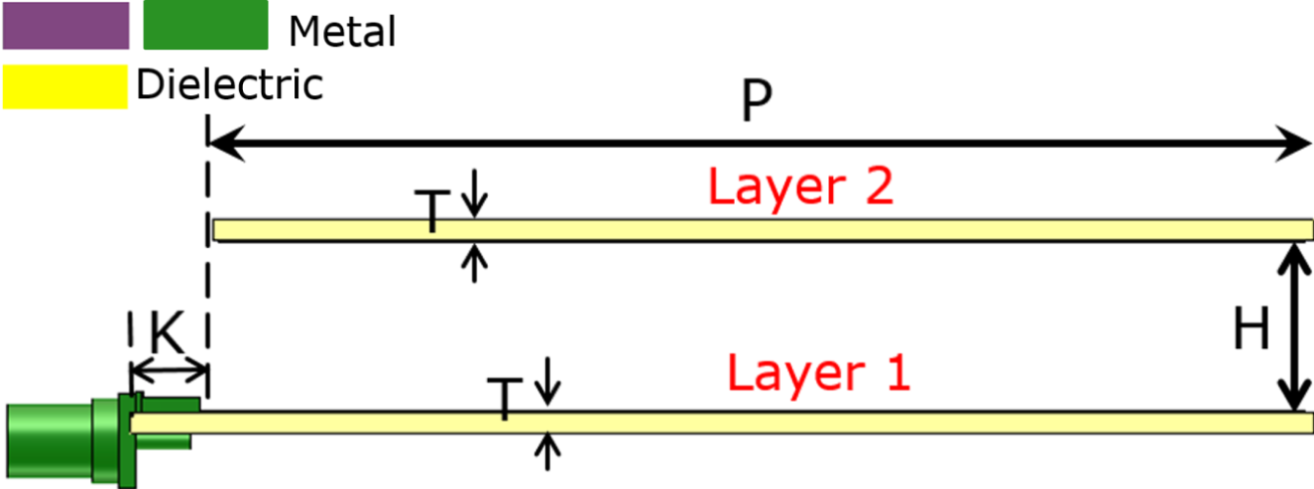}
		\caption{}
		\label{fig:fig1(c)}
	\end{subfigure}
	\begin{subfigure}[b]{0.25\textwidth}
		\includegraphics[width = \textwidth]{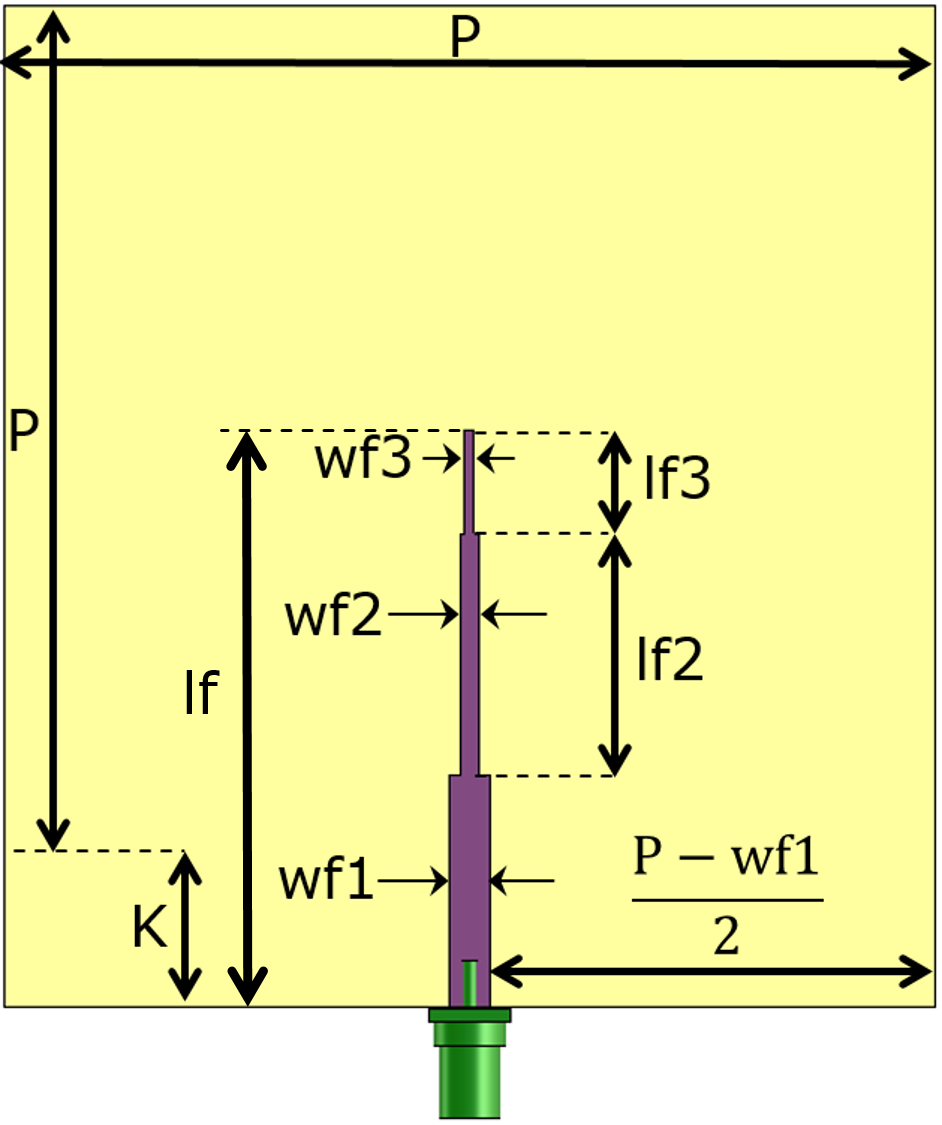}
		\caption{}
		\label{fig:fig1(d)}
	\end{subfigure}
	\begin{subfigure}[b]{0.25\textwidth}
		\includegraphics[width = \textwidth]{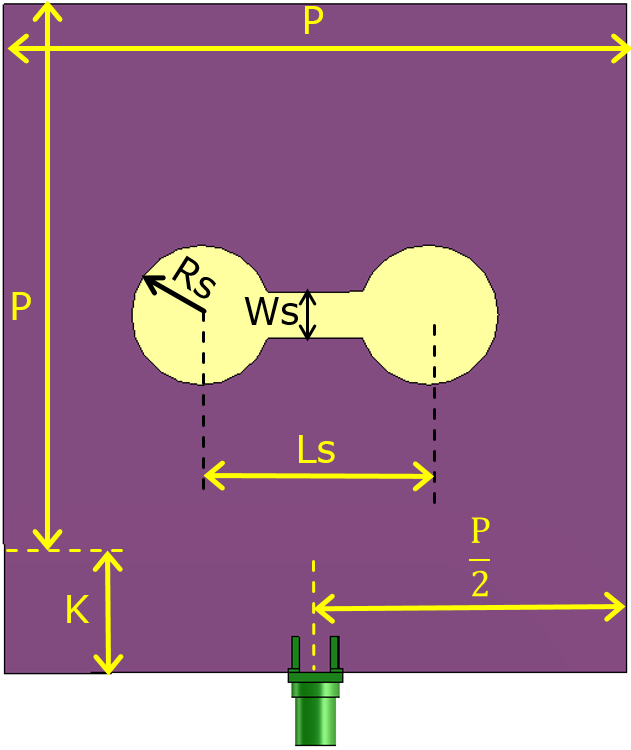}
		\caption{}
		\label{fig:fig1(e)}
	\end{subfigure}
	\begin{subfigure}[b]{0.3\textwidth}
		\includegraphics[width = \textwidth]{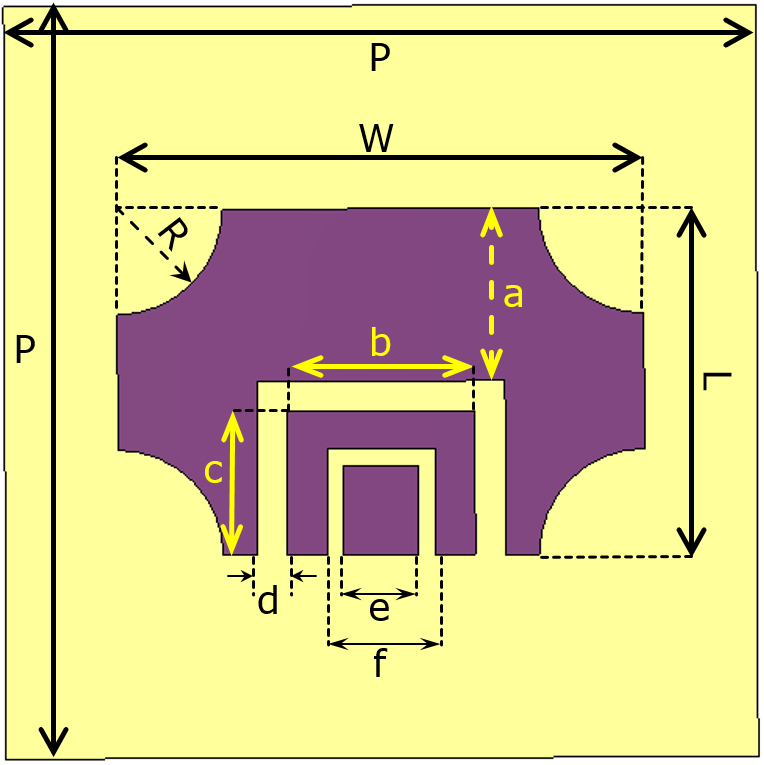}
		\caption{}
		\label{fig:fig1(f)}
	\end{subfigure}
	\begin{subfigure}[b]{0.3\textwidth}
		\includegraphics[width = \textwidth]{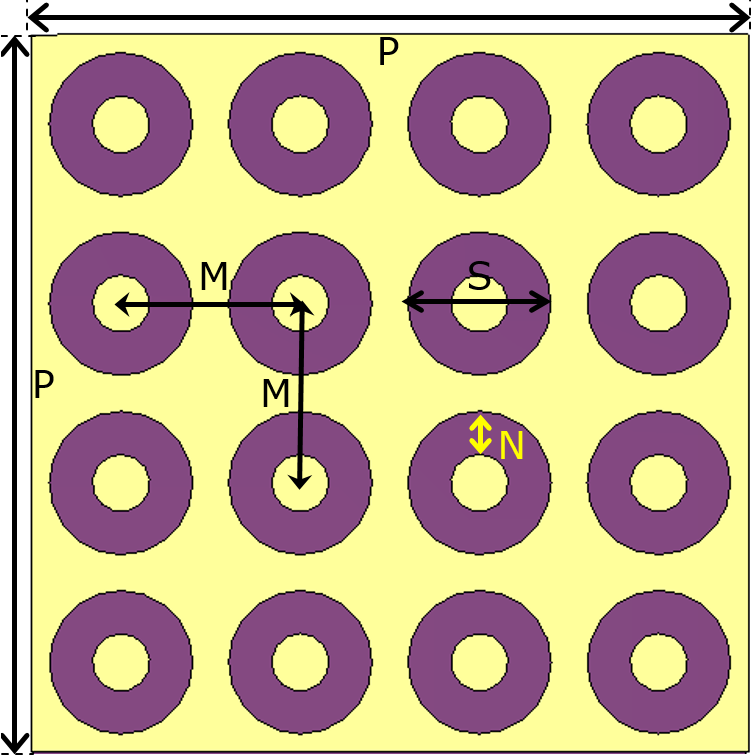}
		\caption{}
		\label{fig:fig1(g)}
	\end{subfigure}
\captionsetup{belowskip=0pt}
\caption{The perspective and schematic views of the radiating element are shown. (a) the truncated rectangular patch with two U-shaped slots and the feed line, (b) the metasurface layer and the dumbbell-shaped slot, (c) the flank view of the schematic, (d) the back face of Layer 1, (e) the front face of Layer 1, (f) the back face of Layer 2, (g) the front face of Layer 2. Note that the perspective views are displayed with realistic dimensions, but images of the schematic views are resized to be more discernable. Table.\ref{tab:Table1} provides the realistic dimensions of the radiating element determined by the parameters. Besides, the yellow parts represent the dielectric, and the purple and green parts represent the metal.}

\end{figure}
\begin{table*}[b]
\caption{\label{tab:Table1} The values of the parameters displayed in Figs.{\ref{fig:fig1(c)}-\ref{fig:fig1(g)}}.}
\centering
 \scalebox{0.8}{
\begin{tabular}{lllllllllllll}
\hline
H (mm) & T (mm) & P (mm) & K (mm) & Wf1 (mm) & Wf2 (mm) & Wf3 (mm) & lf (mm) & lf2 (mm) & lf3 (mm) & Ws (mm) & Ls (mm)& Rs (mm)\\
\hline
12.5 & 1.524 & 80 & 6 & 3.575 & 1.5317 & 1.476& 49.591 & 18.21 & 0.614 & 11.712& 17.13 & 4.8767 \\
\hline
 R (mm) & W (mm) & L (mm) & a (mm) & b (mm) & c (mm) & d (mm) & e (mm) & f (mm) & M (mm) & S (mm)& N (mm)& --\\
\hline
 7.425& 44.1467 & 18.0627 & 6.69 & 9.276 & 10.755 & 0.618 & 2.178 & 2.85& 20 & 8.621 & 0.623& -- \\
\hline
\end{tabular}
}
\end{table*}

\begin{figure}[!t]
\centering
	\begin{subfigure}[b]{0.215\textwidth}
		\includegraphics[width = \textwidth]{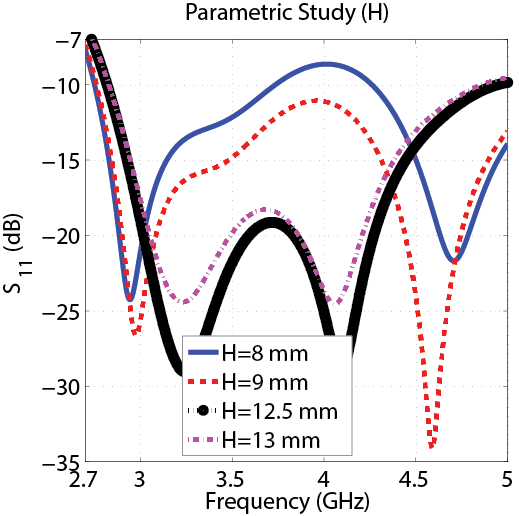}
		\caption{}
		\label{fig:fig2(a)}
	\end{subfigure}
	\begin{subfigure}[b]{0.215\textwidth}
		\includegraphics[width = \textwidth]{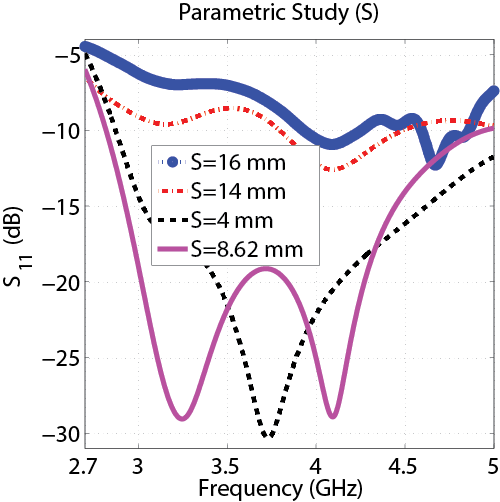}
		\caption{}
		\label{fig:fig2(b)}
	\end{subfigure}
	\begin{subfigure}[b]{0.215\textwidth}
		\includegraphics[width = \textwidth]{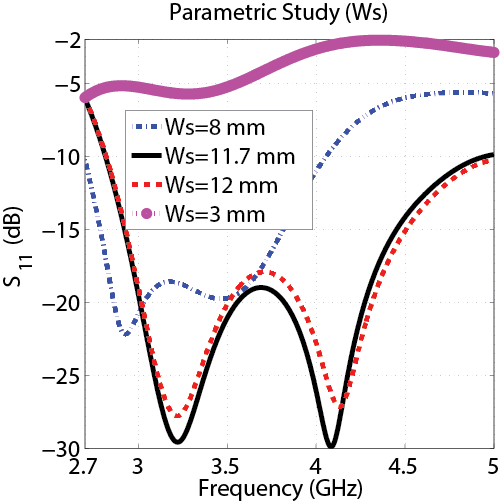}
		\caption{}
		\label{fig:fig2(c)}
	\end{subfigure}
	\begin{subfigure}[b]{0.227\textwidth}
		\includegraphics[width = \textwidth]{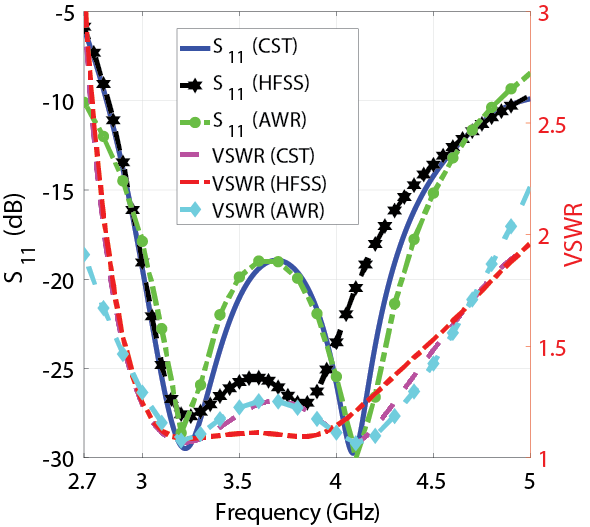}
		\caption{}
		\label{fig:fig2(d)}
	\end{subfigure}
	\begin{subfigure}[b]{0.22\textwidth}
		\includegraphics[width = \textwidth]{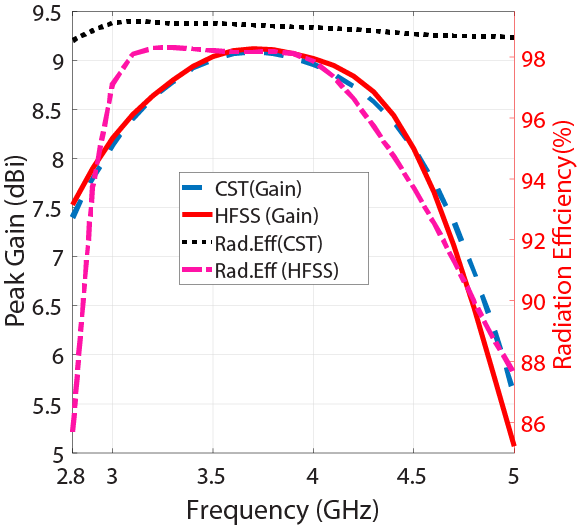}
		\caption{}
		\label{fig:fig2(e)}
	\end{subfigure}
	\begin{subfigure}[b]{0.22\textwidth}
		\includegraphics[width = \textwidth]{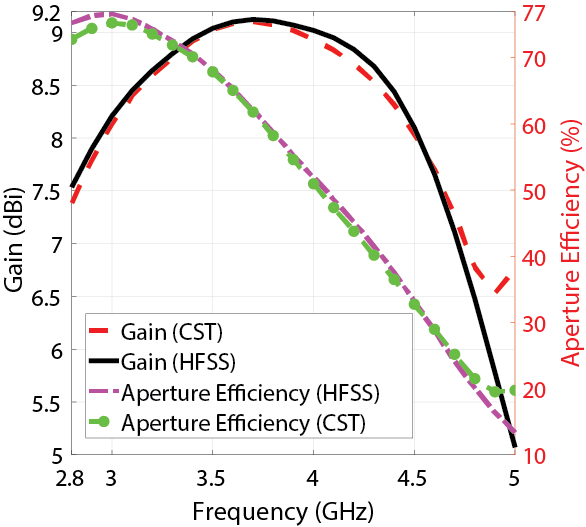}
		\caption{}
		\label{fig:fig2(f)}
	\end{subfigure}
	\begin{subfigure}[b]{0.25\textwidth}
		\includegraphics[width = \textwidth]{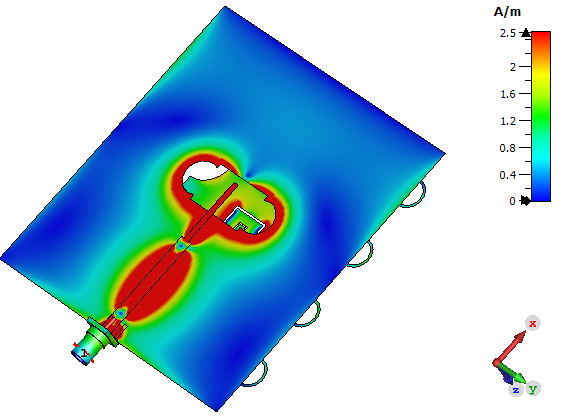}
		\caption{}
		\label{fig:fig2(g)}
	\end{subfigure}
	\begin{subfigure}[b]{0.25\textwidth}
		\includegraphics[width = \textwidth]{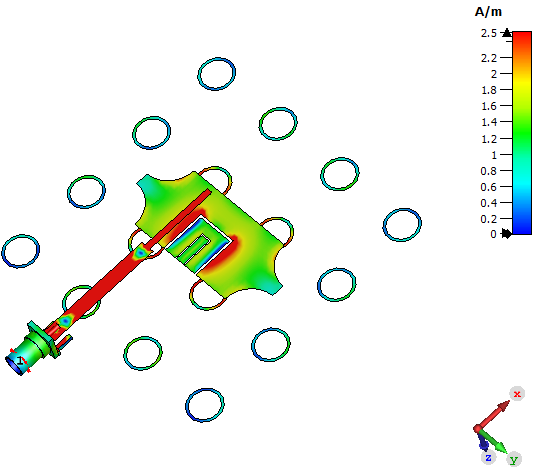}
		\caption{}
		\label{fig:fig2(h)}
	\end{subfigure}
\captionsetup{belowskip=0pt}
\caption{The CST and HFSS simulation results, (a) Simulation results  for various "H", (b) Simulation results for various "S", (c) Simulation results for various "Ws", (d) The simulation $S_{11}$ and VSWR versus frequency (CST, HFSS, and AWR), (e) The simulation peak gain values and the radiation efficiency in terms of the frequency, (f) The simulation peak gain values and aperture efficiency regarding frequency, (g) The surface current distribution of the radiating element from the feed line view, and (h) The surface current distribution from the bottom view (the substrates and the ground plane are hidden to increase the visibility.)}
\end{figure}

\begin{figure}[!t]
\centering
\includegraphics[width=1\linewidth]{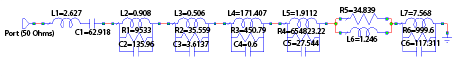}
\caption{The use of lumped elements to approximate the performance of the radiating element. Note that capacitances, resistors, and inductors are in Picofarad, Ohms, and Nanohenry, respectively.}
\label{fig:fig3}
\captionsetup{belowskip=0pt}
\end{figure}

\begin{figure}[!t]
\centering
	\begin{subfigure}[b]{0.2\textwidth}
		\includegraphics[width = \textwidth]{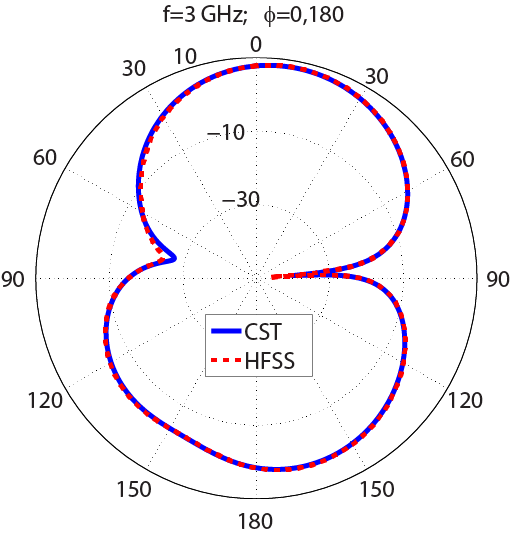}
		\caption{}
		\label{fig:fig4(a)}
	\end{subfigure}
	\begin{subfigure}[b]{0.20\textwidth}
		\includegraphics[width = \textwidth]{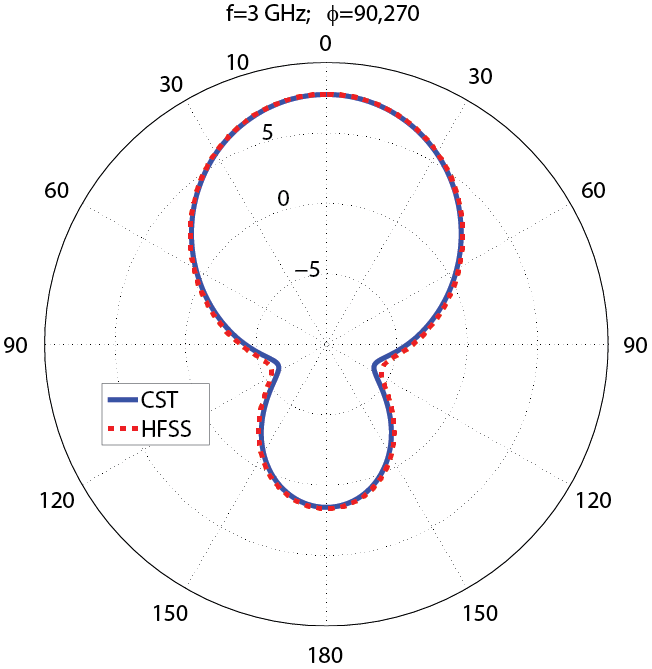}
		\caption{}
		\label{fig:fig4(b)}
	\end{subfigure}
	\begin{subfigure}[b]{0.2\textwidth}
		\includegraphics[width = \textwidth]{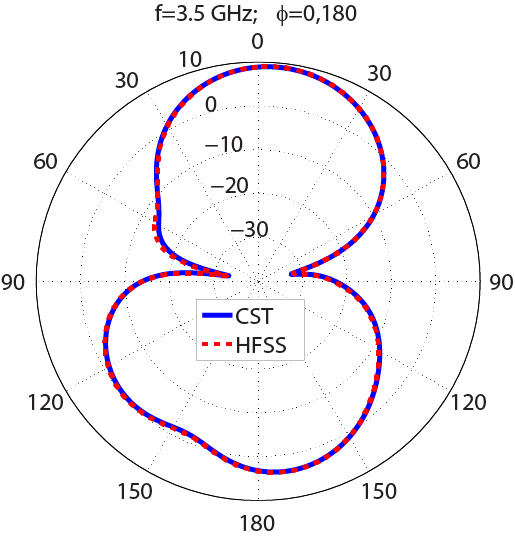}
		\caption{}
		\label{fig:fig4(c)}
	\end{subfigure}
	\begin{subfigure}[b]{0.2\textwidth}
		\includegraphics[width = \textwidth]{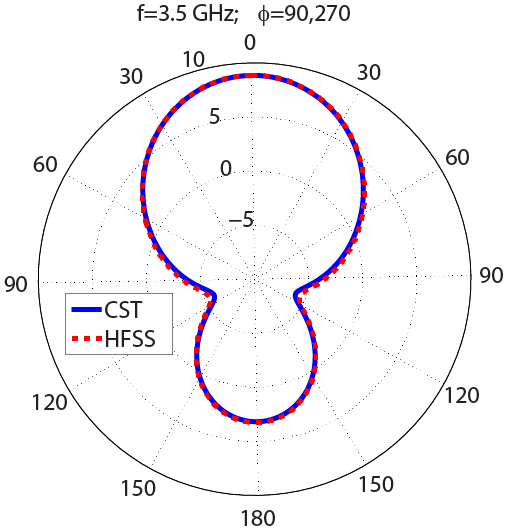}
		\caption{}
		\label{fig:fig4(d)}
	\end{subfigure}
	\begin{subfigure}[b]{0.2\textwidth}
		\includegraphics[width = \textwidth]{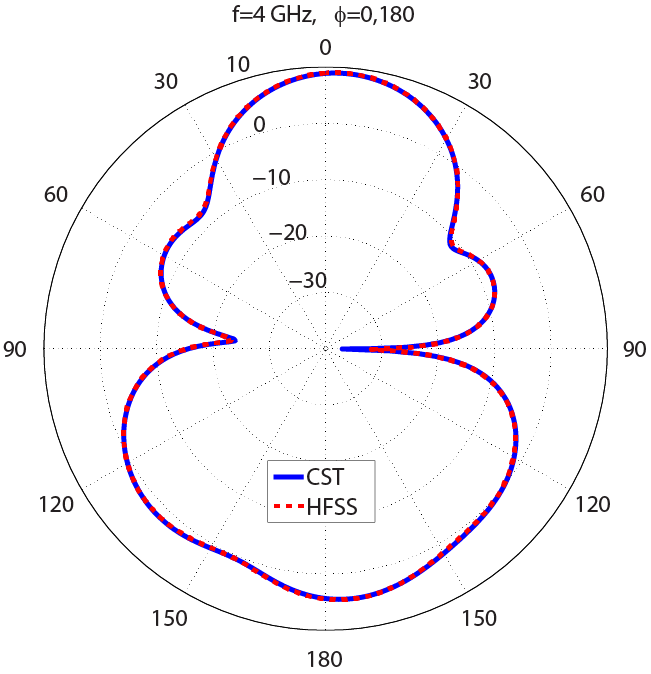}
		\caption{}
		\label{fig:fig4(e)}
	\end{subfigure}
	\begin{subfigure}[b]{0.2\textwidth}
		\includegraphics[width = \textwidth]{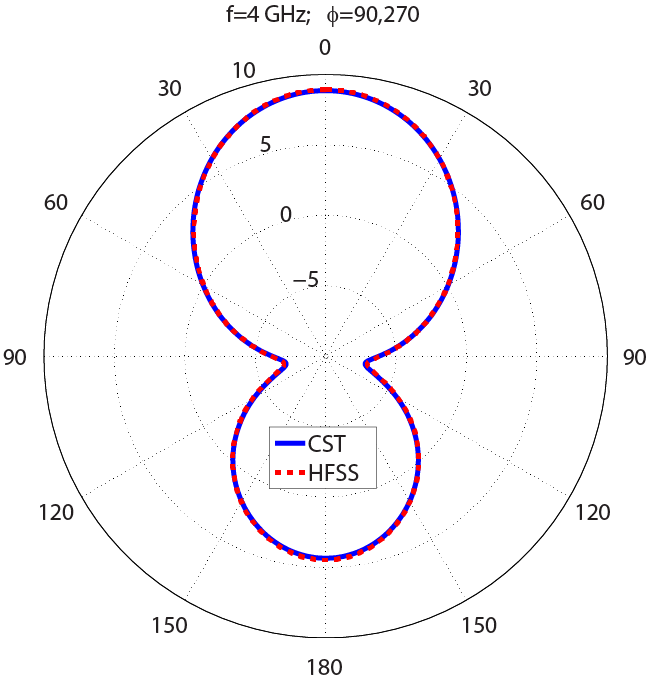}
		\caption{}
		\label{fig:fig4(f)}
	\end{subfigure}
\begin{subfigure}[b]{0.2\textwidth}
		\includegraphics[width = \textwidth]{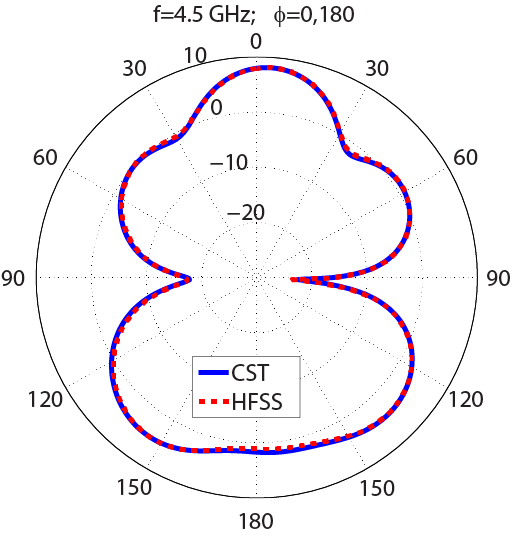}
		\caption{}
		\label{fig:fig4(g)}
	\end{subfigure}
	\begin{subfigure}[b]{0.2\textwidth}
		\includegraphics[width = \textwidth]{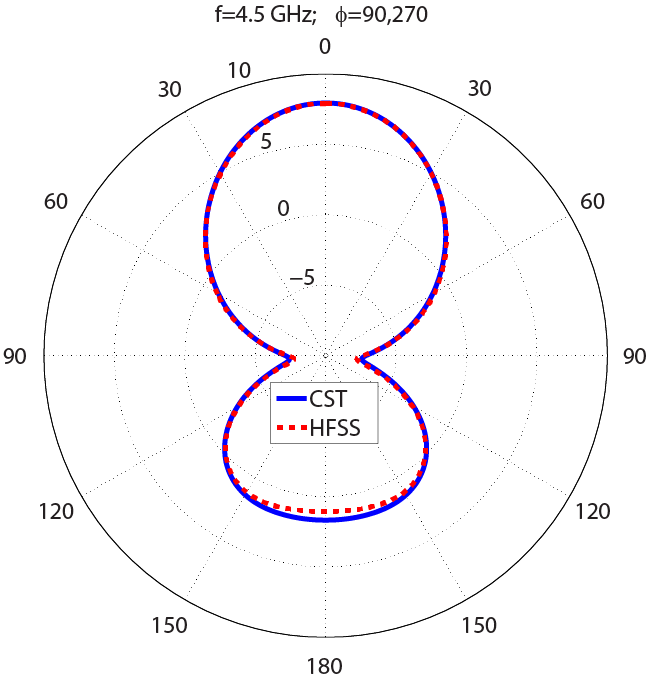}
		\caption{}
		\label{fig:fig4(h)}
	\end{subfigure}
\captionsetup{belowskip=0pt}
\caption{The CST and HFSS and H plane patterns of the radiating element at the following frequencies: (a) 3 GHz ($\phi$=$0^\circ$;$180^\circ$), (b) 3 GHz ($\phi$=$90^\circ$;$270^\circ$), (c) 3.5 GHz ($\phi$=$0^\circ$;$180^\circ$), (d) 3.5 GHz ($\phi$=$90^\circ$;$270^\circ$), (e) 4 GHz ($\phi$=$0^\circ$;$180^\circ$), (f) 4 GHz ($\phi$=$90^\circ$;$270^\circ$), (g) 4.5 GHz ($\phi$=$0^\circ$;$180^\circ$), and (h) 4.5 GHz ($\phi$=$90^\circ$;$270^\circ$).}
\end{figure}

  This section presents a radiating element with dimensions of $86\times80$ $(mm)^{2}$ that can be applied to create a MIMO antenna capable of satisfying the 5G requisites. The perspective views and schematic of the proposed radiating element are displayed in Figs.{\ref{fig:fig1(a)}-\ref{fig:fig1(g)}}. This structure uses the aperture-coupled feeding technique, a slotted truncated radiating patch, and a metasurface layer to guarantee wideband and high-gain performance. As shown in Figs.{\ref{fig:fig1(a)}-\ref{fig:fig1(c)}}, the radiating element consists of two dielectric layers made by Rogers 4003C ($\epsilon_{r}$=3.55, $tan(\gamma)$=0.0027, $t_{sub}=1.5 mm$), designated by layer 1 (feeding layer) and layer 2 (radiating layer). The feed line is printed on the back face of layer 1, and the ground plane with a dumbbell-shaped slot is printed on the front face of this layer, as seen in Figs.{\ref{fig:fig1(d)}-\ref{fig:fig1(e)}}. The ground plane not only provides the ground for the 50 ohms SMA connector but also isolates the feed network from the radiating layer, providing a nearly independent optimization process of two layers and decreasing the mutual coupling between the radiating elements in the MIMO designs. The truncated radiating patch with two U-shaped slots is on the back face of layer 2, and the metasurface, consisting of $4\times4$ ring elements, is on the top of this layer, as illustrated in Figs.{\ref{fig:fig1(f)}-\ref{fig:fig1(g)}}. The truncating rate "R" is used to tune the impedance of the rectangular patch, facilitating the impedance matching. In addition, the height of the air gap between layers has a significant impact on the -10 dB impedance bandwidth. EM waves are coupled to the radiating layer through the dumbbell-shaped slot and illuminate the radiating patch, which employs two U-shaped slots to enhance the bandwidth by increasing the number of resonances. The rectangular patch radiates EM power toward the space and produces surface waves in layer 2, which induces the metasurface to resonate at a specific frequency, increasing the bandwidth. The frequency of these resonances can be tuned by changing the width and length of the rectangular patch, the length of the U-shaped slots, and the gap between them, as well as changing the number and diameter of the rings of the metasurface. The dimensions of the dumbbell-shaped slot are also very critical, as the slot manipulates the coupling power toward layer 2. It is worth mentioning that the dumbbell-shaped slot is chosen instead of the typical rectangular slot, providing better manipulation of the coupled energy and impedance matching by introducing an extra tuning parameter, "Rs," as depicted in Fig.\ref{fig:fig1(e)}. The values of the parameters shown in Figs.{\ref{fig:fig1(c)}-\ref{fig:fig1(g)}}, indicating the dimensions of the radiating element, are declared in Table.\ref{tab:Table1}. These values are obtained from the optimization process of the CST software using the PSO (Particle Swarm Optimization) and TRF (Trust Region Framework) to achieve the desired wideband and high-gain performance. It is useful to inspect some of the parameters of the radiating element. The simulation results obtained from varying the values of three crucial parameters: the air gap between layers ("H"), the width of the dumbbell-shaped slot ("Ws"), and the diameter of the $4\times4$ rings of the metasurface ("S"), are displayed in Figs.{\ref{fig:fig2(a)}-\ref{fig:fig2(c)}}. As seen in Figs.{\ref{fig:fig2(a)}-\ref{fig:fig2(c)}}, H=12.5, S=8.6, and Ws=11.7 mm are the best choices.
 HFSS and the time domain solver of CST simulate the optimized radiating element, applying the FEM (Finite Element Method) and the FIT (Finite Integration Technique) approaches, respectively. As reflected in Fig.{\ref{fig:fig2(d)}, the radiating element experiences below -10 dB $S_{11}$ and VSWR<2 from 2.83 to 4.95 GHz, including the most occupied and in-used frequency bands of the 5G spectrum (3 to 4 GHz and 4.5 to 5 GHz). The simulation results show that the radiating element achieves lower than -19 dB $S_{11}$ over 3 to 4.3 GHz. As seen in Fig.\ref{fig:fig2(e)}, the simulated peak gain values calculated by the CST and HFSS software vary from 5.64 to 9.09 dBi and 5.429 to 9.124 dBi, respectively, in the operational bandwidth (2.8-5 GHz). The obtained 8.1 to 9.1 dBi and 5.5 to 8.2 dBi utmost gain values from 3 to 4 GHz and 4.5 to 5 GHz accentuate the potential of designing a high-gain MIMO antenna for 5G applications. Moreover, Fig.\ref{fig:fig2(e)} shows that the radiating element attains between 86 to 98\% and above 98\% radiation efficiency from 2.8 to 5 GHz according to HFSS and CST simulation results, respectively. One of the crucial features of an antenna is its dimensions, which should be small to be compatible with many applications. However, according to $G=10\times{\log_{10}{{\frac{4\pi{A\eta}}{\lambda^{2}}}}}$, where "A" is the antenna aperture, "G" is the gain in (dB), and "$\eta$" is the aperture efficiency, the aperture must not be chosen so small that the antenna becomes unable to produce the desired gain Ref.\citen{ref46}. For example, if the goal is to obtain nearly 9 dBi gain by making an unrealistic assumption of $\eta=1$, the dimensions should be $0.64\lambda^{2}$. In reality, $\eta{<1}$, so the dimensions must be greater than $0.64\lambda^{2}$. The aperture efficiency of the proposed radiating element varies from 77 to 15\% and 77 to 50\% over 2.8 to 5 GHz and 3 to 4 GHz, respectively, as seen in Fig.\ref{fig:fig2(f)}. For instance, the antenna achieves 67\% aperture efficiency and 9 dBi utmost gain with dimensions equal to $0.9\lambda^{2}$ at 3.5 GHz. Overall, the proposed radiating element shows high aperture efficiency performance and its dimensions are correctly chosen to obtain peak gain values of around 9 dBi. Moreover, the surface current distribution of the proposed radiating element is shown in Figs.{\ref{fig:fig2(g)}-\ref{fig:fig2(h)}} at 3.6 GHz. Fig.{\ref{fig:fig2(g)}} provides the view from the feed line, and Fig.{\ref{fig:fig2(h)}} provides the view from the radiating patch and metasurface. Furthermore, the far-field gain polar diagrams of the radiating element for E and H planes at 3 GHz, 3.5 GHz, 4 GHz, and 4.5 GHz are provided in Figs.{\ref{fig:fig4(a)}-\ref{fig:fig4(h)}}. As seen in the figures, the CST and HFSS far-field gains follow each other. A circuit model consisting of six resonators to approximate the performance of the radiating element is provided in Fig.\ref{fig:fig3}. The AWR software is used to simulate the circuit. The achieved $S_{11}$ and VSWR agree well with the CST simulation results, according to Fig.{\ref{fig:fig2(d)}}. It is worth mentioning that the resonators values of the lumped elements are obtained from the optimization process of AWR software. As seen in Figs.{\ref{fig:fig2(d)}-\ref{fig:fig2(f)}} and Figs.{\ref{fig:fig4(a)}-\ref{fig:fig4(h)}} there are some differences
 between the CST and HFSS simulation results, which are mainly due to the application of different computation methods.

\section*{MIMO Antenna Design}

\begin{figure}[!b]
\centering
	\begin{subfigure}[b]{0.49\textwidth}
		\includegraphics[width = \textwidth]{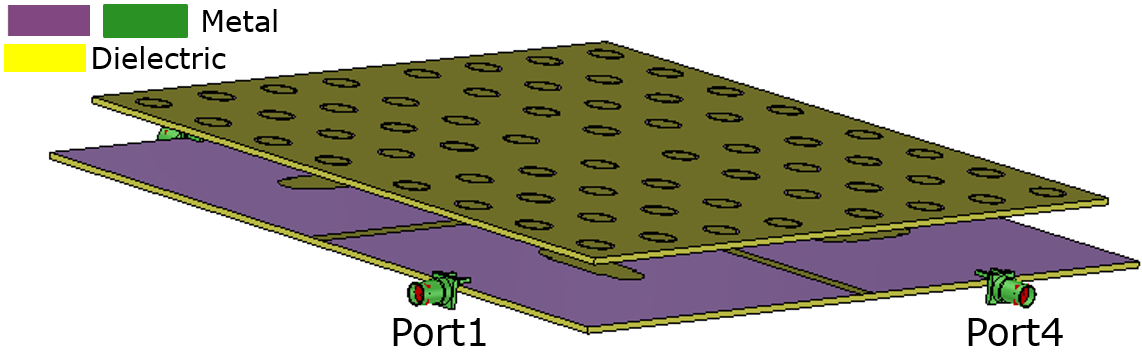}
		\caption{}
		\label{fig:fig5a}
	\end{subfigure}
	\begin{subfigure}[b]{0.49\textwidth}
		\includegraphics[width = \textwidth]{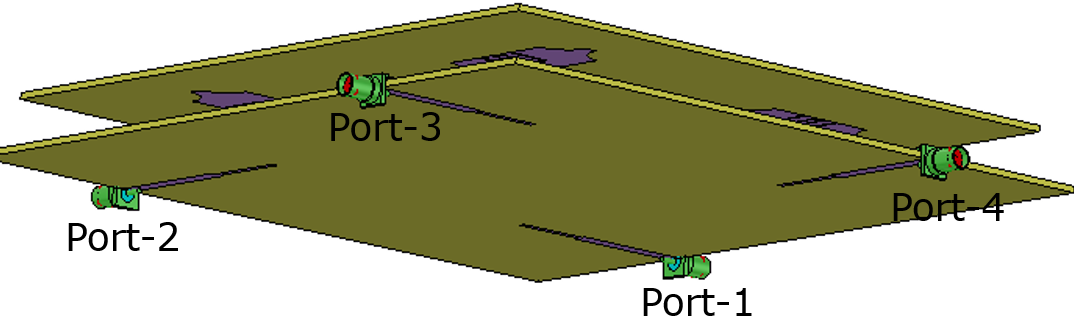}
		\caption{}
		\label{fig:fig5b}
	\end{subfigure}
\captionsetup{belowskip=0pt}
\caption{The MIMO structure without the reflector, (a) the perspective view, and (b) the perspective view. Note that the green and purple parts represent metal and the yellow parts represent the dielectric.}
\end{figure}

\begin{figure}[!t]
\centering
	\begin{subfigure}[b]{0.6\textwidth}
		\includegraphics[width = \textwidth]{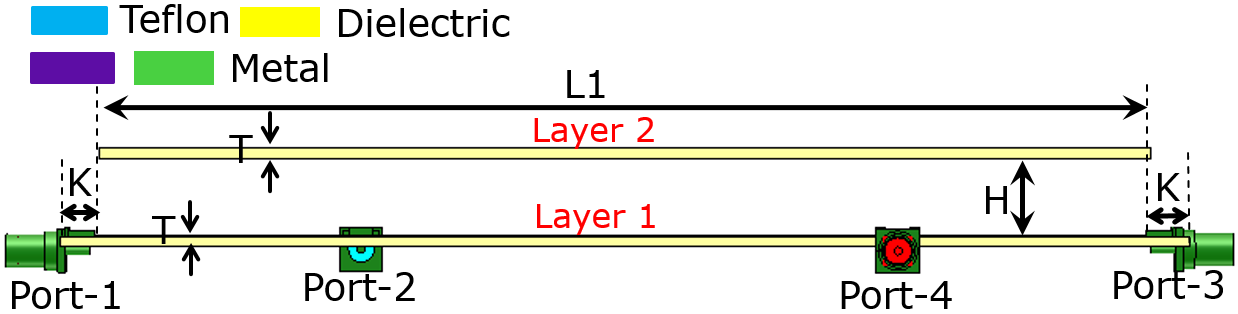}
		\caption{}
		\label{fig:fig6(a)}
	\end{subfigure}
	\begin{subfigure}[b]{0.3\textwidth}
		\includegraphics[width = \textwidth]{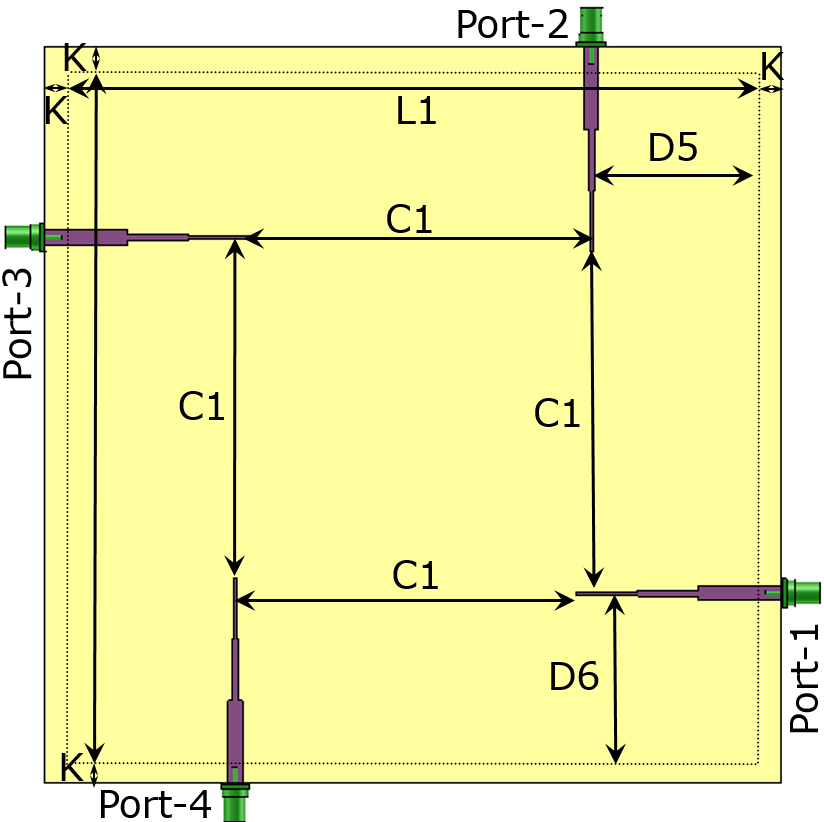}
		\caption{}
		\label{fig:fig6(b)}
	\end{subfigure}
	\begin{subfigure}[b]{0.32\textwidth}
		\includegraphics[width = \textwidth]{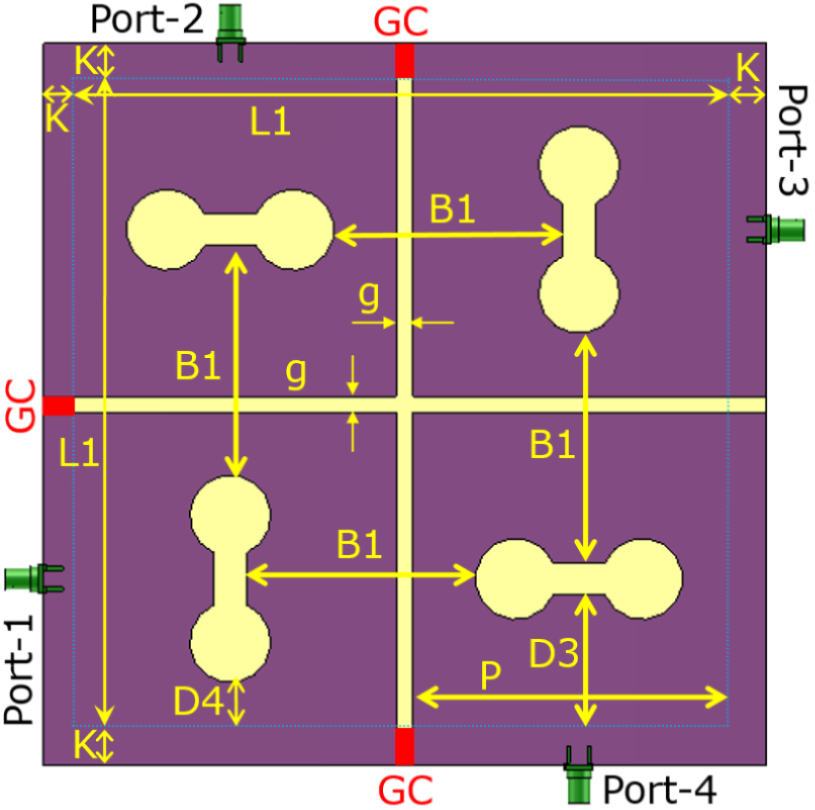}
		\caption{}
		\label{fig:fig6(c)}
	\end{subfigure}
	\begin{subfigure}[b]{0.30\textwidth}
		\includegraphics[width = \textwidth]{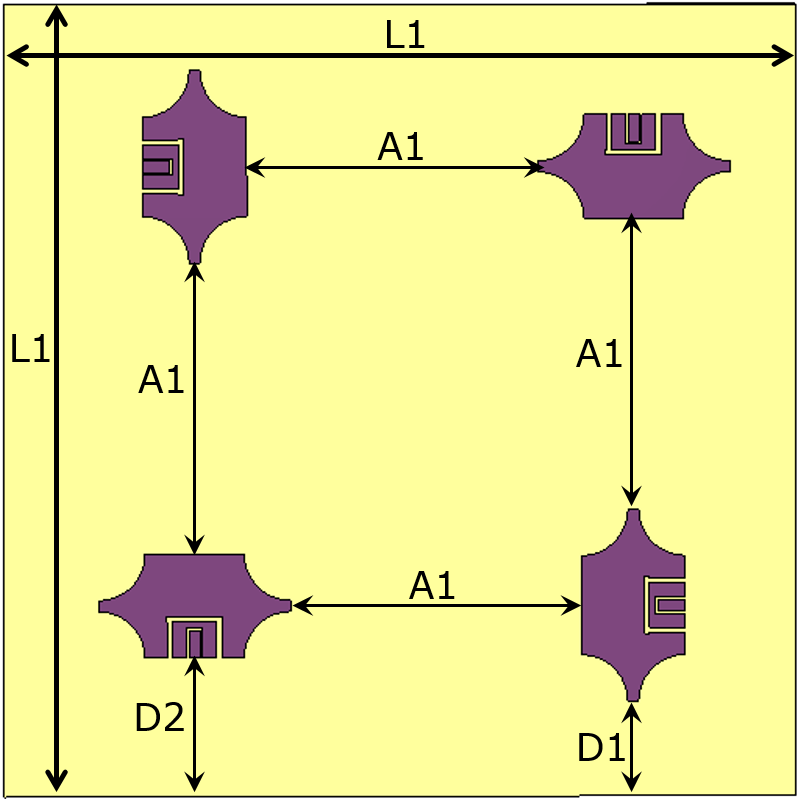}
		\caption{}
		\label{fig:fig6(d)}
	\end{subfigure}
	\begin{subfigure}[b]{0.32\textwidth}
		\includegraphics[width = \textwidth]{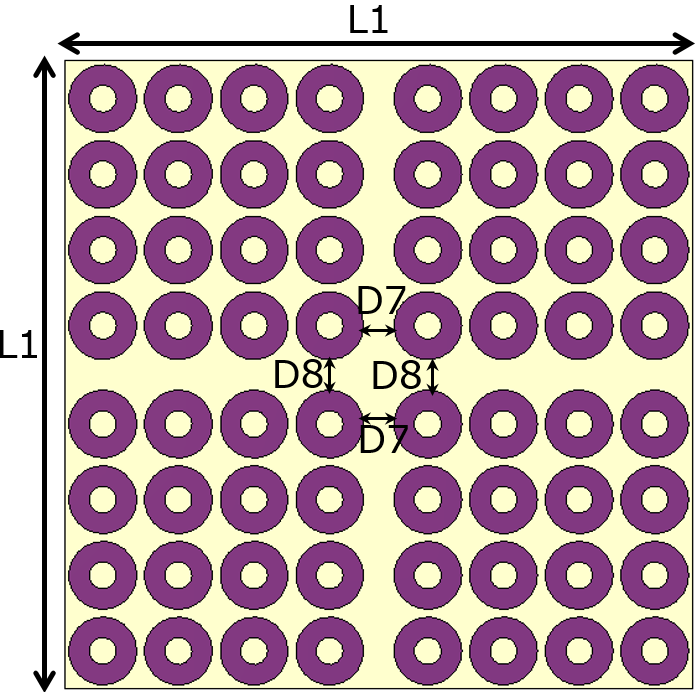}
		\caption{}
		\label{fig:fig6(e)}
	\end{subfigure}
\captionsetup{belowskip=0pt}
\caption{The schematic view of the MIMO antenna is displayed. (a) the flank view, (b) the back face of Layer 1, (c) the front face of Layer 1 (Note that the three small rectangular metal connectors at the head and tail of the vertical slot and the head of the horizontal slot, designated by "GC" and displayed in red color, are exclusively for Antenna\_{3} to create the connected ground.) (d) the back face of Layer 2, (e) the front face of Layer 2. Note that images are resized to increase visibility. The realistic dimensions of the MIMO antenna determined by the parameters are provided in Table.\ref{tab:Table2}. Besides, the purple and green parts represent metal, and the yellow and blue parts represent Rogers 4003C and Teflon, respectively.}
\end{figure}

This section develops a $2\times2$ MIMO antenna using the radiating element displayed in Figs.{\ref{fig:fig1(a)}-\ref{fig:fig1(b)}}. The three-dimensional (3D) and schematic views of the MIMO antenna are illustrated in Figs.{\ref{fig:fig5a}-\ref{fig:fig5b}} and Figs.{\ref{fig:fig6(a)}-\ref{fig:fig6(e)}}. As depicted in the figures, the radiating elements constructing the MIMO antenna are placed like a $2\times2$ matrix with $0^\circ$, $90^\circ$, $180^\circ$, and $270^\circ$ rotation angles to produce orthogonal electromagnetic waves, diminishing the coupling between elements. In addition, using the aperture-coupled feeding technique and embedding the decoupling structure (creating a space of 6 mm between the radiating elements and carving two horizontal and vertical slots with 4 mm widths on the ground plane of the MIMO structure), as depicted in Fig.\ref{fig:fig6(c)}, minimize the coupling effects. As seen in Figs.{\ref{fig:fig6(c)}}, three small rectangular metal connectors, designated with "GC" and displayed in red color, are placed at the head and tail of the vertical slot and the head of the horizontal slot on the ground. The role of these metal rectangular metal connectors is to connect the ground because some of the 5G applications need a MIMO antenna with a connected ground. However, Antenna\_1, Antenna\_2, and Antenna\_4 do not have a connected ground, and only Antenna\_{3} embeds these three rectangular metal connectors. Table.\ref{tab:Table2} provides values for the parameters shown in Figs.{\ref{fig:fig6(a)}-\ref{fig:fig6(e)}}, which indicate the dimensions of the MIMO antenna. The HFSS and the time domain solver of CST simulate Antenna\_{1} and Antenna\_{2}, but Antenna\_{3} and Antenna\_{4} are merely simulated by CST. According to Fig.\ref{fig:fig7(a)}, the $S_{11}$ of Antenna\_{1} is below -10 dB from 2.8 to 4.7 GHz and below -18 dB from 3 to 4.2 GHz. The CST and HFSS results show that the isolation between ports varies from -63 to -35 dB and -56 to -34 dB from 2.8, respectively, in the operational bandwidth (2.8 to 4.7 GHz). Therefore, the presented MIMO structure enjoys magnificent isolation levels between ports. It is worth mentioning that the symmetry of the proposed design leads to identical results for other ports, so it is redundant to show $S_{22}$, $S_{12}$, $S_{32}$, $S_{42}$, $S_{13}$, $S_{23}$, $S_{33}$, $S_{43}$, $S_{14}$, $S_{24}$, $S_{34}$, and $S_{44}$ in Fig.\ref{fig:fig7(a)}. Fig.\ref{fig:fig7(b)} depicts the maximum gain and radiation efficiency values as a function of the frequency. According to the CST and HFSS results, the highest gain values change from 5.97 to 9.41 dBi and 6.55 to 9.7 dBi by changing the frequency from 2.8 to 4.7 GHz, respectively, as seen in Fig.\ref{fig:fig7(b)}. The utmost gain values of the antenna run from 8.49 to 9.41 dBi and 6.23 to 6.3 dBi by changing the frequency from 3 to 4 GHz and 4.5 to 4.7 GH, respectively, according to CST results, and they go from 8.66 to 9.7 dBi and 6.54 to 6.8 dBi for 3 to 4 GHz and 4.5 to 4.7 GHz, respectively, according to HFSS results. In addition, the presented design obtains 86 to 98.2\% and above 98\% radiation efficiency from 2.8 to 4.7 GHz, according to HFSS and CST simulation results, respectively.

\begin{table*}[!t]
\caption{\label{tab:Table2} Assigning values to parameters displayed in Figs.{\ref{fig:fig6(a)}-\ref{fig:fig6(e)}}.}
\centering
 \scalebox{0.8}{
\begin{tabular}{lllllllll}
\hline
H (mm) & K (mm) & T (mm) & L1 (mm) & D1 (mm) & D2 (mm) & D3 (mm) & D4 (mm) & D5 (mm)\\
\hline
12.5 & 6 & 1.524 & 166 & 17.926 & 30.97 & 34.144 & 24.68 & 42.98\\
\hline
D6 (mm) & D7 (mm) & D8 (mm) & C1 (mm)& B1 (mm) & A1 (mm) & g (mm) & P (mm)& ---\\
\hline
42.98 & 17.37& 17.37& 81.67 & 64.82 & 54.895 & 4& 80& ---\\
\hline
\end{tabular}
}
\end{table*}

\par Diversity gain (DG) and Envelope Correlation Coefficient (ECC) are essential factors determining how independent the radiating elements of a MIMO antenna work. Due to the symmetry of the presented design, the simulation results of ECC and diversity gain for other ports are addressed. The CST results for ECC and DG are computed based on the far-field results. In contrast, the HFSS results are calculated according to the scattering parameters using Eqs.{\ref{eq1}-\ref{eq2}} Refs.\citen{ref47,ref48}. Obtaining less than $5\times10^{-5}$ ECC and nearly 10 dB DG from 3 to 4.7 GHz ensures that the radiating elements of the MIMO antenna work independently, and Fig.\ref{fig:fig7(c)} reflects it. It is beneficial to compute how much power a specific radiating element receives in a multi-way communication environment, which is called MEG or (Mean Effective Gain), and is obtained from Eq.{\ref{eq3}} Ref.\citen{ref48}. As illustrated in Fig.\ref{fig:fig7(d)}, $-8.5<MEG_{1}<-6$ from 2.7 to 4.7 GHz, which is within the acceptable range (-12 to -3 dB) Ref.\citen{ref48}. It is worth mentioning that the MEG values for all ports are the same due to the symmetry of the proposed structure. The presented 4T4R MIMO structure can significantly increase the channel capacity by providing four data streams. However, the mutual coupling between the radiating elements deteriorates the data transmission, which is determined by Channel Capacity Loss (CCL) and obtained from Eq.{\ref{eq4}} Refs.\citen{ref49,ref50}. As seen in Fig.\ref{fig:fig7(d)}, CCL is below 0.4 (bits/s/Hz) from 2.8 to 4.5 GHz, and for the rest of the band, it increases to 1.6 (bits/s/Hz). CCL below 0.4 indicates high-quality data transmission, and above 0.5 indicates low-quality data transmission, so Antenna\_{1} provides high-quality data transmission from 2.8 to 4.5 GHz Ref.\citen{ref48}. Another crucial parameter for a MIMO antenna is its effective operational bandwidth, which is called TARC (Total Active Reflection Coefficient) and is calculated from Eq.{\ref{eq6}}, where $\theta_{1}$, $\theta_{2}$, and $\theta_{3}$ are the phase differences between the excitation ports Ref.\citen{ref51}. As reflected in Fig.\ref{fig:fig7(e)}, the TARC of Antenna\_{1} is below -10 dB from 2.8 to 4.4 GHz and doesn't vary significantly for different phase differences. Overall, considering Figs.{\ref{fig:fig7(a)}-\ref{fig:fig7(e)}}, Antenna\_{1} has the best performance from 2.8 to 4.4 GHz in terms of -10 dB bandwidth, isolation, ECC, DG, TARC, MEG, and CCL. The E-field distribution of Antenna\_{1} is shown in Figs.{\ref{fig:fig7(f)}-\ref{fig:fig7(h)}}. Apparently, the isolation between the radiating elements is significantly high, and they work independently. The far field gain patterns at 3, 3.5, 4, and 4.5 GHz for E and H planes are illustrated in Figs.{\ref{fig:fig8(a)}-\ref{fig:fig8(h)}}. According to the Figs.{\ref{fig:fig8(a)}-\ref{fig:fig8(h)}} and Figs.{\ref{fig:fig7(a)}-\ref{fig:fig7(h)}}, the CST and HFSS simulation results are in good agreement. In addition, it is obvious that Antenna\_{1} suffers from undesired back lobes (3 to 4.75 dBi), which is due to the aperture-coupled feeding technique and must be suppressed.

\begin{figure}[!t]
\centering
	\begin{subfigure}[b]{0.2\textwidth}
		\includegraphics[width = \textwidth]{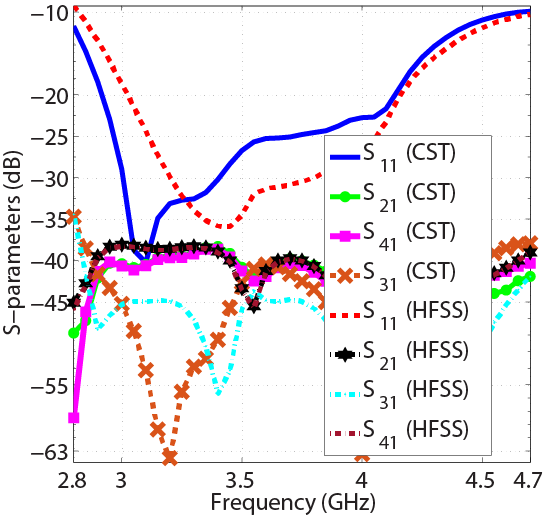}
		\caption{}
		\label{fig:fig7(a)}
	\end{subfigure}
	\begin{subfigure}[b]{0.21\textwidth}
		\includegraphics[width = \textwidth]{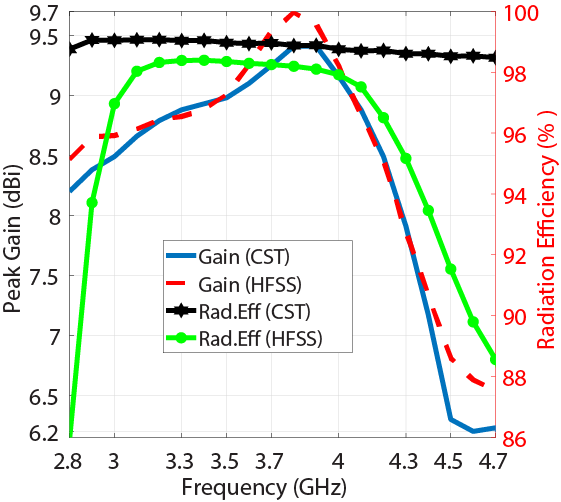}
		\caption{}
		\label{fig:fig7(b)}
	\end{subfigure}
	\begin{subfigure}[b]{0.225\textwidth}
		\includegraphics[width = \textwidth]{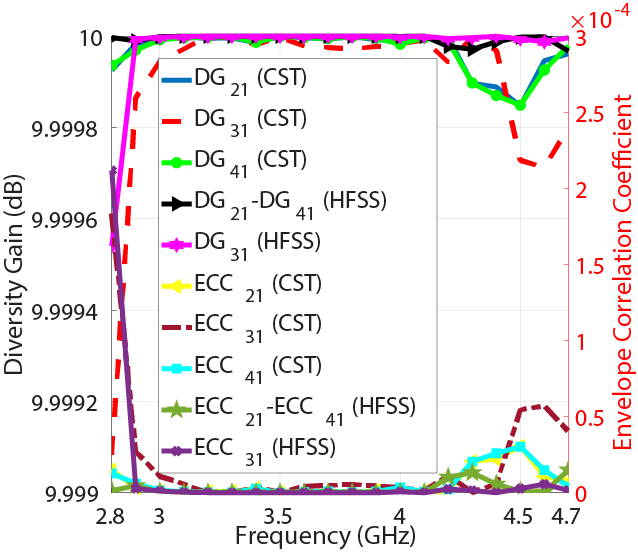}
		\caption{}
		\label{fig:fig7(c)}
	\end{subfigure}
	\begin{subfigure}[b]{0.22\textwidth}
		\includegraphics[width = \textwidth]{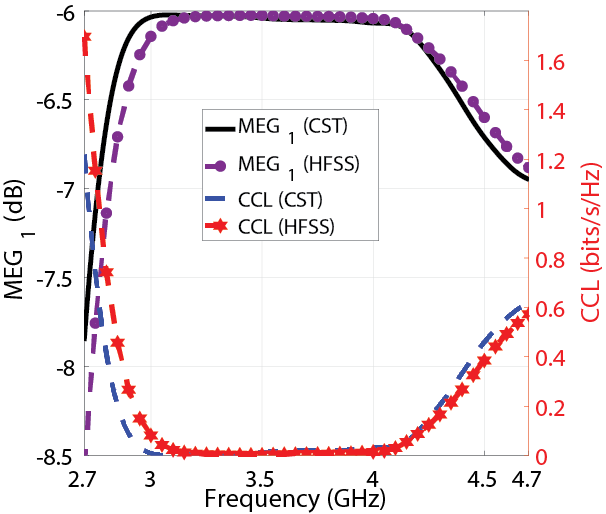}
		\caption{}
		\label{fig:fig7(d)}
	\end{subfigure}
	\begin{subfigure}[b]{0.2\textwidth}
		\includegraphics[width = \textwidth]{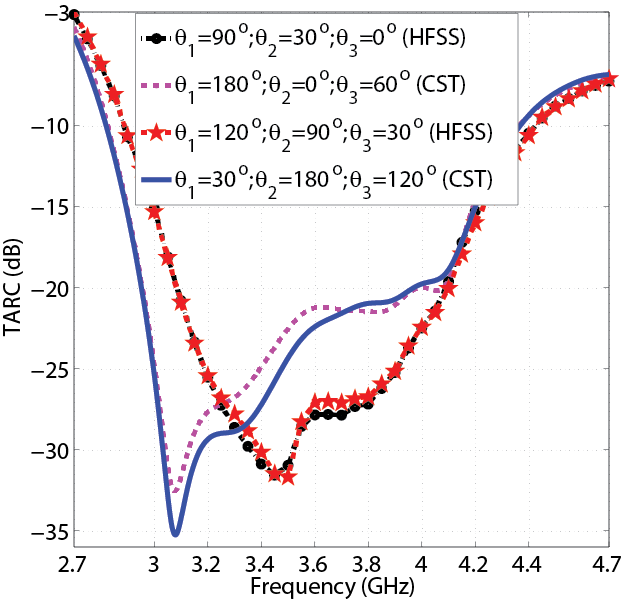}
		\caption{}
		\label{fig:fig7(e)}
	\end{subfigure}
	\begin{subfigure}[b]{0.22\textwidth}
		\includegraphics[width = \textwidth]{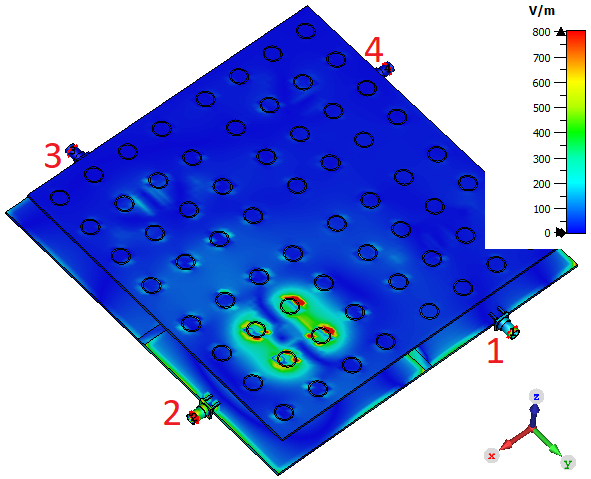}
		\caption{}
		\label{fig:fig7(f)}
	\end{subfigure}
	\begin{subfigure}[b]{0.23\textwidth}
		\includegraphics[width = \textwidth]{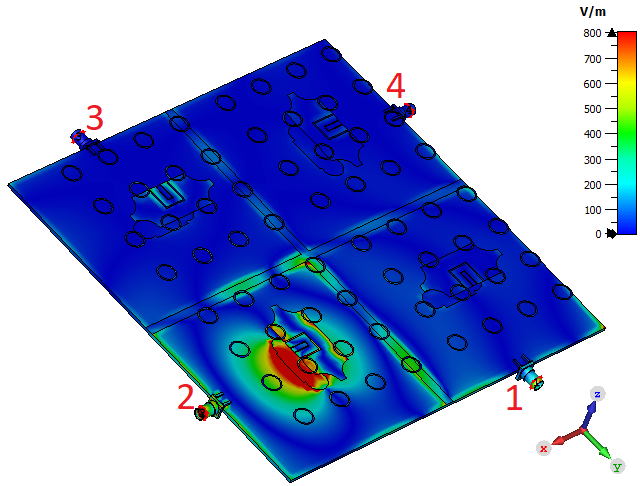}
		\caption{}
		\label{fig:fig7(g)}
	\end{subfigure}
	\begin{subfigure}[b]{0.24\textwidth}
		\includegraphics[width = \textwidth]{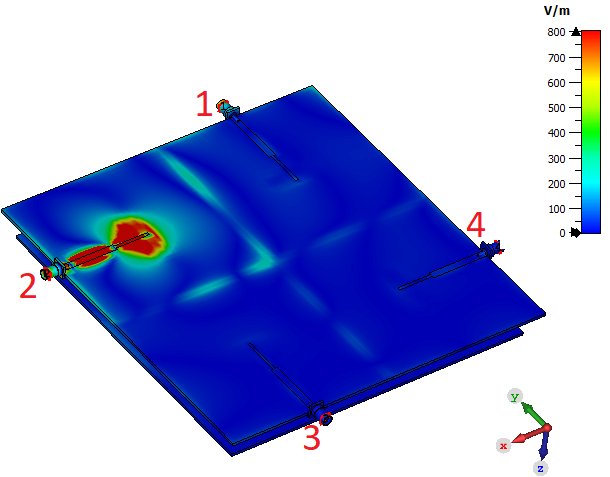}
		\caption{}
		\label{fig:fig7(h)}
	\end{subfigure}
\captionsetup{belowskip=0pt}
\caption{The CST and HFSS simulation results of Antenna\_{1}, (a) The scattering parameters regarding the frequency, (b) The maximum gain values and the radiation efficiency versus the frequency, (c) The ECC and diversity gain versus the frequency, (d) The $MEG_{1}$ and CCL concerning the frequency, (e) The TARC values in terms of the frequency, (f) E-field from the front view at 3.6 GHz, (g) E-field from the front view at 3.6 GHz (the top substrate is hidden to enhance the visibility), and (h) E-field from the feed line view at 3.6 GHz.}
\end{figure}

\begin{figure}[!b]
\centering
	\begin{subfigure}[b]{0.2\textwidth}
		\includegraphics[width = \textwidth]{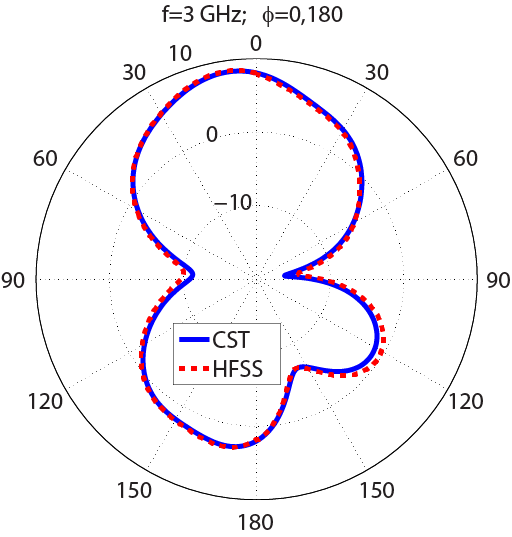}
		\caption{}
		\label{fig:fig8(a)}
	\end{subfigure}
	\begin{subfigure}[b]{0.205\textwidth}
		\includegraphics[width = \textwidth]{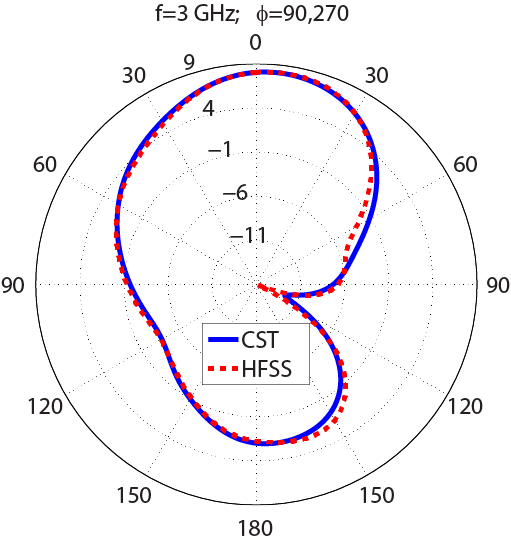}
		\caption{}
		\label{fig:fig8(b)}
	\end{subfigure}
	\begin{subfigure}[b]{0.21\textwidth}
		\includegraphics[width = \textwidth]{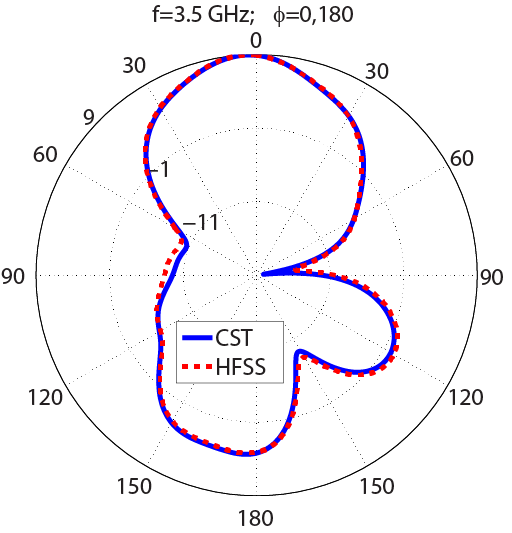}
		\caption{}
		\label{fig:fig8(c)}
	\end{subfigure}
	\begin{subfigure}[b]{0.21\textwidth}
		\includegraphics[width = \textwidth]{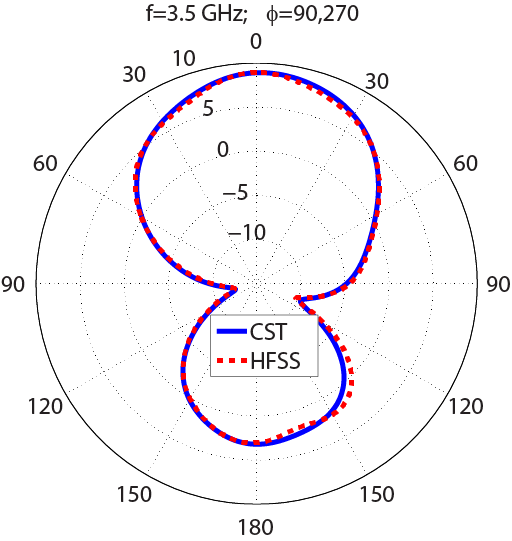}
		\caption{}
		\label{fig:fig8(d)}
	\end{subfigure}
	\begin{subfigure}[b]{0.2\textwidth}
		\includegraphics[width = \textwidth]{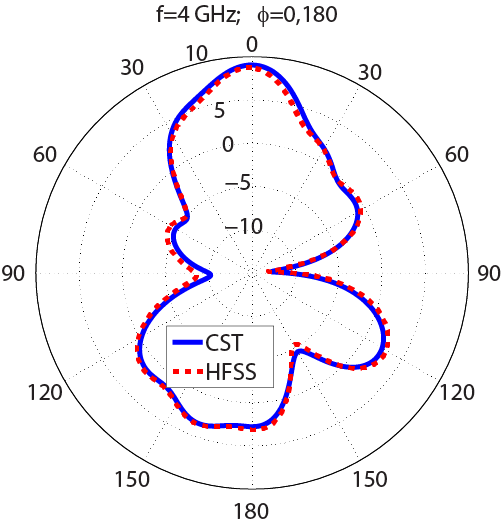}
		\caption{}
		\label{fig:fig8(e)}
	\end{subfigure}
	\begin{subfigure}[b]{0.2\textwidth}
		\includegraphics[width = \textwidth]{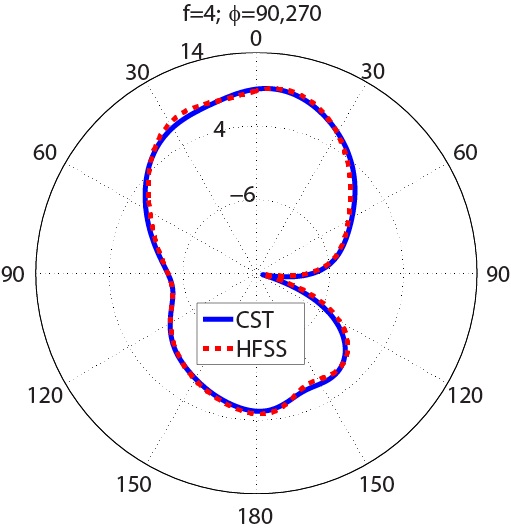}
		\caption{}
		\label{fig:fig8(f)}
	\end{subfigure}
\begin{subfigure}[b]{0.2\textwidth}
		\includegraphics[width = \textwidth]{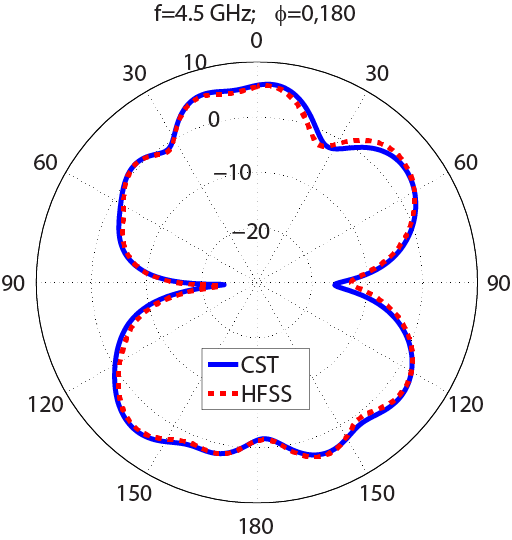}
		\caption{}
		\label{fig:fig8(g)}
	\end{subfigure}
	\begin{subfigure}[b]{0.2\textwidth}
		\includegraphics[width = \textwidth]{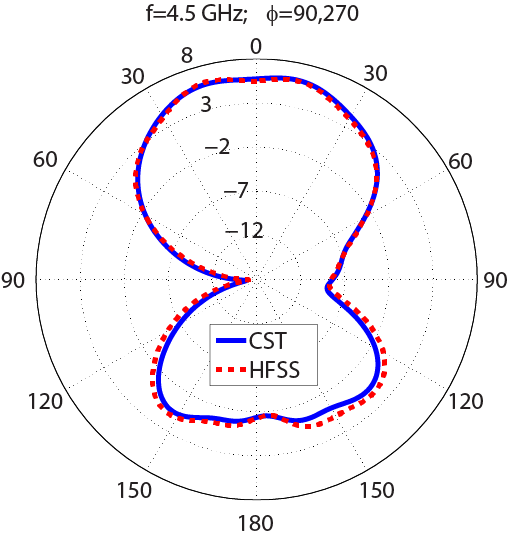}
		\caption{}
		\label{fig:fig8(h)}
	\end{subfigure}
\captionsetup{belowskip=0pt}
\caption{The CST and HFSS E and H plane patterns of the first MIMO antenna at the following frequencies: (a) 3 GHz ($\phi$=$0^\circ$;$180^\circ$), (b) 3 GHz ($\phi$=$90^\circ$;$270^\circ$), (c) 3.5 GHz ($\phi$=$0^\circ$;$180^\circ$), (d) 3.5 GHz ($\phi$=$90^\circ$;$270^\circ$), (e) 4 GHz ($\phi$=$0^\circ$;$180^\circ$), (f) 4 GHz ($\phi$=$90^\circ$;$270^\circ$), (g) 4.5 GHz ($\phi$=$0^\circ$;$180^\circ$), and (h) 4.5 GHz ($\phi$=$90^\circ$;$270^\circ$).}

\end{figure}

\begin{figure}[!b]
\centering
	\begin{subfigure}[b]{0.26\textwidth}
		\includegraphics[width = \textwidth]{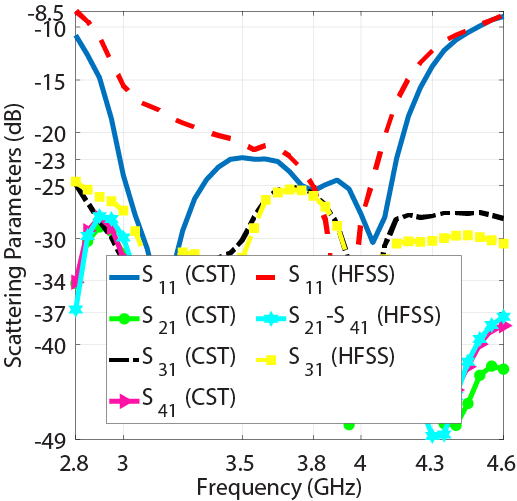}
		\caption{}
		\label{fig:fig9(a)}
	\end{subfigure}
	\begin{subfigure}[b]{0.29\textwidth}
		\includegraphics[width = \textwidth]{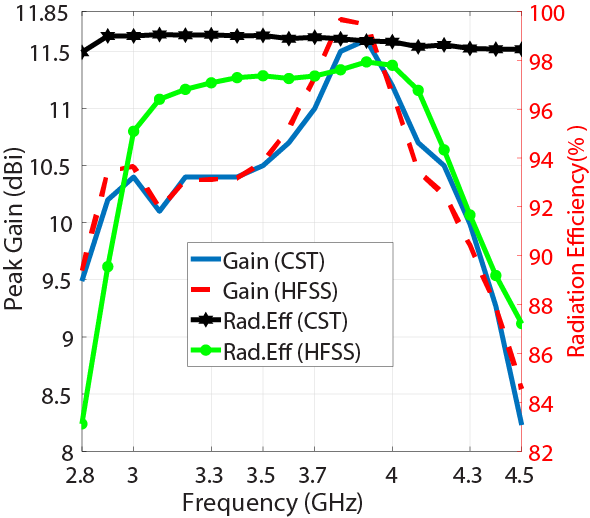}
		\caption{}
		\label{fig:fig9(b)}
	\end{subfigure}
	\begin{subfigure}[b]{0.315\textwidth}
		\includegraphics[width = \textwidth]{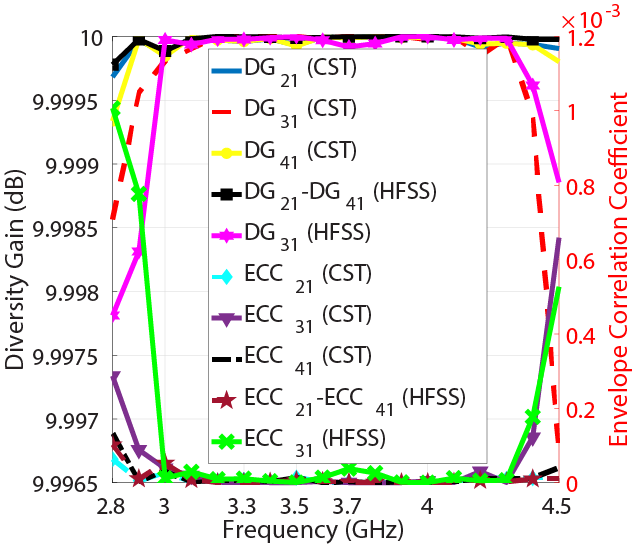}
		\caption{}
		\label{fig:fig9(c)}
	\end{subfigure}
	\begin{subfigure}[b]{0.29\textwidth}
		\includegraphics[width = \textwidth]{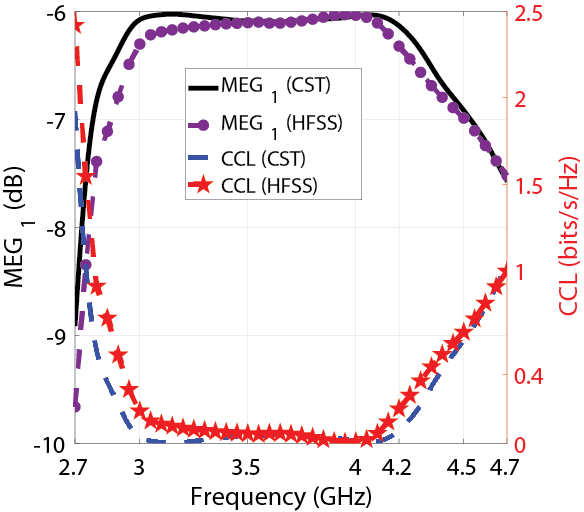}
		\caption{}
		\label{fig:fig9(d)}
	\end{subfigure}
	\begin{subfigure}[b]{0.26\textwidth}
		\includegraphics[width = \textwidth]{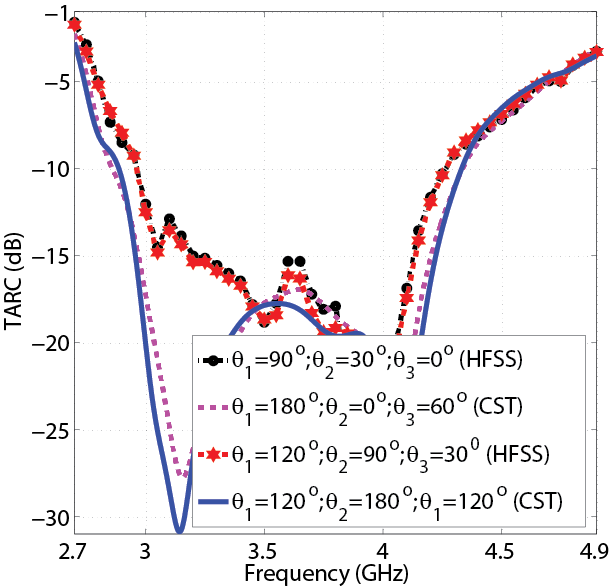}
		\caption{}
		\label{fig:fig9(e)}
	\end{subfigure}
	\begin{subfigure}[b]{0.26\textwidth}
		\includegraphics[width = \textwidth]{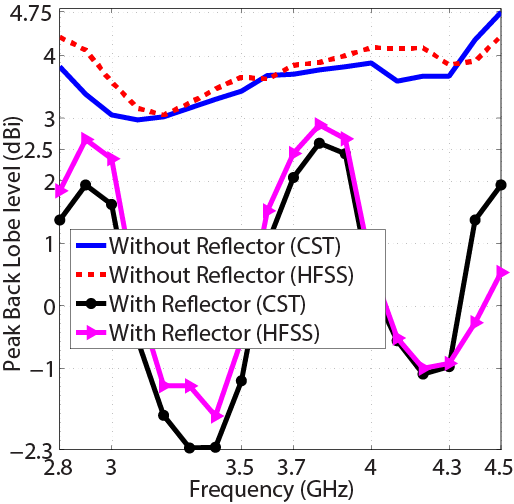}
		\caption{}
		\label{fig:fig9(f)}
	\end{subfigure}
	\begin{subfigure}[b]{0.29\textwidth}
		\includegraphics[width = \textwidth]{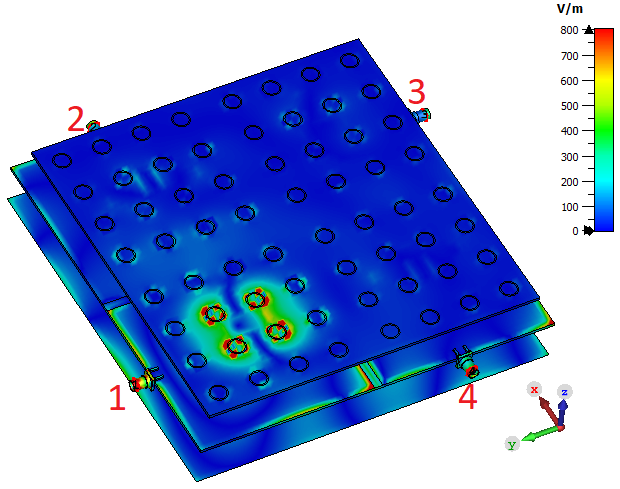}
		\caption{}
		\label{fig:fig9(g)}
	\end{subfigure}
	\begin{subfigure}[b]{0.29\textwidth}
		\includegraphics[width = \textwidth]{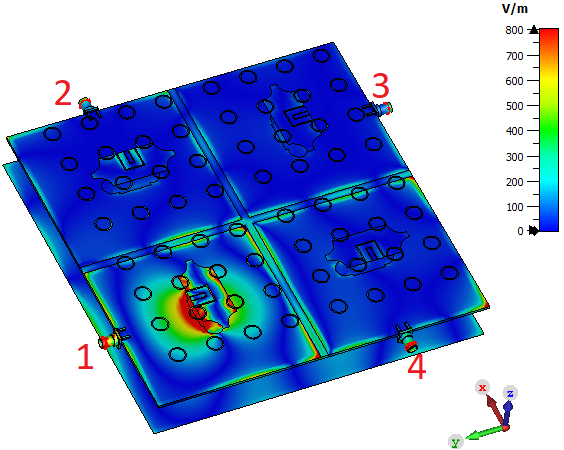}
		\caption{}
		\label{fig:fig9(h)}
	\end{subfigure}
	\begin{subfigure}[b]{0.29\textwidth}
		\includegraphics[width = \textwidth]{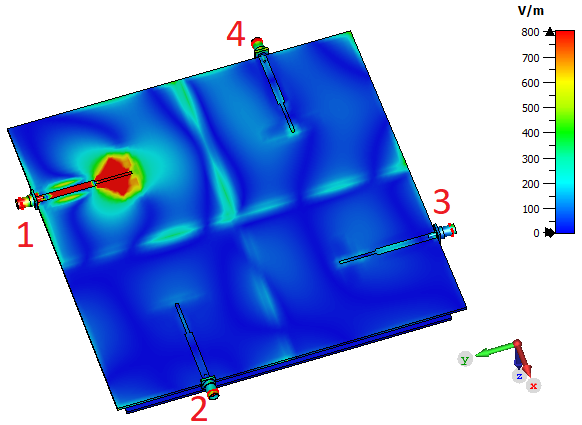}
		\caption{}
		\label{fig:fig9(i)}
	\end{subfigure}

\captionsetup{belowskip=0pt}
\caption{The CST and HFSS simulation results of Antenna\_{2}, (a) The scattering parameters regarding the frequency, (b) The maximum gain values and the radiation efficiency versus the frequency, (c) The ECC and diversity gain versus the frequency, (d) The $MEG_{1}$ and CCL concerning the frequency, (e) The TARC values in terms of the frequency, (f) The simulation back lobes of Antenna\_{1} and Antenna\_{2} concerning the frequency, (g)  E-field from the front view at 3.6 GHz, (h) E-field from the front view at 3.6 GHz (the top substrate is hidden to enhance the visibility), and (i) E-field from the feed line view at 3.6 GHz.}
\end{figure}

\begin{figure}[!t]
\centering
	\begin{subfigure}[b]{0.2\textwidth}
		\includegraphics[width = \textwidth]{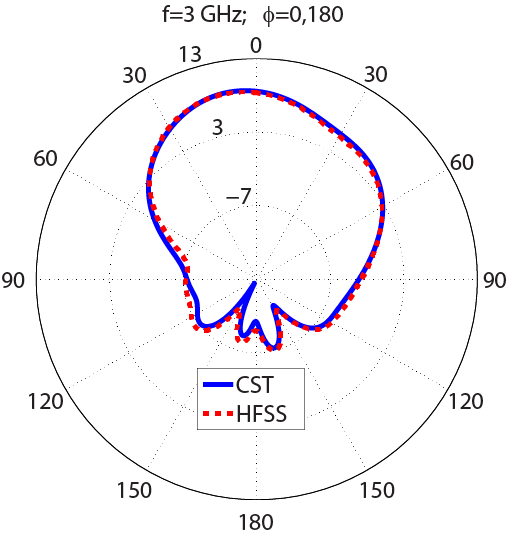}
		\caption{}
		\label{fig:fig10(a)}
	\end{subfigure}
	\begin{subfigure}[b]{0.198\textwidth}
		\includegraphics[width = \textwidth]{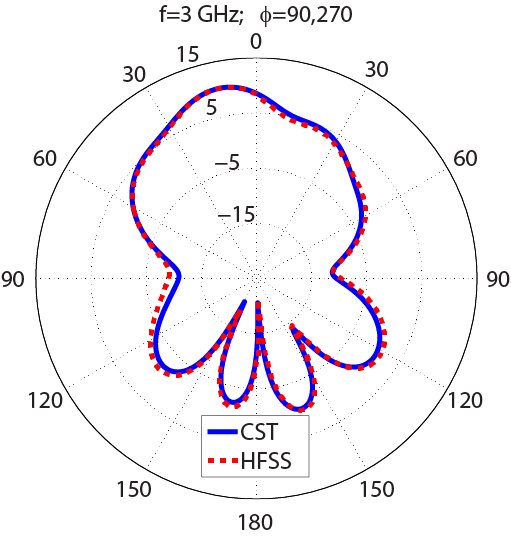}
		\caption{}
		\label{fig:fig10(b)}
	\end{subfigure}
	\begin{subfigure}[b]{0.2\textwidth}
		\includegraphics[width = \textwidth]{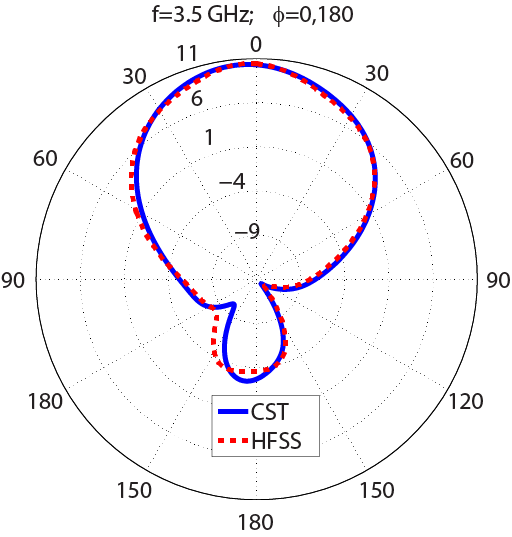}
		\caption{}
		\label{fig:fig10(c)}
	\end{subfigure}
	\begin{subfigure}[b]{0.21\textwidth}
		\includegraphics[width = \textwidth]{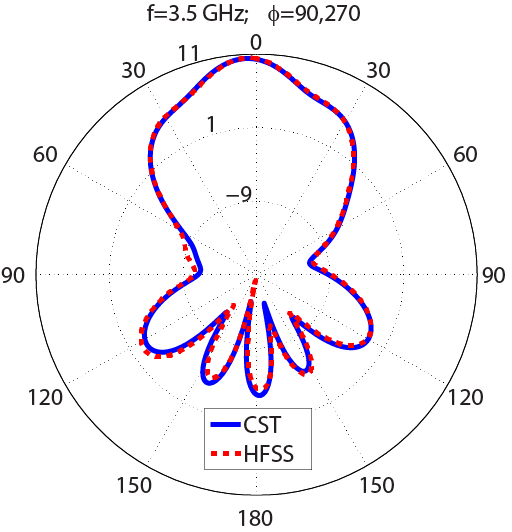}
		\caption{}
		\label{fig:fig10(d)}
	\end{subfigure}
	\begin{subfigure}[b]{0.2\textwidth}
		\includegraphics[width = \textwidth]{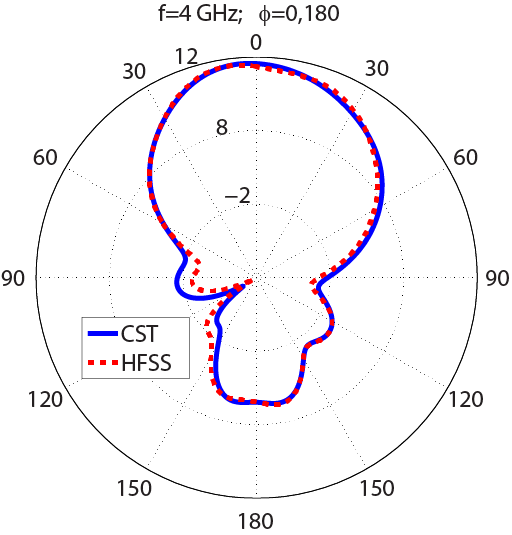}
		\caption{}
		\label{fig:fig10(e)}
	\end{subfigure}
	\begin{subfigure}[b]{0.2\textwidth}
		\includegraphics[width = \textwidth]{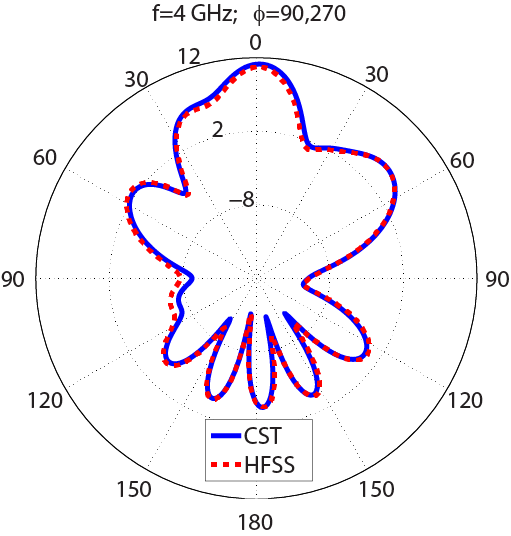}
		\caption{}
		\label{fig:fig10(f)}
	\end{subfigure}
\begin{subfigure}[b]{0.2\textwidth}
		\includegraphics[width = \textwidth]{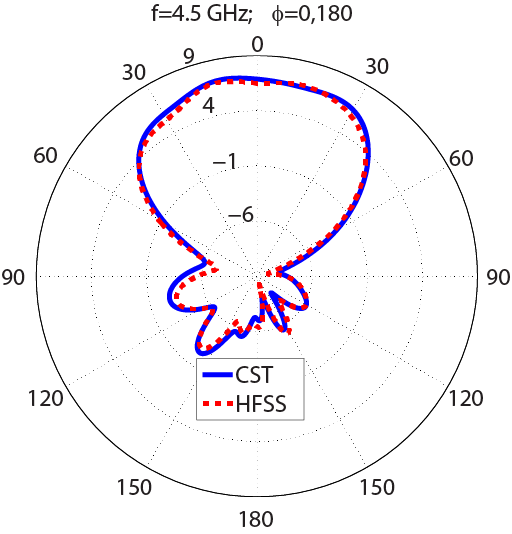}
		\caption{}
		\label{fig:fig10(g)}
	\end{subfigure}
	\begin{subfigure}[b]{0.2\textwidth}
		\includegraphics[width = \textwidth]{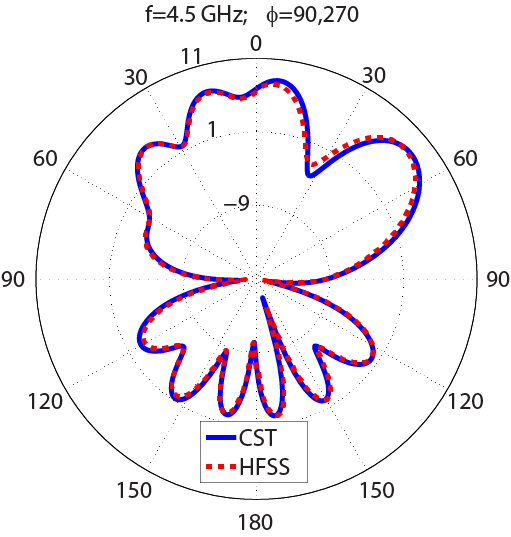}
		\caption{}
		\label{fig:fig10(h)}
	\end{subfigure}
\captionsetup{belowskip=0pt}
\caption{The CST and HFSS E and H plane patterns of Antenna\_{2} at the following frequencies: (a) 3 GHz ($\phi$=$0^\circ$;$180^\circ$), (b) 3 GHz ($\phi$=$90^\circ$;$270^\circ$), (c) 3.5 GHz ($\phi$=$0^\circ$;$180^\circ$), (d) 3.5 GHz ($\phi$=$90^\circ$;$270^\circ$), (e) 4 GHz ($\phi$=$0^\circ$;$180^\circ$), (f) 4 GHz ($\phi$=$90^\circ$;$270^\circ$), (g) 4.5 GHz ($\phi$=$0^\circ$;$180^\circ$), and (h) 4.5 GHz ($\phi$=$90^\circ$;$270^\circ$).}
\end{figure}

\begin{figure}[!t]
\centering
	\begin{subfigure}[b]{0.22\textwidth}
		\includegraphics[width = \textwidth]{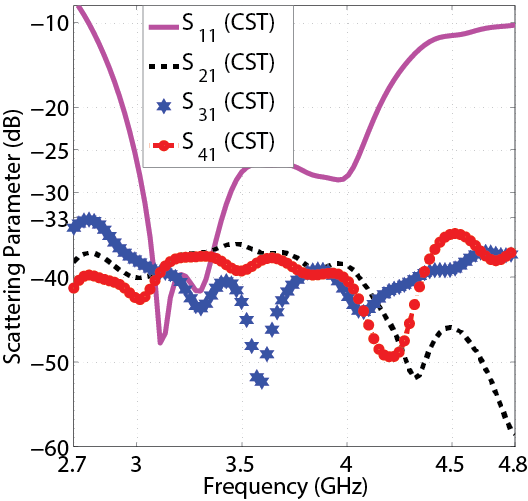}
		\caption{}
		\label{fig:fig11(a)}
	\end{subfigure}
	\begin{subfigure}[b]{0.24\textwidth}
		\includegraphics[width = \textwidth]{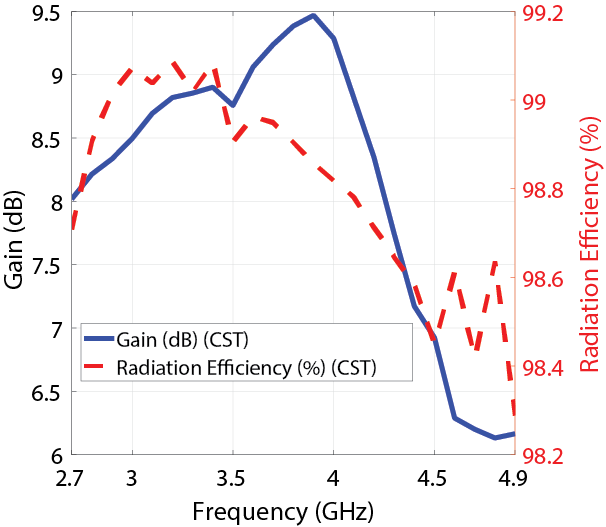}
		\caption{}
		\label{fig:fig11(b)}
	\end{subfigure}
	\begin{subfigure}[b]{0.25\textwidth}
		\includegraphics[width = \textwidth]{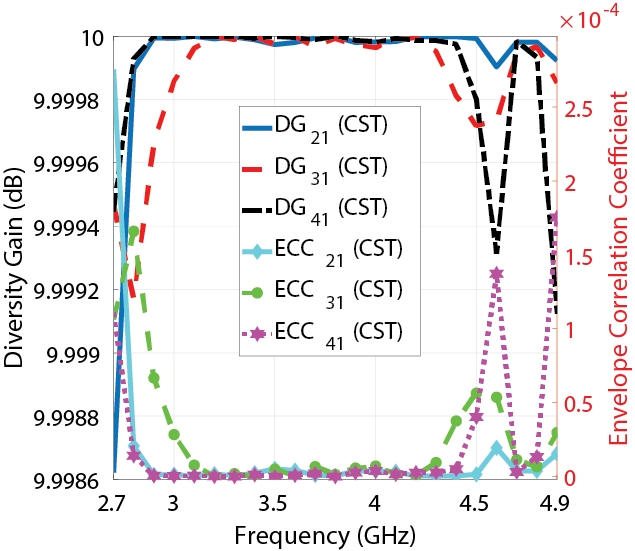}
		\caption{}
		\label{fig:fig11(c)}
	\end{subfigure}
	\begin{subfigure}[b]{0.24\textwidth}
		\includegraphics[width = \textwidth]{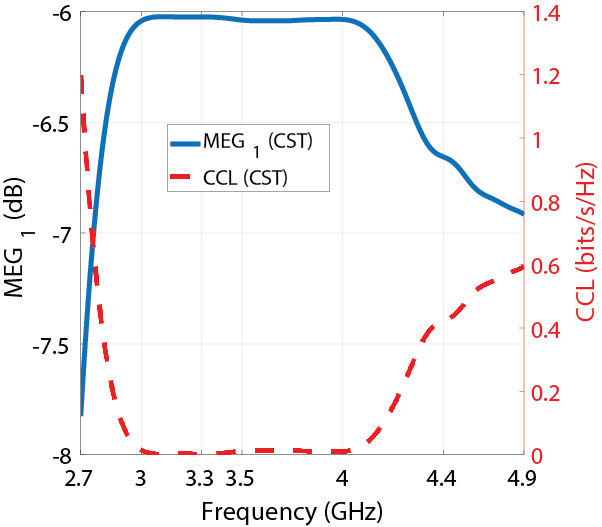}
		\caption{}
		\label{fig:fig11(d)}
	\end{subfigure}
	\begin{subfigure}[b]{0.20\textwidth}
		\includegraphics[width = \textwidth]{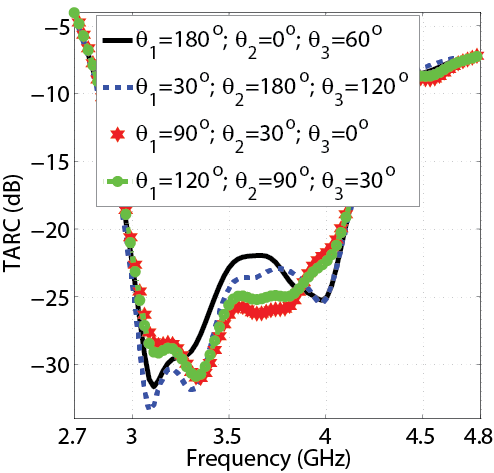}
		\caption{}
		\label{fig:fig11(e)}
	\end{subfigure}
	\begin{subfigure}[b]{0.25\textwidth}
		\includegraphics[width = \textwidth]{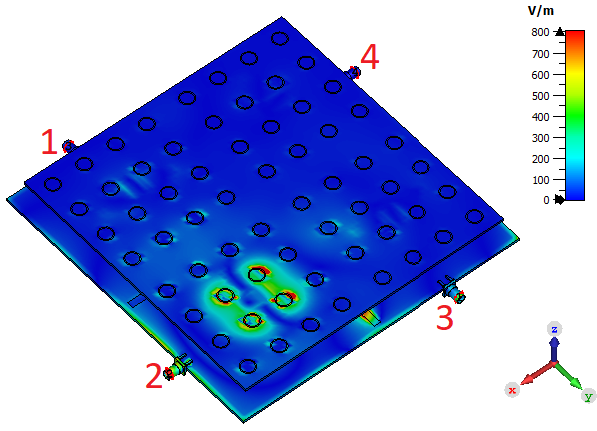}
		\caption{}
		\label{fig:fig11(f)}
	\end{subfigure}
	\begin{subfigure}[b]{0.23\textwidth}
		\includegraphics[width = \textwidth]{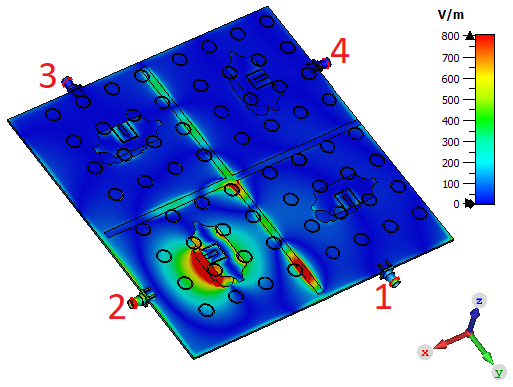}
		\caption{}
		\label{fig:fig11(g)}
	\end{subfigure}
	\begin{subfigure}[b]{0.23\textwidth}
		\includegraphics[width = \textwidth]{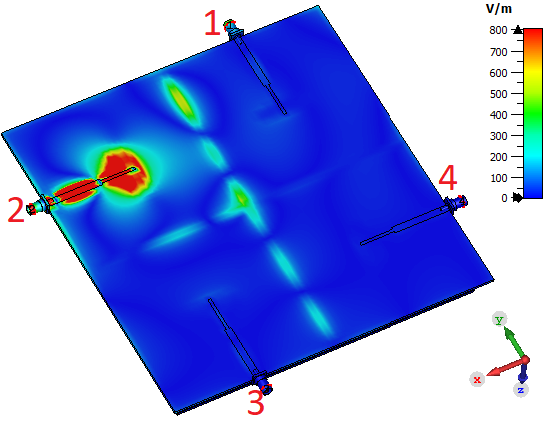}
		\caption{}
		\label{fig:fig11(h)}
	\end{subfigure}

\captionsetup{belowskip=0pt}
\caption{The CST simulation results of Antenna\_{3}, (a) The scattering parameters regarding the frequency, (b) The maximum gain values and the radiation efficiency versus the frequency, (c) The ECC and diversity gain versus the frequency, (d) The $MEG_{1}$ and CCL concerning the frequency, (e) The TARC values in terms of the frequency, (f)  E-field from the front view at 3.6 GHz, (g) E-field from the front view at 3.6 GHz (the top substrate is hidden to enhance the visibility), and (h) E-field from the feed line view at 3.6 GHz.}
\end{figure}

\begin{figure}[!b]
\centering
	\begin{subfigure}[b]{0.22\textwidth}
		\includegraphics[width = \textwidth]{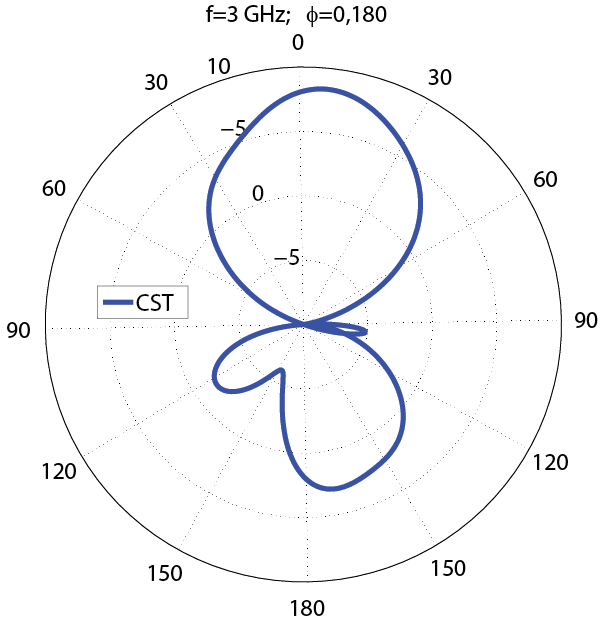}
		\caption{}
		\label{fig:fig12(a)}
	\end{subfigure}
	\begin{subfigure}[b]{0.22\textwidth}
		\includegraphics[width = \textwidth]{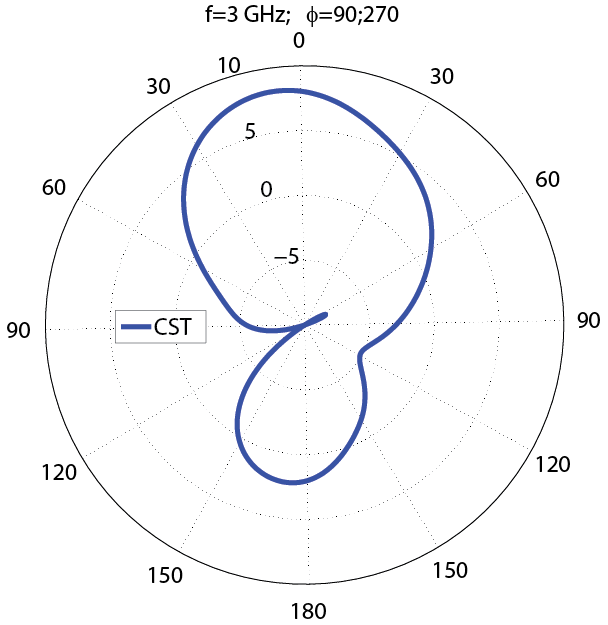}
		\caption{}
		\label{fig:fig12(b)}
	\end{subfigure}
	\begin{subfigure}[b]{0.22\textwidth}
		\includegraphics[width = \textwidth]{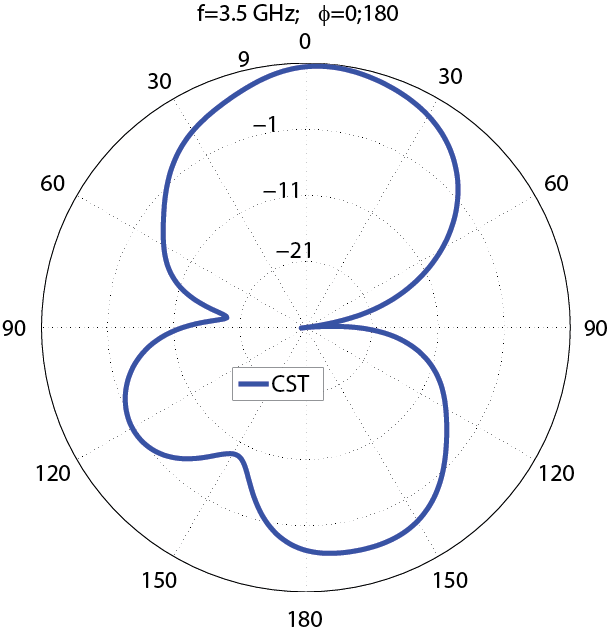}
		\caption{}
		\label{fig:fig12(c)}
	\end{subfigure}
	\begin{subfigure}[b]{0.22\textwidth}
		\includegraphics[width = \textwidth]{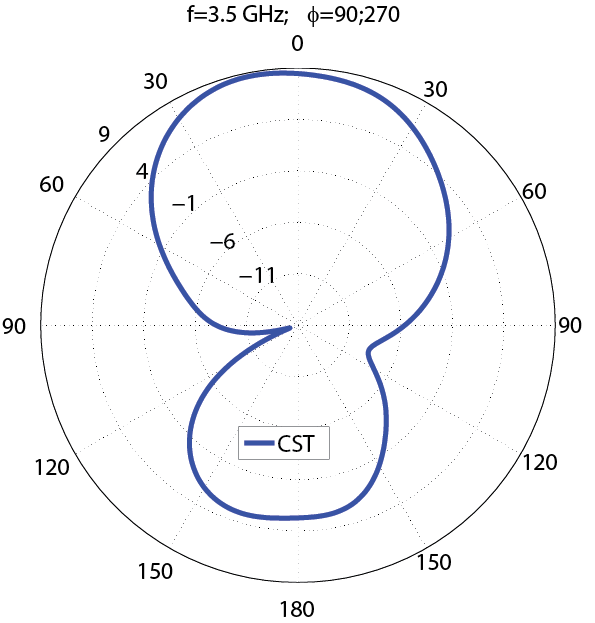}
		\caption{}
		\label{fig:fig12(d)}
	\end{subfigure}
	\begin{subfigure}[b]{0.22\textwidth}
		\includegraphics[width = \textwidth]{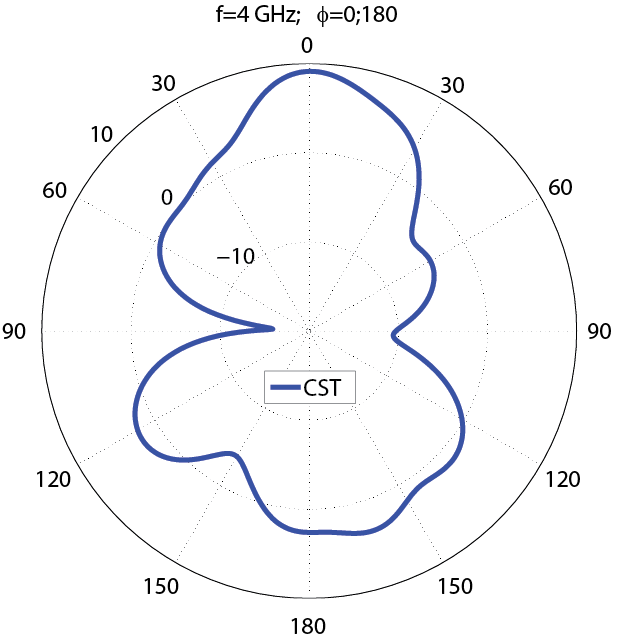}
		\caption{}
		\label{fig:fig12(e)}
	\end{subfigure}
	\begin{subfigure}[b]{0.22\textwidth}
		\includegraphics[width = \textwidth]{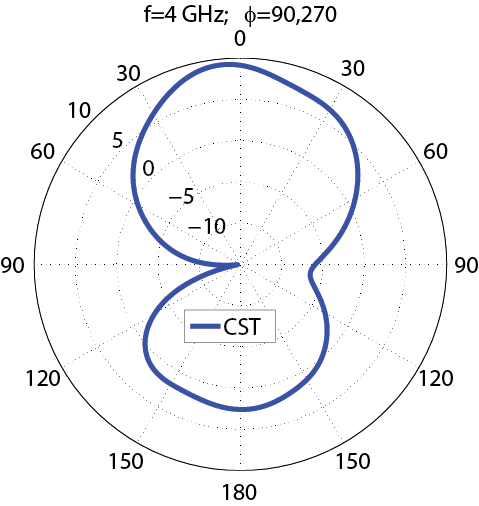}
		\caption{}
		\label{fig:fig12(f)}
	\end{subfigure}
	\begin{subfigure}[b]{0.22\textwidth}
		\includegraphics[width = \textwidth]{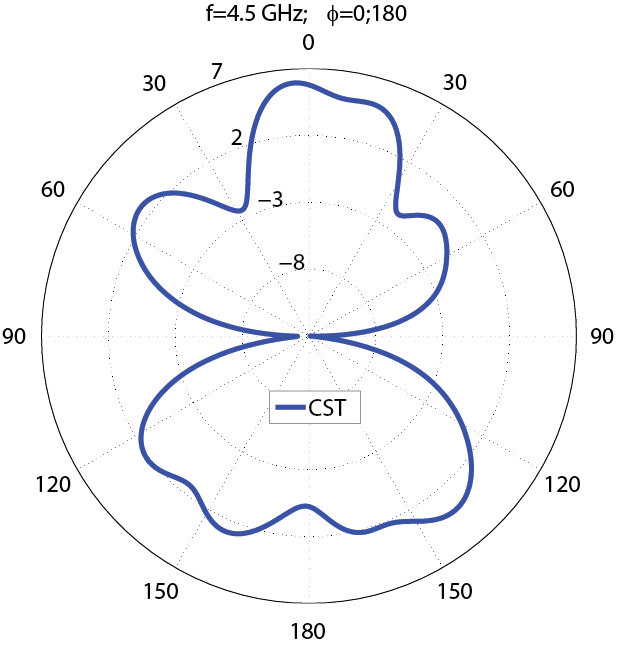}
		\caption{}
		\label{fig:fig12(g)}
	\end{subfigure}
	\begin{subfigure}[b]{0.22\textwidth}
		\includegraphics[width = \textwidth]{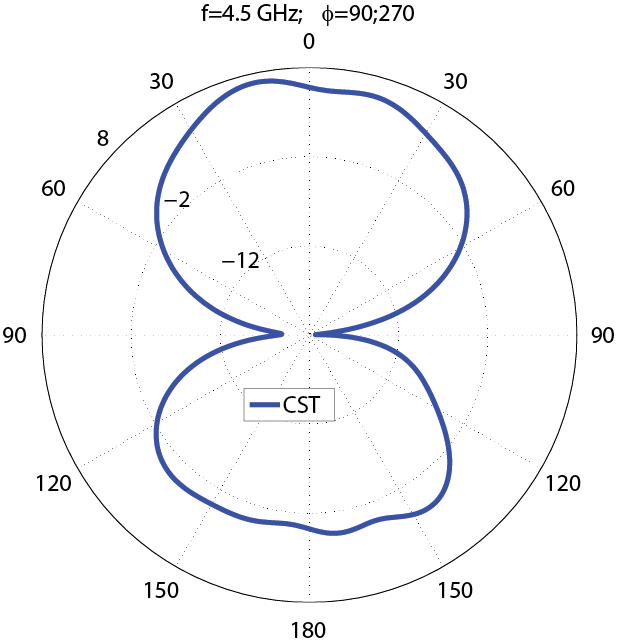}
		\caption{}
		\label{fig:fig12(h)}
	\end{subfigure}

\captionsetup{belowskip=0pt}
\caption{The simulation  E and H plane patterns of Antenna\_{3} at the following frequencies: (a) 3 GHz ($\phi$=$0^\circ$;$180^\circ$), (b) 3 GHz ($\phi$=$90^\circ$;$270^\circ$), (c) 3.5 GHz ($\phi$=$0^\circ$;$180^\circ$), (d) 3.5 GHz ($\phi$=$90^\circ$;$270^\circ$), (e) 4 GHz ($\phi$=$0^\circ$;$180^\circ$), (f) 4 GHz ($\phi$=$90^\circ$;$270^\circ$), (g) 4.5 GHz ($\phi$=$0^\circ$;$180^\circ$), and (h) 4.5 GHz ($\phi$=$90^\circ$;$270^\circ$).}
\end{figure}

\begin{equation}
\rho_{e}=\frac{{|S_{11}^{*}S_{12}+S_{21}^{*}S_{22}|}^{2}}{(1-(|S_{11}|^{2}+|S_{21}|^{2}))(1-(|S_{22}|^{2}+|S_{12}|^{2}))}\label{eq1}
\end{equation}
\begin{equation}
DG=10\times{{\log_{10}(10{\sqrt{1-{\rho_{e}}^{2}}})}}\label{eq2}
\end{equation}

\begin{equation}
MEG_{i}=0.5\times{(1-\Sigma_{j=1}^{4}{|S_{ij}|^{2})}}\label{eq3}
\end{equation}

\begin{align}\label{eq4}
CLL=-(\log_{2}{(\alpha^{R})})\\
\alpha^{R} =
\begin{pmatrix}\nonumber
\alpha_{11} & \alpha_{12}& \alpha_{13} &\alpha_{14} \\
\alpha_{21} & \alpha_{22}& \alpha_{23} &\alpha_{24} \\
\alpha_{31} & \alpha_{32}& \alpha_{33} &\alpha_{34} \\
\alpha_{41} & \alpha_{42}& \alpha_{43} &\alpha_{44}
\end{pmatrix}\\
\alpha_{ii}= 1-(\Sigma_{j=1}^{4}{|S_{ij}|^{2}})\nonumber\\
\alpha_{ij}= -(S_{ii}^{*}S_{ij}+S_{ji}^{*}S_{ij})\nonumber
\end{align}

\begin{align}\label{eq6}
TARC=\sqrt{\frac{|S_{1}|^{2}+|S_{2}|^{2}+|S_{3}|^{2}+|S_{4}|^{2}}{2}}\\
S_{1}=S_{11}+S_{12}e^{\theta_{1}}+S_{13}e^{\theta_{2}}+S_{14}e^{\theta_{3}}\nonumber\\
S_{2}=S_{21}+S_{22}e^{\theta_{1}}+S_{23}e^{\theta_{2}}+S_{24}e^{\theta_{3}}\nonumber\\
S_{3}=S_{31}+S_{32}e^{\theta_{1}}+S_{33}e^{\theta_{2}}+S_{34}e^{\theta_{3}}\nonumber\\
S_{3}=S_{41}+S_{42}e^{\theta_{1}}+S_{43}e^{\theta_{2}}+S_{44}e^{\theta_{3}}\nonumber
\end{align}

 A $178\times178$ $(mm)^{2}$ metal plane is placed 20 mm beneath Antenna\_{1}, creating Antenna\_{2}, to kill the undesired back lobe levels by creating a semi-cavity. Note that the reflector plane increases the antenna size in the simulation step, but in reality, the metal bodies of the vehicles work like a reflector plane, exempting the proposed MIMO antenna from embedding a reflector plane for vehicular communications. The HFSS and CST are applied to simulate Antenna\_{2}. The simulation results indicate that the antenna attains lower than -10 dB $S_{11}$ from 2.85 to 4.5, as shown in Fig.\ref{fig:fig9(a)}. The antenna experiences below -22.36 and -15 dB $S_{11}$ from 3 to 4 GHz, according to CST and HFSS results, respectively. As seen in Fig.\ref{fig:fig9(a)}, the isolation between ports varies from -50 to -25 dB by changing the frequency from 2.8 to 4.5 GHz, which is deteriorated compared to when the reflector is not used, as depicted in Fig.\ref{fig:fig7(a)}. However, the isolation results are still perfect and acceptable. In addition, adding the reflector plane has resulted in losing the impedance matching in one of the 5G frequency bands from 4.5 to 4.7 GHz. According to the CST and HFSS simulation results, the maximum gain values change from 8.23 to 11.6 dBi and 9.58 to 11.73 dBi, respectively, by varying the frequency from 2.8 to 4.5 GHz, as depicted in Fig.\ref{fig:fig9(b)}. Compared with Fig.\ref{fig:fig7(b)}, the metal plate has significantly increased the antenna gain by killing the backward propagation power and converting it to the forward propagation power, which boosts the forward radiation. Moreover, the antenna achieves 82 to 98\% and above 98\% radiation efficiency, according to HFSS and CST simulation results, respectively, in the operational bandwidth. A comparison of the radiation efficiency, as provided in Fig.\ref{fig:fig7(b)}, shows that adding a reflector slightly worsens the efficiency , but the results are still acceptable. As seen in Fig.\ref{fig:fig9(c)}, the antenna achieves almost 10 dB DG and less than 0.001 ECC in the operational bandwidth. Compared with Fig.\ref{fig:fig7(c)}, although adding the reflector has slightly deteriorated ECC and DG values, the radiating elements still work independently, and there is very low mutual coupling between the elements. \par The $MEG_{1}$ values vary from -9.8 to -6 dB, which are within the acceptable range (-12 to -3 dB), as depicted in Fig.\ref{fig:fig9(d)}. Compared with Fig.\ref{fig:fig7(d)}, the minimum range of $MEG_{1}$ has gotten closer to -12 dB (the unacceptable limit), but is still excellent. As reflected in Fig.\ref{fig:fig9(d)}, CCL is below 0.4 (bits/s/Hz) from 2.85 to 4.3 GHz, indicating high-quality data transmission. Apparently, the reflector plane has precluded 4.3 to 4.5 GHz from flawless data transmission, compared with Fig.\ref{fig:fig7(d)}. The TARC values for various phase differences are illustrated in Fig.\ref{fig:fig9(e)}. Varying the phase differences does not result in significant changes in TARC values, and the -10 dB TARC bandwidth encompasses 2.8 to 4.3 GHz. Conclusively, based on Figs.{\ref{fig:fig9(a)}-\ref{fig:fig9(e)}}, Antenna\_{2} has superior performance from 2.85 to 4.3 GHz, regarding -10 dB bandwidth, isolation, ECC, DG, TARC, MEG, and CCL. The E-field distribution of Antenna\_{2} is displayed in Figs.{\ref{fig:fig9(g)}-\ref{fig:fig9(i)}}. Clearly, the isolation between the radiating elements are extremely high. The polar far field gain diagrams of Antenna\_{2} for E and H planes at 3, 3.5, 4, and 4.5 GHz are shown in Figs.{\ref{fig:fig10(a)}-\ref{fig:fig10(h)}}. Comparing with Figs.{\ref{fig:fig7(a)}-\ref{fig:fig7(h)}}, the back lobe levels are reduced, and the gain values are increased significantly. For clarification, the maximum back lobe levels of Antenna\_{1} and Antenna\_{2} are compared in Fig.\ref{fig:fig9(d)}. According to CST and HFSS results, when the reflector plane is used (Antenna\_{2}), the back lobe levels vary from -2.3 to 2.5 dBi and -2 to 3 dBi from 3 to 4.5 GHz, respectively. However, Antenna\_{1} experiences 3 to 4.75 dBi back lobe levels without employing the reflector plane. It is relevant to mention that CST and HFSS results concur with each other, and the differences between the CST and HFSS are due to applying different computational methods (FEM and FIT) for simulating the antenna.
 \par Some applications necessitate MIMO antennas with a connected ground plane. Therefore, three small rectangular metal connectors, designated by "GC," are placed at the head and tail of the vertical slot and the head of the horizontal slot to connect the ground plane, creating Antenna\_{3}, as shown in Fig.\ref{fig:fig6(c)}. The CST software is used to simulate Antenna\_{3}. The simulation results show that Antenna\_{3} attains below -10 dB $S_{11}$ from 2.75 to 4.8, as shown in Fig.\ref{fig:fig11(a)}. The antenna experiences lower than -25 dB $S_{11}$ from 3 to 4 GHz.  As seen in Fig.\ref{fig:fig11(a)}, the isolation between ports changes from -60 to -33 dB from 2.7 to 4.8 GHz. According to Fig.\ref{fig:fig11(b)}, the maximum gain values vary from 6.2 to 9.5 dBi by changing the frequency from 2.7 to 4.9 GHz. Moreover, the antenna achieves 98.3 to 99\% over 2.7 to 4.9 GHz. The antenna achieves virtually 10 dB diversity gain and lower than 0.00022 ECC values, as reflected in Fig.\ref{fig:fig11(c)}. According to Fig.\ref{fig:fig11(d)}, $MEG_{1}$ varies between -7.8 to -6 dB over 2.7 to 4.9 GHz, which is in the acceptable range. Furthermore, CCL is below 0.4 (bits/s/Hz) from 2.75 to 4.35 GHz, where data transmissions are done with a high quality. Fig.\ref{fig:fig11(e)} shows that TARC curves do not change significantly, and the -10 dB bandwidth of TARC includes 2.8 to 4.4 GHz. Antenna\_3 definitely performs superbly from 2.8 to 4.35 GHz, considering -10 dB bandwidth, isolation, ECC, DG, TARC, MEG, and CCL. The E-field distribution of Antenna\_{3} is shown in Figs.{\ref{fig:fig11(f)}-\ref{fig:fig11(h)}}. It is evident that the mutual coupling between the radiating elements is extremely low, and they are completely independent. In addition, the polar far field gain diagrams of Antenna\_{3} for E and H planes at 3, 3.5, 4, and 4.5 GHz are displayed in Figs.{\ref{fig:fig12(a)}-\ref{fig:fig12(h)}}, indicating the high-gain performance of the proposed MIMO antenna.

  \section*{Manufacturing a Prototype}

 \begin{figure}[h! tbp]
\centering
	\begin{subfigure}[b]{0.35\textwidth}
		\includegraphics[width = \textwidth]{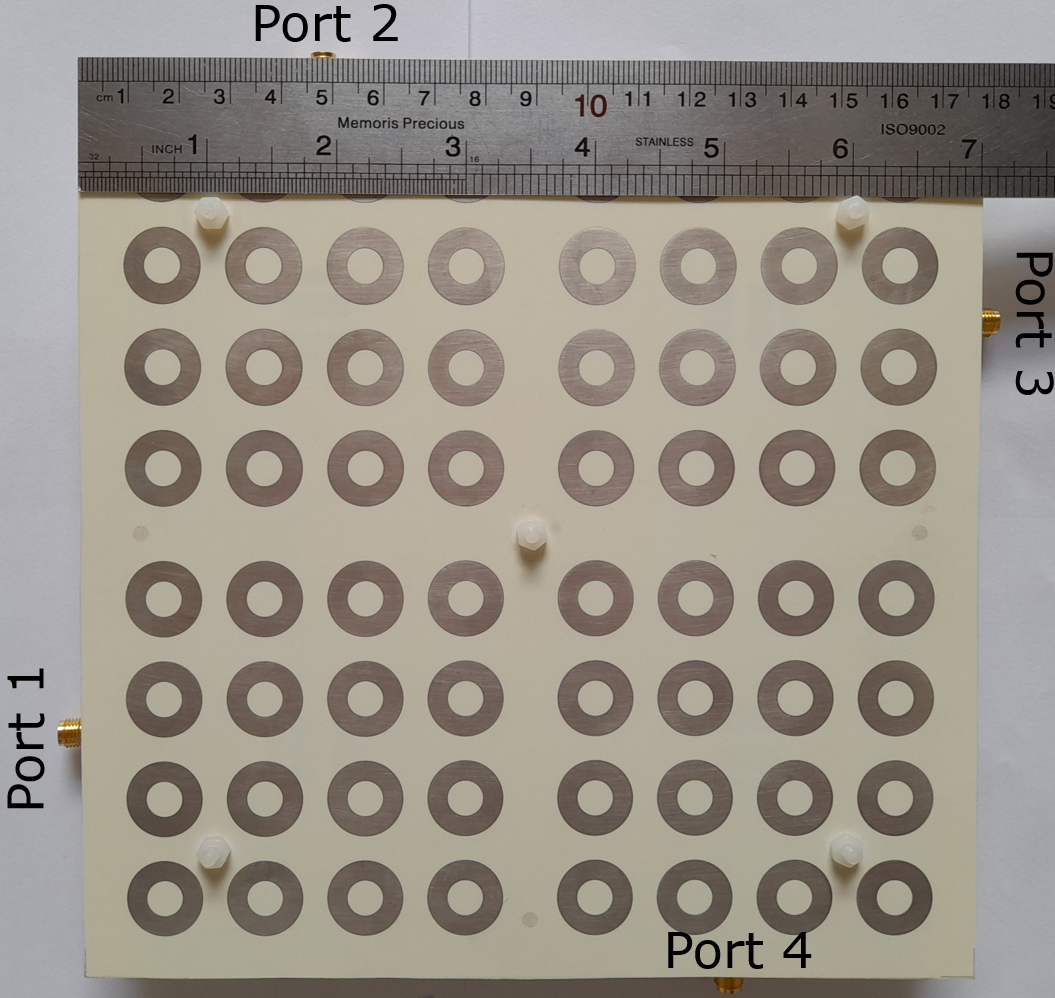}
		\caption{}
		\label{fig:fig13(a)}
	\end{subfigure}
	\begin{subfigure}[b]{0.335\textwidth}
		\includegraphics[width = \textwidth]{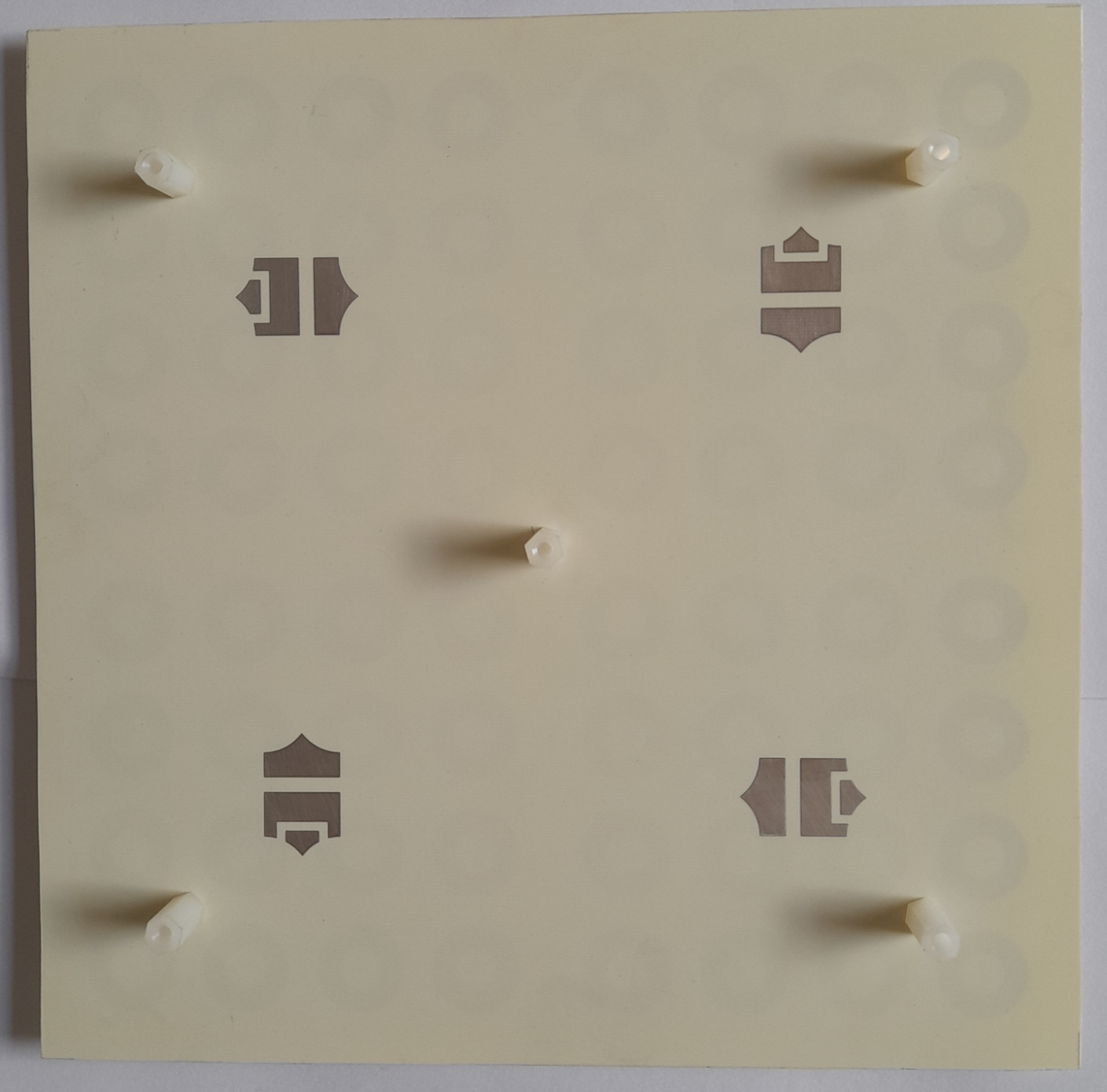}
		\caption{}
		\label{fig:fig13(b)}
	\end{subfigure}
\begin{subfigure}[b]{0.355\textwidth}
		\includegraphics[width = \textwidth]{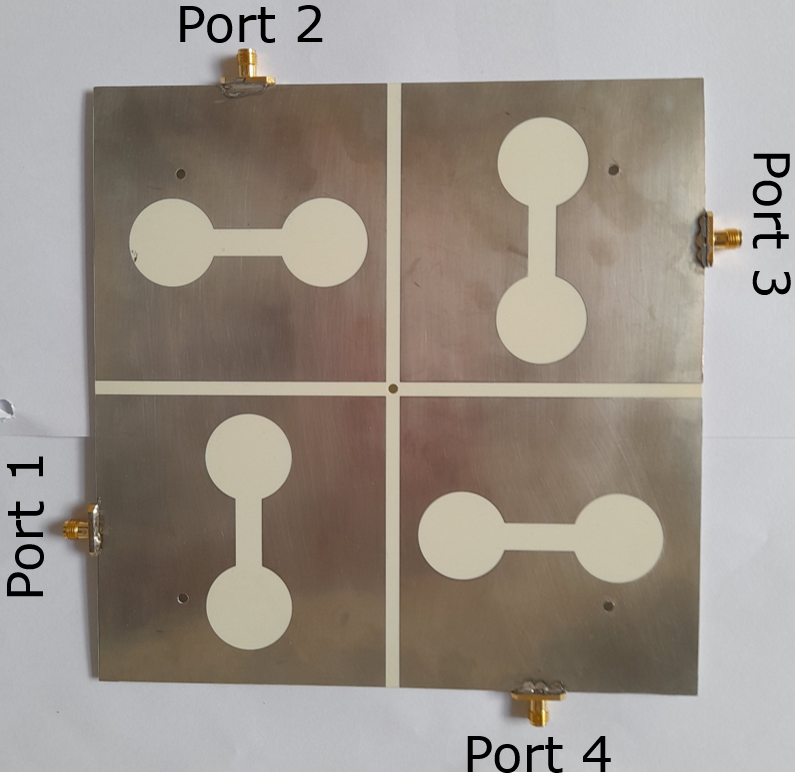}
		\caption{}
		\label{fig:fig13(c)}
	\end{subfigure}
	\begin{subfigure}[b]{0.325\textwidth}
		\includegraphics[width = \textwidth]{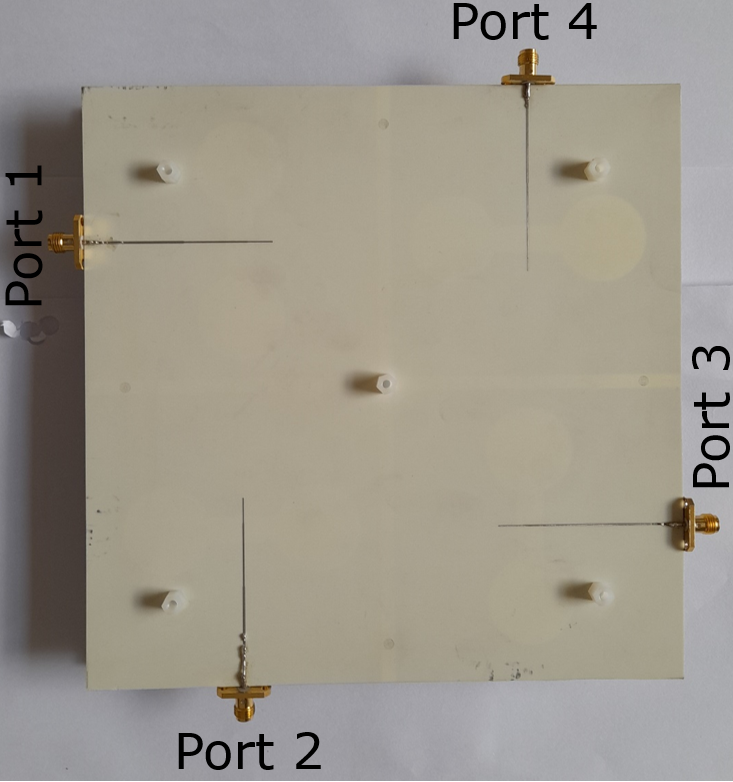}
		\caption{}
		\label{fig:fig13(d)}
	\end{subfigure}
\begin{subfigure}[b]{0.7\textwidth}
		\includegraphics[width = \textwidth]{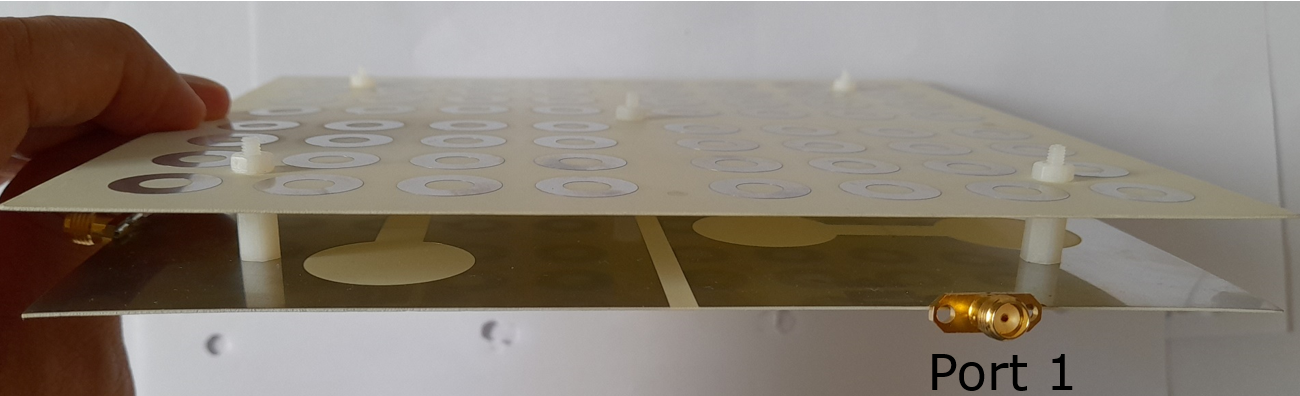}
		\caption{}
		\label{fig:fig13(e)}
	\end{subfigure}
\begin{subfigure}[b]{0.35\textwidth}
		\includegraphics[width = \textwidth]{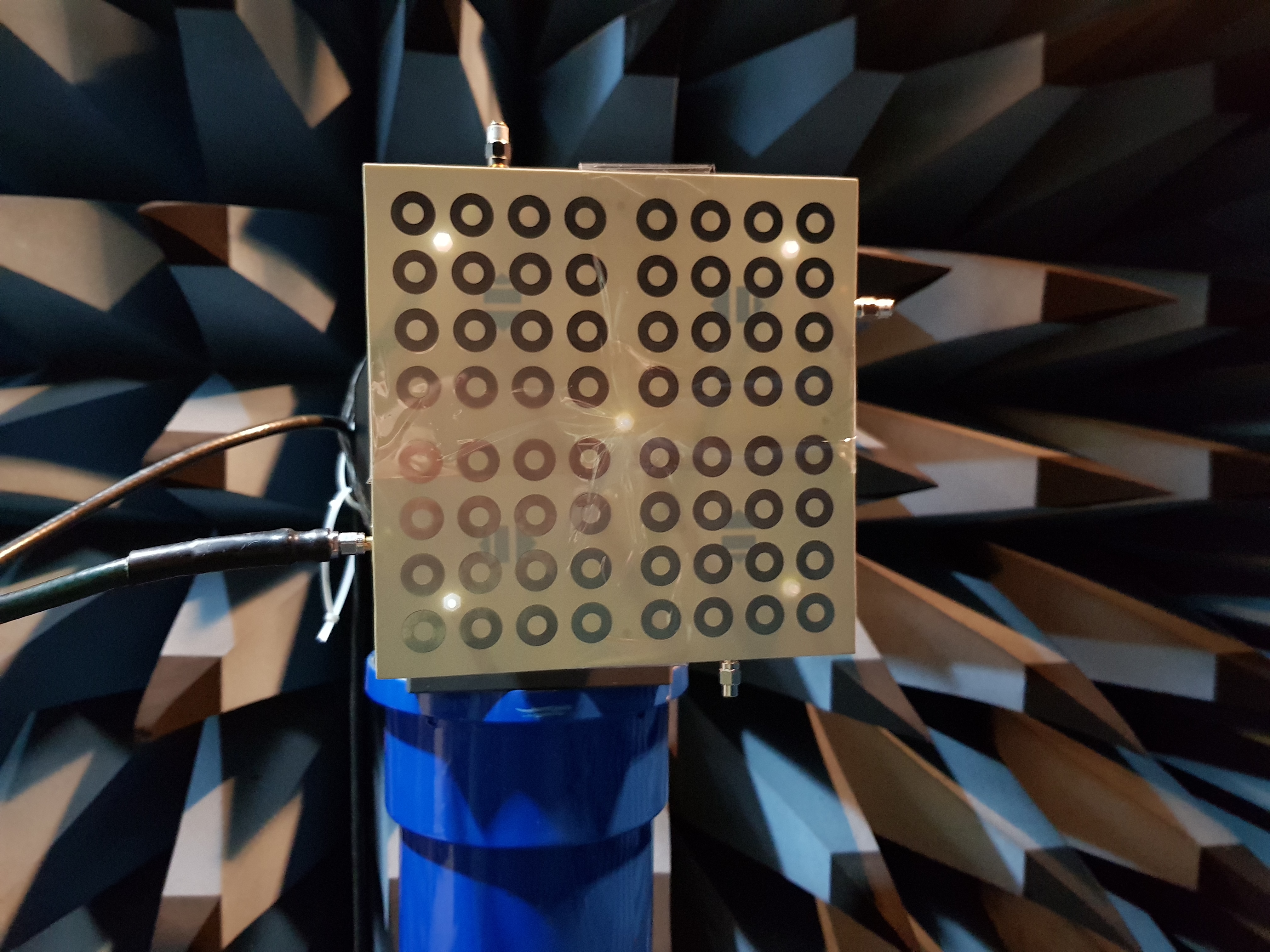}
		\caption{}
		\label{fig:fig13(f)}
	\end{subfigure}
\begin{subfigure}[b]{0.35\textwidth}
		\includegraphics[width = \textwidth]{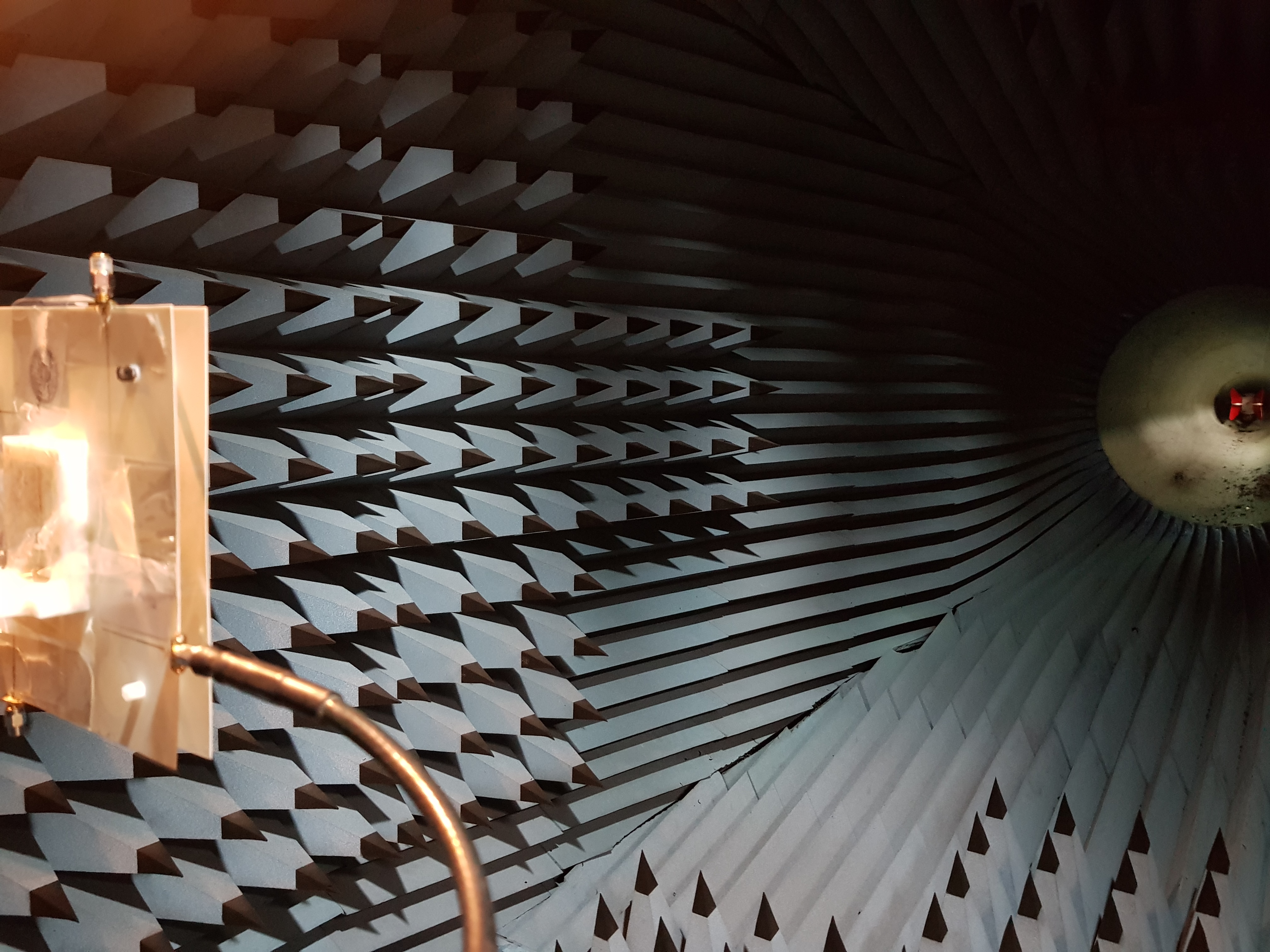}
		\caption{}
		\label{fig:fig13(g)}
	\end{subfigure}
\captionsetup{belowskip=0pt}
\caption{The manufactured antenna (a) the overhead aspect of the second layer (b) the underside aspect of the second layer (c) the overhead aspect of the first layer (d) the underside aspect of the first layer (e) the side view of the antenna and (f) the MIMO antenna in the antenna testing Lab, and (g) the MIMO antenna in the antenna testing Lab .}
\end{figure}

\begin{figure}[h! tbp]
\centering
	\begin{subfigure}[b]{0.21\textwidth}
		\includegraphics[width = \textwidth]{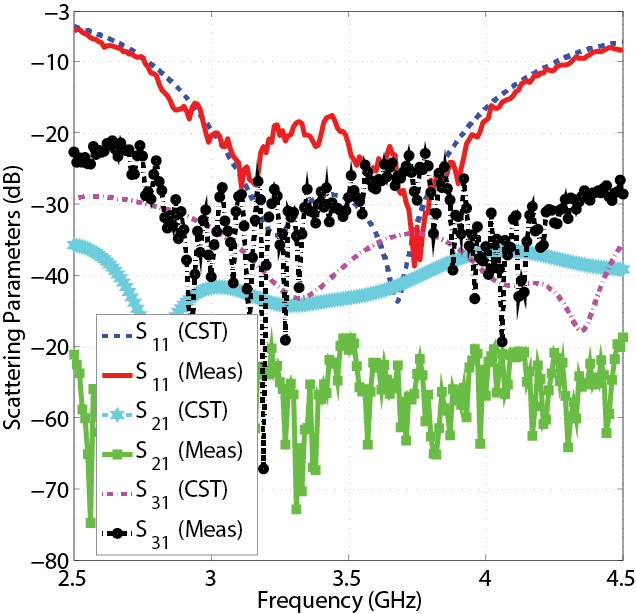}
		\caption{}
		\label{fig:fig14(a)}
	\end{subfigure}
	\begin{subfigure}[b]{0.24\textwidth}
		\includegraphics[width = \textwidth]{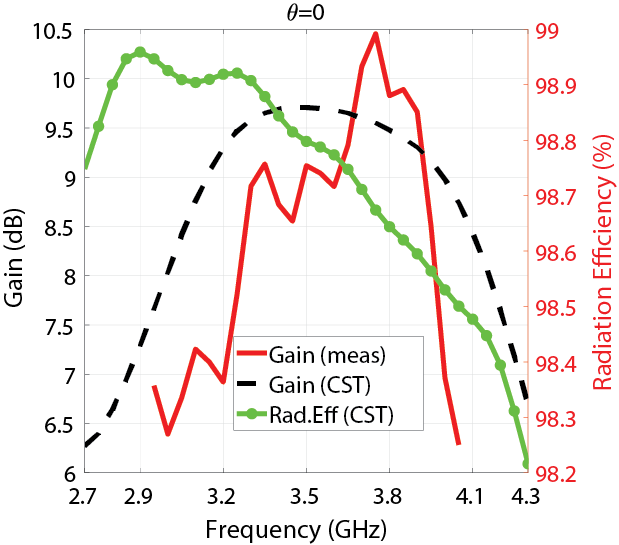}
		\caption{}
		\label{fig:fig14(b)}
	\end{subfigure}
\begin{subfigure}[b]{0.24\textwidth}
		\includegraphics[width = \textwidth]{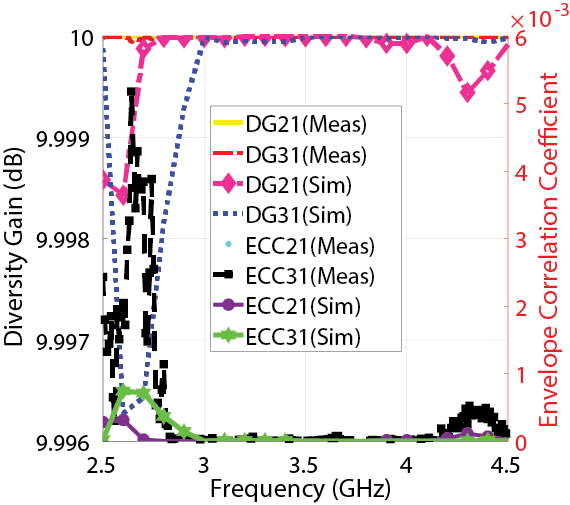}
		\caption{}
		\label{fig:fig14(c)}
	\end{subfigure}
\begin{subfigure}[b]{0.24\textwidth}
		\includegraphics[width = \textwidth]{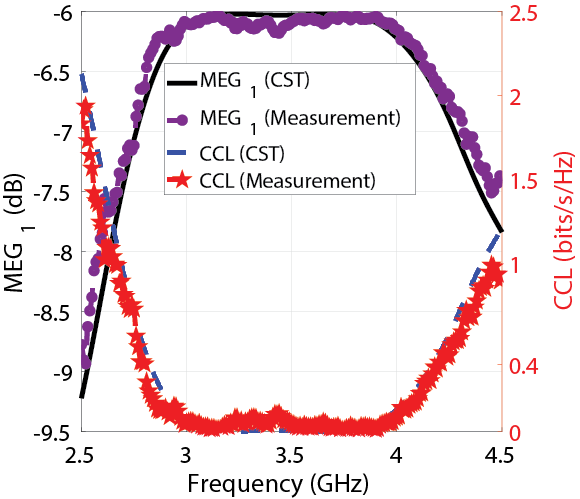}
		\caption{}
		\label{fig:fig14(d)}
	\end{subfigure}
\captionsetup{belowskip=0pt}
\begin{subfigure}[b]{0.21\textwidth}
		\includegraphics[width = \textwidth]{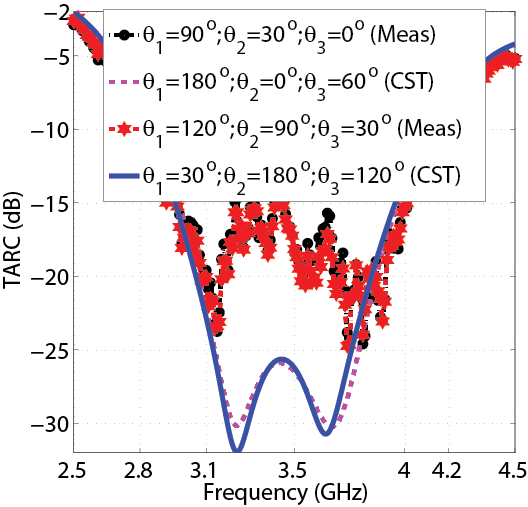}
		\caption{}
		\label{fig:fig14(e)}
	\end{subfigure}
\begin{subfigure}[b]{0.25\textwidth}
		\includegraphics[width = \textwidth]{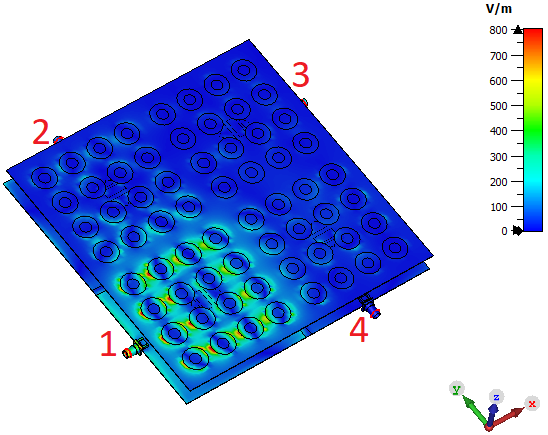}
		\caption{}
		\label{fig:fig14(f)}
	\end{subfigure}
\begin{subfigure}[b]{0.25\textwidth}
		\includegraphics[width = \textwidth]{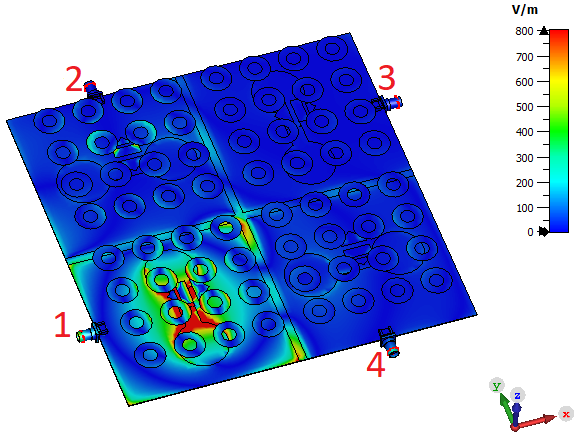}
		\caption{}
		\label{fig:fig14(g)}
	\end{subfigure}
\begin{subfigure}[b]{0.25\textwidth}
		\includegraphics[width = \textwidth]{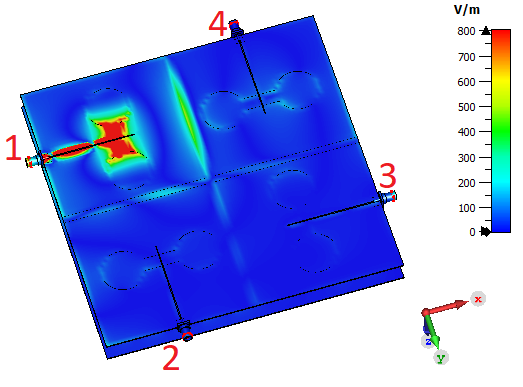}
		\caption{}
		\label{fig:fig14(h)}
	\end{subfigure}

\caption{The simulation and measurement results of Antenna\_{4}, (a) The scattering parameters regarding the frequency, (b) The simulation and measurement gain at ($\theta$=$0^\circ$,$\phi$=$0^\circ$), (c) The simulation and measurement ECC and diversity gain, (d) The simulation and measurement $MEG_{1}$ and CCL concerning the frequency, (e) The simulation and measurement TARC values in terms of the frequency, (f)  E-field from the front view at 3.5 GHz, (g) E-field from the front view at 3.5 GHz (the top substrate is hidden to enhance the visibility), and (h) E-field from the feed line view at 3.5 GHz.}
\end{figure}

\begin{figure}[h! tbp]
\centering
	\begin{subfigure}[b]{0.22\textwidth}
		\includegraphics[width = \textwidth]{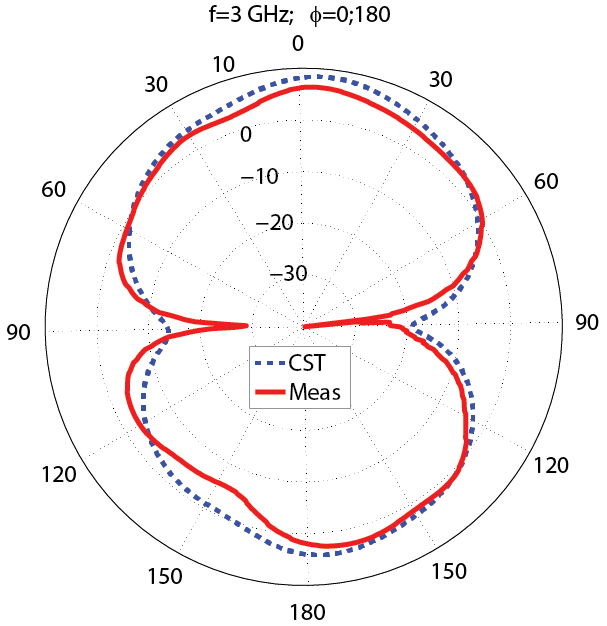}
		\caption{}
		\label{fig:fig15(a)}
	\end{subfigure}
	\begin{subfigure}[b]{0.21\textwidth}
		\includegraphics[width = \textwidth]{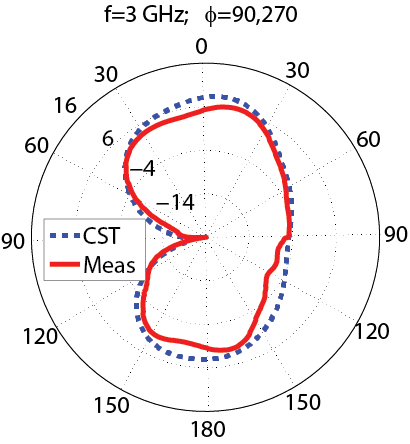}
		\caption{}
		\label{fig:fig15(b)}
	\end{subfigure}
\begin{subfigure}[b]{0.22\textwidth}
		\includegraphics[width = \textwidth]{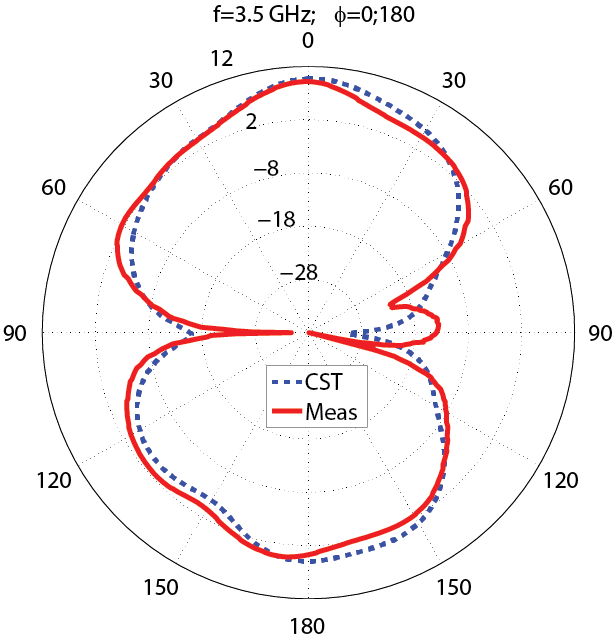}
		\caption{}
		\label{fig:fig15(c)}
	\end{subfigure}
\begin{subfigure}[b]{0.22\textwidth}
		\includegraphics[width = \textwidth]{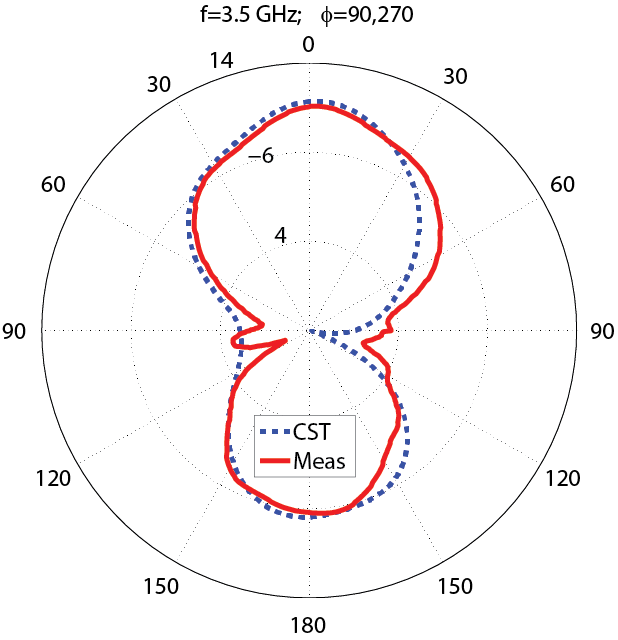}
		\caption{}
		\label{fig:fig15(d)}
	\end{subfigure}
\begin{subfigure}[b]{0.22\textwidth}
		\includegraphics[width = \textwidth]{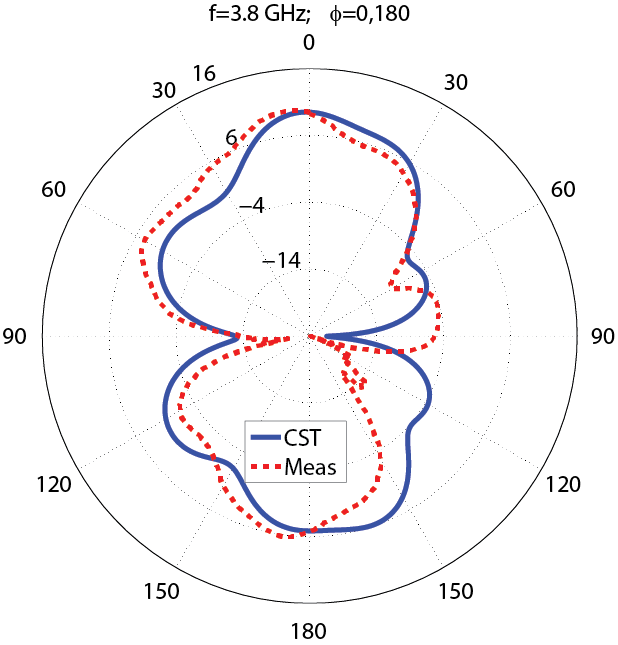}
		\caption{}
		\label{fig:fig15(e)}
	\end{subfigure}
\begin{subfigure}[b]{0.22\textwidth}
		\includegraphics[width = \textwidth]{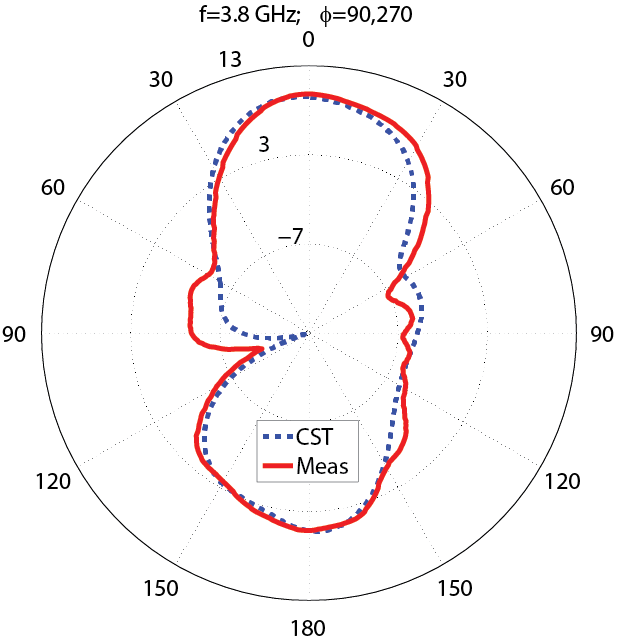}
		\caption{}
		\label{fig:fig15(f)}
	\end{subfigure}
\begin{subfigure}[b]{0.22\textwidth}
		\includegraphics[width = \textwidth]{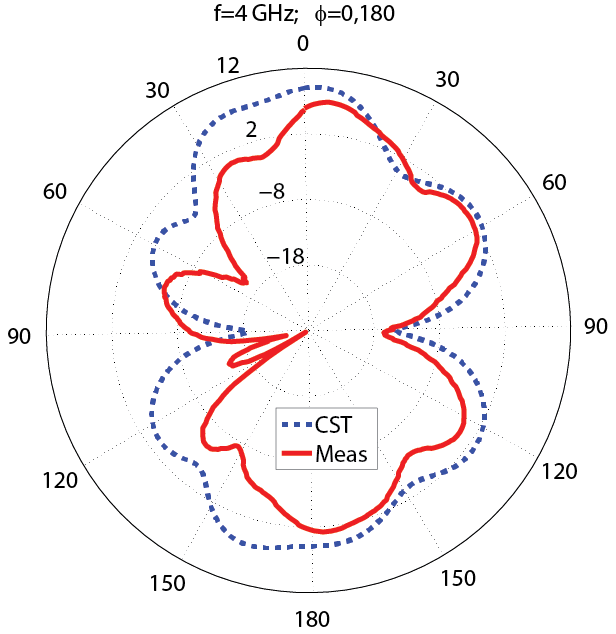}
		\caption{}
		\label{fig:fig15(g)}
	\end{subfigure}
\begin{subfigure}[b]{0.22\textwidth}
		\includegraphics[width = \textwidth]{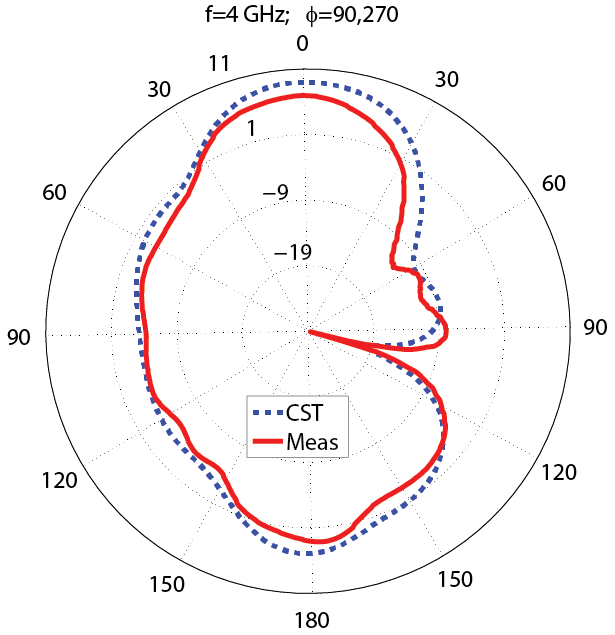}
		\caption{}
		\label{fig:fig15(h)}
	\end{subfigure}
\captionsetup{belowskip=0pt}
\caption{The simulation and measurement E and H plane patterns of the proposed MIMO antenna  at the following frequencies: (a) 3 GHz ($\phi$=$0^\circ$;$180^\circ$), (b) 3 GHz ($\phi$=$90^\circ$;$270^\circ$), (c) 3.5 GHz ($\phi$=$0^\circ$;$180^\circ$), (d) 3.5 GHz ($\phi$=$90^\circ$;$270^\circ$), (e) 3.8 GHz ($\phi$=$0^\circ$;$180^\circ$), (f) 3.8 GHz ($\phi$=$90^\circ$;$270^\circ$), (g) 4 GHz ($\phi$=$0^\circ$;$180^\circ$), and (h) 4 GHz ($\phi$=$90^\circ$;$270^\circ$).}
\end{figure}

  As the only available dielectric in our laboratory is Rogers 4350B ($\epsilon_{r}$=3.66, $tan(\gamma)$=0.0037, $t_{sub}=0.508 mm$), which has a different height from the dielectric material Antenna\_{1}, Antenna\_{2}, and Antenna\_{3} are made, the dimensions of the above antennas are re-optimized to determine the dimensions of Antenna\_{4}. The dimensions of Antenna\_{4} are provided in Table.\ref{tab:Table3}, concerning Figs.{\ref{fig:fig1(a)}-\ref{fig:fig1(g)}} and Figs.{\ref{fig:fig6(a)}-\ref{fig:fig6(e)}}. The CST software is used to simulate Antenna\_{4}.  As a proof of principle, a prototype of Antenna\_{4} comprised of two $178\times178$ ${mm}^2$ Rogers RO4350B dielectric layers with an air gap of 13 mm is machined, and its measurements are performed in an anechoic chamber. The overhead aspect of the second layer of the manufactured antenna, the underside aspect of the second layer, the overhead aspect of the first layer, the underside aspect of the first layer, the side view of the antenna, and the antenna under test in the anechoic chamber are displayed in Figs.{\ref{fig:fig13(a)}-\ref{fig:fig13(g)}}, respectively. The measurement and simulation scattering parameters of the antenna are shown in Fig.{\ref{fig:fig14(a)}. According to this figure, the measurement and simulation $S_{11}$ values are less than -10 dB from 2.76 to 4.3 GHz, including the middle sub 6 GHz 5G frequency spectrum. The measurement $S_{11}$ curve goes below -17 dB from 3 to 4 GHz, while the simulation values are inferior to -20 dB.
  According to the measurement and simulation results, the antenna experiences below -23 and -30 dB isolation at its -10 dB impedance bandwidth, respectively. Hence, the proposed MIMO design can provide four independent paths for sending and receiving data. The simulation and measurement gain values at $(\theta=0^\circ,\phi=0^\circ)$ and the simulation radiation efficiency are displayed in Fig.{\ref{fig:fig14(b)}. The simulation gain varies from 6.7 to 9.7 dBi by changing the frequency from 2.8 to 4.2 GHz, underlying the simulation radiation efficiency superior to 98.4\%. The measurement gain changes from 6.28 to 10.45 dBi from 3 to 4 GHz. As seen in Fig.{\ref{fig:fig14(c)}, the ECC values are less than 0.001, and the diversity gain is almost 10 dB in the -10 dB impedance bandwidth. They indicate how highly independent the radiating elements work. It is worth mentioning that the measurement diversity gain and ECC coefficients are obtained from Eqs.{\ref{eq1}-\ref{eq2}} by substituting the measurement scattering parameters. $MEG_{1}$ values vary from -9.2 to -6 dB from 2.5 to 4.5 GHz, which is in the acceptable range (-12 to -3 dB), as depicted in Fig.{\ref{fig:fig14(d)}}. In addition, according to Fig.{\ref{fig:fig14(d)}}, CCL values are below 0.4 (bits/s/Hz) over 2.8 to 4.2, indicating superb data transmission. The measurement and simulation TARC values versus the frequency for various phase differences are illustrated in Fig.{\ref{fig:fig14(e)}}. The measurement and simulation TARC do not change notably by varying the phase differences and the -10 dB TARC bandwidth 2.8 to 4.15 GHz. Conclusively, based on -10 dB Figs.{\ref{fig:fig14(a)}-\ref{fig:fig14(e)}}, Antenna\_{3} has the best performance from 2.8 to 4.15 GHz in terms of bandwidth, isolation, ECC, DG, TARC, MEG, and CCL. The E-field distribution of Antenna\_{4} is depicted in  Figs.{\ref{fig:fig14(f)}-\ref{fig:fig14(h)}}. Evidently, the mutual coupling between the radiating elements is very low, and they work independently.  The measurement and simulation polar patterns are illustrated in Figs.{\ref{fig:fig15(a)}-\ref{fig:fig15(h)}}. Differences can be seen between the measurement and simulation results, mainly due to the insertion loss of the SMA connectors, manufacturing tolerances of the antenna, the accuracy rate of the antenna measurement equipment, and approximations of the simulation model.

   \par An analogy between the achievements of Antenna\_{4} and other studies is provided in Table.\ref{tab:Table4}.
   Concerning the polarization, Antenna\_{4} radiates LP waves, while Refs.\citen{ref45,ref46} design CP MIMO antennas that can notably survive multi-path interferences and fading. Regarding the bandwidth, the above antennas perform better than Refs.\citen{ref26,ref27,ref28,ref29,ref30,ref31,ref34,ref36}. As the purpose of this study has been to bring forward a high gain and highly isolated MIMO design to operate from 3 to 4 GHz, the middle band of the 5G spectrum where many technologies offer their 5G services, other studies that work in a broader frequency spectrum do not belittle the achievements of the proposed MIMO structure. Concerning the gain, Antenna\_{4} performs better than the studies mentioned in Table.\ref{tab:Table4} except for \citen{ref52}. As 5G needs high gain antennas to maintain the signal quality, increase its robustness against noise signals, increase its reliability, send more directed power to avoid the loss of the achievable penetration and transfer data over long distances, the proposed MIMO structure can be considered as a remedy for these critical issues. Concerning the isolation, Antenna\_{4} benefits from higher isolation than other studies except for Refs.\citen{ref27,ref29,ref30,ref32,ref34,ref36,ref37,ref53}. Regarding the number of radiating elements, Antenna\_{4} has four radiating elements, which can send and receive data, creating four data streams. As the antenna experiences high isolation between its radiating elements, the provided data streams are nearly independent and can potentially increase the channel capacity of the radiating element by 400\%. However, Refs.\citen{ref26,ref27,ref28,ref35,ref36,ref37} only provides up to two independent data streams and can potentially increase the channel capacity of their radiating elements by 200\%. Concerning the ECC, the manufactured antenna has lower ECC values than other studies except Refs.\citen{ref32,ref34}. The maximum ECC values of Antenna\_[4} and Refs.\citen{ref32,ref34} are the same. It shows that the proposed MIMO structure enjoys low mutual couplings between their radiating elements, achieving high channel capacity, one of the fundamental requisites of the 5G technology. Concerning the radiation efficiency, the simulation radiation efficiency of Antenna\_{4} is superior to other studies. Regarding the dimensions, the aperture of the proposed MIMO antenna is larger than that of other studies. The dimension of a MIMO antenna is determined by the dimension of its radiating element. As elaborated before, according to $G=10\times{\log_{10}{{\frac{4\pi{A\eta}}{\lambda^{2}}}}}$, where "A" is the antenna aperture, "G" is the gain in (dB), and "$\eta$" is the aperture efficiency Ref.\citen{ref46}. The aperture efficiency of the proposed radiating element changes from 77 to 15\% and 77 to 50\% over 2.8 to 5 GHz and 3 to 4 GHz, respectively, as seen in Fig.\ref{fig:fig2(f)}, which are high. For example, the radiating element obtains 67\% aperture efficiency and 9 dBi utmost gain with dimensions equal to $0.9\lambda^{2}$ at 3.5 GHz. Therefore, the dimensions of the radiating element are correctly chosen to achieve high gain and aperture efficiency valeus, and the dimensions of Antenna\_{4} are not large although Antenna\_{4} is bigger than other studies. Furthermore, the height of Antenna\_{4} is larger than other studies. This is mainly due to the use of the aperture-coupled feeding technique, which introduces the air gap. However, the applied feeding mechanism isolates the feeding layer from the radiating layer, facilitating the optimization process and boosting the isolation between ports in the MIMO structure. Conclusively, based on the above achievements, the proposed MIMO designs are among the best candidates for 5G applications, such as vehicular communications, smart industries, and IoT applications, and the above comparison verifies it.

\begin{table*}[!t]
\caption{\label{tab:Table3} The values of the parameters displayed in Figs.{\ref{fig:fig2(a)}-\ref{fig:fig2(e)}} and Figs.{\ref{fig:fig6(a)}-\ref{fig:fig6(e)}}, indicating the dimensions of the manufactured antenna.}
\centering
 \scalebox{0.75}{
\begin{tabular}{lllllllllllll}
\hline
H (mm) & T (mm) & P (mm) & K (mm) & Wf1 (mm) & Wf2 (mm) & Wf3 (mm) & lf (mm) & lf2 (mm) & lf3 (mm) & Ws (mm) & Ls (mm)& Rs (mm)\\
\hline
13 & 0.508 & 86 & 0 & 0.683 & 0.453 & 0.327& 55.61 & 18.81 & 7.51 & 7.86& 46.16 & 12.93 \\
\hline
 R (mm) & W (mm) & L (mm) & a (mm) & b (mm) & c (mm) & d (mm) & e (mm) & f (mm) & M (mm) & S (mm)& N (mm)& L1 (mm)\\
\hline
 8.22& 31 & 13.29 & 7.72 & 13.29 & 11.09 & 2.51 & 5.89 & 8.91& 20 & 15.2 & 4.07& 178 \\
\hline
 D1 (mm) & D2 (mm) & D3 (mm) & D4 (mm) & D5 (mm) & D6 (mm) & D7 (mm) & D8 (mm) & C1 (mm) & B1 (mm) & A1 (mm)& g (mm)& -- \\
\hline
 39.35& 35.33 & 42.068 & 6.95 & 45.83 & 45.83 & 10.78 & 10.78 & 76.55& 46.048 & 68.69 & 4& -- \\
 \hline
\end{tabular}
}
\end{table*}

 \begin{table*}[!t]
\caption{\label{tab:Table4} Achievement analogies between the presented MIMO antennas and other studies. }
\centering
 \scalebox{0.55}{
\begin{tabular}{llllllllll}
\hline
Ref & Polarization&-10 dB Bandwidth (GHz) &  Gain range (dBi) & Isolation (dB) & Num.Ports & Num. Radiating Elements & ECC & Rad.Eff (\%)Size ($mm\times{mm}$$\times{mm}$) \\
\hline
Ref.\citen{ref20} &LP& 3.6-5.3;6.4-10 & 1 to 7 & <-15 & 4 & 4 & <0.0015&---& $60\times60$$\times1.6$\\
\hline
Ref.\citen{ref21} &LP& 2.7-5.1;5.9-12 & 2.5 to 6 & <-17 & 4 & 4 & ---&---& $50\times39.8$$\times1.524$\\
\hline
Ref.\citen{ref22} &LP& 3.1-10.6 & -3 to 4 & <-20 & 4 & 4 & <0.2&92-96& $40\times43$$\times1$\\
\hline
Ref.\citen{ref26} &LP& 1.66-2.17 & 2.5 to 2.9 & <-10 & 2 & 2 & <0.23&>96& $68\times98$$\times1.524$\\
\hline
Ref.\citen{ref27} &LP& 5.8 & 2.5 to 2.9 & <-70 & 2 & 2 & <0.03&---& $7.3\times5.8$$\times1.6$\\
\hline
Ref.\citen{ref28} &LP& 3.3-4.5 & 6.5 to 7.5 & <-15 & 4 & 1 & <0.3&84-92& $\pi(75)^{2}$$\times10.4$\\
\hline
Ref.\citen{ref29} &LP& 2.4 & 2.4 & <-25 & 4 & 4 & <0.03&77& $26\times26$$\times0.8$\\
\hline
Ref.\citen{ref30} &LP& 2.4 & 2.84 & <-58.87 & 4 & 4 & <0.0054&90& $45.5\times45.5$$\times1.96$\\
\hline
Ref.\citen{ref31}&LP& 3.4-3.7 & --- & <-15 & 8 & 4 & <0.03&50-68& $130\times70$$\times0.8$\\
\hline
Ref.\citen{ref32}&LP& 2.4-2.52;3.66-4;4.62-5.54 & 1 & <-30.5 & 4 & 4 & <0.001&85& $23.5\times83$$\times1.6$\\
\hline
Ref.\citen{ref34}&LP& 3.27-3.82 & 8.7 & <-32 & 4 & 4 & <0.001&92-96& $146\times146$$\times3.048$\\
\hline
Ref.\citen{ref35}&LP& 4.65-4.97;4.67-4.94 & 1.83 and 1.65 & <-15 & 2 & 2 & <0.02& 48-53;58-59& $70\times25$$\times1.48$\\
\hline
Ref.\citen{ref36}&LP& 3.7-4.3 & 3 to 4.1 & <-25 & 2 & 2 & ---&---& $64\times36$$\times2.4$\\
\hline
Ref.\citen{ref37}&LP& 3.296-5.962 & -1 to 6.22 & <-50 & 2 & 2 & 0.05& 42-85& $30\times18$$\times1.6$\\
\hline
Ref.\citen{ref44}&LP&3.34-5.01&{3 to 4}& <-20 & {4} & {4} & {<0.017}& {68}& {$55\times55$$\times0.2$}\\
&&{8.9-9.2}&&&&&&&\\
\hline
{Ref.\citen{ref45}}&{LP/CP}& {2.38-2.62 (LP);}{3.3-4.4 (CP)}& {4 to 4.7} & {<-20} & {4} & {4} {red}{ <0.04}&{85}& {$45\times38$$\times0.2$}\\
&&{4.98-5.9 (LP)}&&&&&&&\\
\hline
{Ref.\citen{ref46}}&{CP}&{7.9-9.59}&{2 to 3.5}&{<-18}&{4}&{4}&{<0.01}&{70-78}& {$36\times27$$\times1.6$}\\
\hline
{Ref.\citen{ref52}}&{LP}&{23-31}&{5 to 10}&{<-21}&{4}&{4}&{<0.0012}&{>87}& {$20\times20$$\times0.88$}\\
\hline
{Ref.\citen{ref53}}&{LP}&{1.9-20}&{4.5 to 8}&{<-25.5}&{4}&{4}&{<0.002}&{77-89}& {$41\times41$$\times1.6$}\\
\hline
Third Antenna &LP&2.76-4.3 & 6.2 to 10.5 (3-4 GHz) & <-23 & 4 & 4 & 0.001& >98 (Simulation)& {$178\times178$$\times14$}\\

\hline
\end{tabular}
}
\end{table*}

\section*{Conclusion}
This paper suggests a metasurface-based radiating element with optimized dimensions to achieve a wideband, high gain and highly efficient performance in the sub 6 GHz 5G frequency spectrum. It uses the aperture-coupled feeding technique with a dumbbell-shaped slot, a truncated square patch with two U-shaped slots, and a metasurface layer. Four $178\times178\times14$ $(mm)^{3}$ 4T4R MIMO antennas are designed to operate from 2.8 to 4.3 GHz, where most 5G technologies offer their services based on the proposed radiating element. These elements are placed with $90^\circ$ successive rotations and 6 mm gaps. Moreover, two horizontal and vertical strip slots are carved on the ground of the MIMO antennas to create the decoupling structure. Based on the simulation results, the antennas achieve high gain, diversity gain, and very low mutual coupling and ECC. Antenna\_{4} is manufactured and measured as proof of material. According to the measurement results, which concur well with the simulation results, the proposed MIMO structure achieves 6.28 to 10.45 dBi (3 to 4 GHz), below -23 dB isolation, 0.001 ECC, $-9.2<MEG_{1}<-6 dB$, CCL<0.4 (Bits,s,Hz)from 2.8 to 4.2 GHz. The proposed MIMO designs can materialize reliable wireless communication with increased channel capacity, high data rate, signal quality, and data throughput, low latency, multi-pass fading effects, power, and penetration loss. The analogy of the achievements of the proposed antennas with other studies shows that the presented antennas are among the best choices for 5G applications, such as vehicular communications (e.g., rooftop antennas of cars or trains), smart factories, and IoT applications.

\section*{Data Availability}
Data underlying the results will be available upon request.

\section*{Additional information}

Authors proclaim no competing of interest.

\section*{Author Contributions statement}

This project was done under the guidance of Prof. Oraizi as the supervisor. Dr. Homayoon Oraizi suggested the project. Mr. Salehi, as the first author, proposed the method and performed the simulations. All authors analyzed the measurement and simulation results. In addition, they reviewed and revised the manuscript.

\bibliography{sample}

\end{document}